%% file: mnras.tex
\documentclass[fleqn,usenatbib]{mnras}

\usepackage{newtxtext,newtxmath}

\usepackage[T1]{fontenc}

\DeclareRobustCommand{\VAN}[3]{#2}
\let\VANthebibliography\thebibliography
\def\thebibliography{\DeclareRobustCommand{\VAN}[3]{##3}\VANthebibliography}

\usepackage{graphicx}	
\usepackage{graphicx}	
\usepackage{amsmath}	
\usepackage{mathtools}
\usepackage{xspace}
\usepackage{graphicx}
\usepackage{graphbox}
\usepackage{bbm}
\usepackage{soul}
\usepackage{float}
\usepackage{booktabs}
\usepackage{tabularx}
\usepackage{multirow}
\usepackage{braket}
\usepackage{threeparttable} 

\newcolumntype{I}{@{\extracolsep{\fill}}c}

\newcommand{\gs}{\textsc{GravSphere}\xspace}
\newcommand{\gnn}{\textsc{GraphNPE}\xspace}
\newcommand{\latte}{Latte\xspace}
\newcommand{\rockstar}{\textsc{rockstar}\xspace}

\newcommand\Tstrut{\rule{0pt}{2.6ex}}         
\newcommand\Bstrut{\rule[-1.6ex]{0pt}{0pt}}   

\newcommand{\msun}{\ensuremath{\mathrm{M}_\odot}\xspace}

\newcommand{\mf}{\texttt{m12f}\xspace}
\newcommand{\mm}{\texttt{m12m}\xspace}

\newcommand{\mfsidmt}{\texttt{m12f\_SIDM10}\xspace}
\newcommand{\mmsidmt}{\texttt{m12m\_SIDM10}\xspace}
\newcommand{\vmax}{\ensuremath{V_\mathrm{max}}\xspace}

\newcommand{\mhl}{\ensuremath{M_{r_{\star, 1/2}}}\xspace}
\newcommand{\rhl}{\ensuremath{r_{\star, 1/2}}\xspace}
\newcommand{\mvir}{\ensuremath{M_\mathrm{200m}}\xspace}
\newcommand{\vlos}{\ensuremath{v_\mathrm{los}}\xspace}
\newcommand{\sigmalos}{\ensuremath{\sigma_\mathrm{los}}\xspace}
\newcommand{\rhodm}{\ensuremath{\rho_\mathrm{dm}}\xspace}
\newcommand{\Sigmastar}{\ensuremath{\Sigma_\star}\xspace}
\newcommand{\mvirp}{\ensuremath{M_\mathrm{200m}^\mathrm{peak}}\xspace}
\newcommand{\nperi}{\ensuremath{N_\mathrm{peri}}\xspace}
\newcommand{\rperi}{\ensuremath{r_\mathrm{peri}}\xspace}
\newcommand{\tperi}{\ensuremath{t_\mathrm{peri}}\xspace}
\newcommand{\rcurr}{\ensuremath{r_\mathrm{curr}}\xspace}

\newcommand{\modot}{\ensuremath{\mathrm{M_\odot}}\xspace}
\newcommand{\kms}{\ensuremath{\mathrm{km/s}}\xspace}
\newcommand*\diff{\mathop{}\!\mathrm{d}}

\title[GraphNPE on FIRE]{Trial by FIRE: Probing the dark matter density profile of dwarf galaxies with GraphNPE}

\author[Nguyen et al.]{Tri Nguyen,$^{1, 2, 3, 4, 5}$\thanks{E-mail: trivtnguyen@northwestern.edu}
Justin Read,$^{6}$
Lina Necib,$^{4, 5}$
Siddharth Mishra-Sharma$^{3, 5, 7, 8}$\thanks{Currently at Anthropic; worked performed while at MIT/IAIFI.},
\newauthor
Claude-André Faucher-Giguère$^{1, 2, 9}$,
Andrew Wetzel$^{10}$,
and Tjitske K. Starkenburg$^{1, 2, 9}$
\\
$^{1}$Center for Interdisciplinary Exploration and Research in Astrophysics, Northwestern University, 1800 Sherman Ave, Evanston, IL 60201\\
$^{2}$NSF-Simons AI Institute for the Sky, 172 E. Chestnut St., Chicago, IL 60611, USA\\
$^{3}$Department of Physics, Massachusetts Institute of Technology, Cambridge, MA 02139, USA\\
$^{4}$Kavli Institute for Astrophysics and Space Research, Massachusetts Institute of Technology, Cambridge, MA 02139, USA\\
$^{5}$NSF AI Institute for Artificial Intelligence and Fundamental Interactions, Cambridge, MA 02139, USA\\
$^{6}$Department of Physics, University of Surrey, Guildford, GU2 7XH, UK\\
$^{7}$Center for Theoretical Physics, Massachusetts Institute of Technology, Cambridge, MA 02139, USA\\
$^{8}$Department of Physics, Harvard University, Cambridge, MA 02138, USA\\
$^{9}$Department of Physics \& Astronomy, Northwestern University, 2145 Sheridan Road, Tech F165, Evanston, IL 60208, USA\\
$^{10}$Department of Physics \& Astronomy, University of California, Davis, CA 95616, USA\\
}

\date{Accepted XXX. Received YYY; in original form ZZZ}

\pubyear{\the\year{}}

\begin{document}
\label{firstpage}
\pagerange{\pageref{firstpage}--\pageref{lastpage}}
\maketitle

\begin{abstract}
The Dark Matter (DM) distribution in dwarf galaxies provides crucial insights into both structure formation and the particle nature of DM. 
GraphNPE (Graph Neural Posterior Estimator), first introduced in Nguyen et al. (2023), is a novel simulation-based inference framework that combines graph neural networks and normalizing flows to infer the DM density profile from line-of-sight stellar velocities. 
Here, we apply GraphNPE to satellite dwarf galaxies in the FIRE-2 Latte simulation suite of Milky Way-mass halos, testing it against both Cold and Self-Interacting DM scenarios.
Our method demonstrates superior precision compared to conventional Jeans-based approaches, recovering DM density profiles to within the 95\% confidence level even in systems with as few as 30 tracers. 
Moreover, we present the first evaluation of mass modeling methods in constraining two key parameters from realistic simulations: the peak circular velocity, \vmax, and the peak virial mass, \mvirp.
Using only line-of-sight velocities, GraphNPE can reliably recover both \vmax and \mvirp within our quoted uncertainties, including those experiencing tidal effects ($\gtrsim 63\%$ of systems are recovered within our 68\% confidence intervals and $\gtrsim 92\%$ within our $95\%$ confidence intervals).
The method achieves $10-20\%$ accuracy in \vmax recovery, while \mvirp is recovered to $0.1-0.4 \, \mathrm{dex}$ accuracy. 
This work establishes GraphNPE as a robust tool for inferring DM density profiles in dwarf galaxies, offering promising avenues for constraining DM models.
The framework's potential extends beyond this study, as it can be adapted to non-spherical and disequilibrium models, showcasing the broader utility of simulation-based inference and graph-based learning in astrophysics.
\end{abstract}


\section{Introduction}
\label{section:intro}

Dark matter (DM) makes up about 85\% of matter in the Universe, playing an integral role in the formation and evolution of galaxies and clusters~\citep{1978MNRAS.183..341W, 1997ApJ...490..493N, 2000ApJ...538..528S, 2003ApJ...588..680D, 2005Natur.435..629S}.
Despite its significance, the particle nature of DM remains among the greatest outstanding questions in modern astrophysics and cosmology. 
While the standard Lambda Cold Dark Matter ($\Lambda$CDM) paradigm has successfully explained the large-scale structure of the Universe, such as the fluctuations in the cosmic microwave background (CMB) ~\citep[e.g.][]{2013ApJS..208...20B, 2020A&A...641A...6P} and the distribution of galaxy clusters~\citep[e.g.][]{2008ApJ...688..709T, 2010ApJ...708..645R}, discrepancies arise at the small scales.
Observations of dwarf galaxies in the Local Group have uncovered small-scale challenges, including the core-cusp problem~\citep{1994ApJ...427L...1F, 1994Natur.370..629M, 1996MNRAS.283L..72N, 2004MNRAS.351..903G, 2005MNRAS.356..107R, 2005AJ....129.2119S, 2011AJ....141..193O, 2014Natur.506..171P,  2015AJ....149..180O, 2019MNRAS.484.1401R, 2024MNRAS.535.1015D}, missing satellites problem~\citep{1999ApJ...522...82K, 1999ApJ...524L..19M, 2019MNRAS.487.5799R}, too-big-to-fail~\citep{2011MNRAS.415L..40B}, satellite plane~\citep{1976RGOB..182..241K, 1976MNRAS.174..695L}, and diversity of shapes problem~\citep{Oman2015, Creasey2017, Zavala2019b, Hayashi2020}~\citep[for a review, refer to ][]{2017ARA&A..55..343B}.
These challenges can be partially addressed by a more complete understanding of baryonic processes~\citep{2010Natur.463..203G, 2012MNRAS.421.3464P, 2017MNRAS.471.3547F, 2017MNRAS.471.1709G, 2018PhRvL.121u1302K, 2018MNRAS.478..548S, 2019MNRAS.487.1380G, 2019MNRAS.483.1314B, 2022NatAs...6..897S}, environmental effects such as tidal interaction with the Milky Way and other galaxies~\citep{2011MNRAS.415L..40B, 2006MNRAS.369.1021M, 2006MNRAS.367..387R}, and survey selection effects~\citep{2008ApJ...688..277T, 2020ApJ...893...47D}. 
However, open questions still remain~\citep{2015ApJ...815...19P, 2020MNRAS.495...58S}, motivating alternative theories of DM such as self-interaction~\citep{2000PhRvL..84.3760S}. 

Characterized by their high mass-to-light ratio, small size, and proximity to the Galaxy, the Milky Way's satellite galaxies present an exciting frontier for exploring and testing the particle nature of DM.
Key observables, including the satellite's inner DM density profile, their mass function, luminosity function, and phase-space distribution, can offer stringent constraints on DM properties, such as self-interaction cross-section~\citep{2013MNRAS.430...81R, 2016PhRvL.116d1302K, 2018PhR...730....1T, 2018MNRAS.481..860R, 2021MNRAS.503..920C, 2022MNRAS.511.1757T, 2023A&A...678A..38J}, DM dissipation~\citep{2010JCAP...05..021K,2013PhRvD..87j3515C,2015PhLB..748...61F, 2015PhRvD..91b3512F, 2016JCAP...07..013F, 2021MNRAS.506.4421S,2024ApJ...966..131S,2024ApJ...967...21G,2024arXiv240815317R}, DM-baryon interaction~\citep{2019ApJ...878L..32N}, and mass of warm and fuzzy DM~\citep{2001ApJ...556...93B, 2014MNRAS.442.2487K, 2014MNRAS.439..300L, 2016MNRAS.455..318B, 2021MNRAS.501.1188A, 2021ApJ...912L...3H, 2022arXiv220305750D, 2024arXiv240918917T}.

The inner density profile of DM in a galaxy is affected by two sources: the DM particle properties and interactions, and the baryonic processes.
In the absence of baryons, cold DM models are expected to exhibit a Navarro-Frenk-White (NFW) profile, where the inner profile is $r^{-1}$ \citep{1997ApJ...490..493N}. 
Self-interacting DM (SIDM) models, however, largely predict a cored profile, where $r \propto r^0 \propto \mathtt{constant}$ in the inner galaxy\footnote{In some cases, there are scenarios where gravothermal collapse occurs, leading to a strong cusp in the center of galaxies with SIDM~\citep[e.g.][]{2002ApJ...568..475B, 2011MNRAS.415.1125K, 2020PhRvD.101f3009N}.}~\citep{2018PhR...730....1T}. 
Dissipation can instead lead to stronger cusps, for example \citet{2021MNRAS.506.4421S} found cusps that scale as $r^{-1.5}$. 

Baryonic processes, such as feedback, can also significantly influence the density profiles~\citep[e.g.][]{2014MNRAS.437..415D, 2014MNRAS.441.2986D, 2019MNRAS.484.1401R, 2024MNRAS.535.1015D}.
Feedback can lead to a diverse range of cored and cuspy profiles, depending on where a galaxy is in its cycle of gas inflows and outflows~\citep{2016ApJ...820..131E, 2017ApJ...835..193E, 2025MNRAS.536..314M}.
However, if the gas is stripped, such as in Milky Way satellites, the galaxy can be ``frozen into'' a cored state.
This effect is particularly relevant for a specific mass range, including classical dwarfs and Milky Way-mass galaxies~\citep[e.g.][]{2020MNRAS.497.2393L}. 
Consequently, dwarf galaxies are the ideal laboratory for investigating the inner profile affected solely by the particle nature of DM~\citep[e.g.][]{2021A&A...651A..80Z, 2021arXiv211209374Z}.
This is particularly true for ultra-faint dwarfs ($M_V \gtrsim -7.7$), which experience weaker stellar feedback. 

Apart from serving as a direct probe of the particle nature of DM through its inner slope of the DM density profile, Dwarf galaxies are an excellent target for indirect detection of DM.
They are relatively close, making the annihilation/decay rates of DM into standard model particles quite high and more easily detectable, and generally devoid of astrophysical backgrounds that might mimic DM signals (e.g., the potential population of millisecond pulsars near the Galactic Center muddling the waters as to whether DM is the source of the Galactic center excess~\citep{2011JCAP...03..051C, 2011PhRvL.107x1302A, 2011PhRvL.107x1303G, 2012APh....37...26M, 2014PhRvD..89d2001A, 2015MNRAS.451.2524M, 2015PhRvL.115w1301A, 2015PhRvD..91h3535G, 2015ApJ...801...74G, 2015arXiv150203081A, 2019arXiv190408430L, 2024AAS...24331503A}.

Given the critical information that is encoded in the density profile of DM, one needs to work through recovering such profiles, particularly in dwarf galaxies where the effect of baryons is minimal. 
To tackle this problem, one could recover the underlying gravitational potential of the galaxy through the dynamics of its stars.
Traditionally, this has been done using the Jeans dynamical modeling techniques~\citep{1915MNRAS..76...70J, 2015MNRAS.446.3002B}, which relate the velocity dispersion of tracer stars within these galaxies to their DM-dominated gravitational potential. 
Jeans methods are straightforward to implement. 
Most models assume spherical symmetry, which has been demonstrated to be a reasonable approximation on non-spherical mocks with less than $\sim 10,000$ tracer stars~\citep{2017MNRAS.471.4541R, 2020MNRAS.498..144G}.
Some models have explored axisymmetric fits to nearby dwarfs~\citep{2020ApJ...904...45H, 2023ApJ...953..185H}.
Even highly tidally disrupting dwarfs have been shown to be successfully modeled, providing contaminating tidal debris can be correctly distinguished from bound stars~\citep{2018MNRAS.481..860R, 2024MNRAS.535.1015D}.

A long-standing challenge for Jeans methods, applied to just line-of-sight velocity data, has been \textit{mass-anisotropy degeneracy}~\citep{1990AJ.....99.1548M, 2002MNRAS.330..778W, 2003MNRAS.343..401L, 2009MNRAS.395...76D, 2017MNRAS.471.4541R, 2017ApJ...835..193E, 2020MNRAS.498..144G}.
This refers to a degeneracy between the enclosed mass, that we want to infer, $M(<r)$, and the velocity anisotropy parameter $\beta(r)$ ($\beta = 0, 1, -\infty$ correspond to an isotropic, radial anisotropy, and tangential anisotropy velocity dispersion, respectively; \citealt{2008gady.book.....B}). 
However, this degeneracy can be broken in a number of ways.
Firstly, where available, proper motion data can be used~\citep{2007ApJ...657L...1S, 2017MNRAS.471.4541R, 2018ApJ...860...56S, 2020A&A...633A..36M, 2024ApJ...970....1V}.
Secondly, we can use higher order moments of the velocity distribution, for example through ``Virial Shape Parameters''~\citep{1990AJ.....99.1548M, 2009MNRAS.394L.102L, 2014MNRAS.441.1584R, 2017MNRAS.471.4541R, 2013MNRAS.429.3079M}.
Thirdly, we can simultaneously fit multiple tracer populations moving within the same gravitational potential~\citep[e.g.][]{2011ApJ...742...20W, 2012MNRAS.419..184A, 2016MNRAS.463.1117Z, 2017MNRAS.471.4541R}. 
And finally, we can move away from Jeans modeling to Schwarzschild~\citep[e.g.][]{2013A&A...558A..35B, 2014ApJ...791L...3B, 2019MNRAS.482.5241K} or distribution function modeling~\citep{2002MNRAS.330..778W, 2018MNRAS.480..927P, 2021MNRAS.501..978R}.
These latter methods simultaneously fit all moments of the velocity distribution function, $f(v)$, by assuming some functional form for $f$ (or a range of functional forms).

Distribution function methods maximize the information content in the data, while providing a natural way to incorporate survey selection functions and binary star contamination~\citep[e.g.][]{2018AJ....156..257S, 2022ApJ...929...77K}.
However, previously proposed methods require models that are expensive to compute, limiting the parameter space that can be explored, especially as we move beyond spherical models. 
Traditional distribution function methods also provide no natural route to moving beyond the assumption of pseudo-dynamical equilibrium. 

In this paper, we further improve and test a new method, \gnn, a simulation-based inference (SBI) framework that can massively accelerate distribution function modeling. 
First introduced by~\citealt{2023PhRvD.107d3015N} (hereafter N23), \gnn employs normalizing flows~\citep{2016arXiv160506376P, 2019arXiv191202762P} and graph neural networks~\citep{4700287, 2016arXiv160902907K, 2017arXiv170401212G, 2018arXiv180601261B} to parametrically model the density profiles. 
SBI uses complex simulations to implicitly model the likelihood, thus directly incorporating these simulations into the inference process ~\citep[e.g.][]{2020PNAS..11730055C}.
In this work, we train \gnn on idealized simulations of dwarf galaxies that are spherical and in equilibrium. 
However, we emphasize that the framework can be readily adapted to analyze non-spherical or disequilibrium mock datasets, which we plan to explore in future studies.

Mass modeling methods have been extensively tested on mock data, including realistic triaxial mocks~\citep{2017MNRAS.471.4541R}, tidally stripped mocks~\citep{2018MNRAS.481..860R, 2024MNRAS.535.1015D}, and cosmologically realistic satellites~\citep{2020MNRAS.498..144G}. 
\citet{2020MNRAS.498..144G} applied the \gs, a higher-order moment Jeans modeling code~\citep{2017MNRAS.471.4541R}, to satellites from the APOSTLE simulations~\citep{2016MNRAS.457.1931S}, examining the impact of tidal stripping on the recovery of mass profiles. 
Their focus was on recovering central DM densities, cusp slopes, and half-stellar radius mass $\mhl$. 

In this work, we test \gnn on similarly realistic dwarf galaxies data in Milky Way-like environments, using the \latte suite \citep{2016ApJ...827L..23W} from the Feedback In Realistic Environments\footnote{\url{https://fire.northwestern.edu/}} (FIRE-2; \citealt{2018MNRAS.480..800H}) simulations.
We will use the cold DM suite of \citet{2016ApJ...827L..23W, 2023ApJS..265...44W} as well as the self-interacting DM suite from~\citet{2021MNRAS.507..720S, 2022MNRAS.516.2389V,  2024ApJ...974..223A}. 
In doing so, we aim to assess the robustness of \gnn against phenomena such as tidal interactions and validate their reliability in a realistic astrophysical environment, for both cored and cuspy profiles.
Moving forward, the ultimate goal is to apply \gnn to real observational datasets of dwarf galaxies in the Local Group. 

We also investigate the recovery of the peak circular velocity of halos, \vmax, and their peak virial mass, \mvirp--both of key interest for constraining cosmological models~\citep[e.g.][]{1995ApJ...447L..25B, 2001MNRAS.321..559B, 2003MNRAS.345..923V, 2016MNRAS.462..893R, 2019MNRAS.487.5799R, 2021arXiv210609050K}. 
This is the first study to evaluate how well mass modeling methods can constrain \vmax and \mvirp for simulated dwarf galaxies in realistic environments. 
This test is particularly timely given that two state-of-the-art mass modeling methods in the literature, \textsc{CJAM} (Jeans Anisotropic Multi-Gaussian Expansion;~\citealt{2013MNRAS.436.2598W}) and \gs, yield quite different inferences when applied to the same data~\citep{2021arXiv211209374Z}.

This paper is structured as follows. 
Section~\ref{section:simulation} provides an overview of the \latte suite of FIRE-2 simulations and the procedure for constructing the dwarf galaxy dataset. 
Section~\ref{section:method} details the forward modeling approach and the machine learning architecture of \gnn. 
Section~\ref{section:comparison} evaluates \gnn against two Jeans-based methods on a subset of FIRE-2 dwarf galaxies: a simple Jeans model using an unbinned Gaussian likelihood (Section~\ref{section:jeans_gnn}) and \gs (Section~\ref{section:gs_gnn}). 
Section~\ref{section:tidal} presents the \gnn result on the full FIRE-2 dataset, under varying degrees of tidal effects.
Section~\ref{section:cdm_sidm} provides a brief comparison of the performance between the CDM and SIDM samples.
Section~\ref{section:discussion} discusses differences between the simulations and real observations (Section~\ref{section:observation}) and compares \gnn with other mass modeling methods (Section~\ref{section:df}).
Finally, Section~\ref{section:conclusion} concludes the paper.

\section{FIRE-2 Suites of Simulation}
\label{section:simulation}

\subsection{FIRE-2 Simulations}

In this paper, we aim to test the validity of \gnn on more realistic scenarios than previously done in N23, with the end goal of applying \gnn to the Milky Way's satellites. To that end, we use the satellite population of the Milky Way-mass simulated galaxies as part of the \latte suite~\citep{2016ApJ...827L..23W}. 
The \latte suite is a set of zoom-in simulations of Milky Way-mass galaxies, in which the hydrodynamics is based on the FIRE-2 physics~\citep{2018MNRAS.480..800H}. 
In order to ensure robustness of the results with varying density profiles, we will use the CDM galaxies~\citep{2016ApJ...827L..23W, 2023ApJS..265...44W}, along with SIDM galaxies presented in~\citep{2021MNRAS.507..720S, 2022MNRAS.516.2389V, 2024ApJ...974..223A}. 
We discuss each of these properties in turn. 

\textbf{FIRE-2 physics:} The simulations are run using the GIZMO\footnote{\url{https://bitbucket.org/phopkins/gizmo-public}} code base~\citep{2014ascl.soft10003H, 2015MNRAS.450...53H}, which utilizes a mesh-free, finite-mass (MFM) hydrodynamic solver, along with a version of the Tree-PM gravity solver from GADGET-3~\citep{2005MNRAS.364.1105S}.
MFM is a mesh-free Lagrangian finite-volume Godunov method that provides adaptive spatial resolution while maintaining conservations of mass, energy, momentum, and angular momentum.
It integrates the advantages of both smoothed particle hydrodynamics (SPH) and traditional grid-based methods, enabling adaptive resolution and the accurate capture of fluid dynamics in highly complex astrophysical phenomena. 
The FIRE-2 physics model~\citep{2018MNRAS.480..800H} additionally includes star formation~\citep{2013MNRAS.432.2647H}, stellar feedback from radiation pressure, supernovae Ia and II, stellar winds, photoelectric heating, and photoionization, along with a detailed treatment of cooling, heating, UV background~\citep{2009ApJ...703.1416F}, metal mixing in the interstellar medium~\citep{2017MNRAS.471..144S,2018MNRAS.474.2194E} and enrichment from stars and supernovae~\citep{2016MNRAS.456.2140M}.

\textbf{\latte simulations:} The \latte suite, first introduced in~\citet{2016ApJ...827L..23W}, consists of isolated Milky Way-mass halos with $M_\mathrm{200m} = 1-2\times10^{12} \, \msun$ at $z=0$~\citep{2018ApJ...869...12S,2023ApJS..265...44W}.
These halos are selected from a periodic box with a length of $85.5 \, \mathrm{Mpc}$ as part of the AGORA project~\citep{2014ApJS..210...14K}. 
The simulations start at $z=99$, with initial conditions generated using the MUlti-Scale Initial Conditions code (MUSIC; \citealt{2011MNRAS.415.2101H}).

\textbf{Cosmology:} The simulations adopt a standard $\Lambda\mathrm{CDM}$ cosmology with parameters consistent with the Planck 2018 results~\citep{2020A&A...641A...6P}.
Specifically, the hosts are run with AGORA cosmology ($\Omega_\Lambda = 0.728$, $\Omega_m = 0.272$, $\Omega_b = 0.0455$, $\sigma_8 = 0.807$, $n_s = 0.961$, $h = 0.702$).
Star particles have an initial mass resolution of $7070 \, \msun$ and a spatial resolution of $4 \, \mathrm{pc}$.
DM particles have a mass resolution of $35,000 \, \msun$ and a spatial resolution of $40 \, \mathrm{pc}$~\citep{2023ApJS..265...44W}.

\textbf{Dark Matter Models:} 
To make sure that we span a variety of inner profiles, we include in our analysis simulations with Cold Dark Matter (CDM) as well as Self-Interacting Dark Matter (SIDM)~\citep[see e.g.][for a review]{2018PhR...730....1T}. 
We analyze the fiducial CDM simulations from the FIRE-2 public data release~\citep{2023ApJS..265...44W} and the SIDM simulations presented in~\citet{2021MNRAS.507..720S, 2022MNRAS.516.2389V, 2024ApJ...974..223A}.\footnote{Prior SIDM simulations using the FIRE-2 galaxy formation model include simulations of $10^{10}\,\modot$ halos~\citep{2017MNRAS.472.2945R, 2019MNRAS.490..962F} and dissipative SIDM scenarios~\citep{2021MNRAS.506.4421S, 2024ApJ...966..131S}; however, these are not included in this study.}
For the SIDM simulations, we use two hosts, \mf and \mm, each simulated with two constant DM self-interaction cross-sections of $\sigma / m = 1$ and $10 \, \mathrm{cm^2 /g}$.
DM self-interaction is implemented using the Monte Carlo scattering algorithm in~\citet{2013MNRAS.430...81R, 2013MNRAS.430..105P}, which determines the probability of interaction using a spline kernel with adaptive smoothing length~\citep{1985A&A...149..135M}.
All scattering events are modeled as elastic and isotropic in the center-of-mass frame.

As discussed in \citet{2021MNRAS.507..720S, 2022MNRAS.516.2389V}, the SIDM simulations are run with a modified version of the FIRE-2 physics model.
Specifically, these simulations ignore the thermal-to-kinetic energy conversion in the unresolved Sedov-Taylor shock expansion phase from mass loss in massive stars.
This leads to reduced star formation rates and lower stellar masses in the simulated galaxies, ultimately also increasing the number of satellites, as shown in Table~\ref{tab:simulations}.
In general, this is advantageous for our study, as it provides another robustness test for our model; however, it complicates direct comparisons between CDM and SIDM scenarios (Section~\ref{section:cdm_sidm}).

\input{table_simulation}

\subsection{Test samples of dwarf galaxies}

Given that our goal is to apply \gnn on Milky Way satellite galaxies, we will focus our studies on the satellites of the Milky Way-mass simulated galaxies. 
In FIRE-2, halos and subhalos are identified using a modified version of the \rockstar\footnote{\url{https://bitbucket.org/awetzel/rockstar-galaxies}} six-dimensional halo finder~\citep{2013ApJ...762..109B}, which accounts for multi-mass and multi-species particles.
For each subhalo, we use the DM particles assigned by \rockstar and identify its associated star particles.
We use a similar procedure to \citet{2019ApJ...883...27N} to assign member star particles: we require that the star particles lie within the current $r_\mathrm{200m}$ (the radius within which the average density is 200 times the mean matter density of the Universe) of the subhalo, and have velocities within $3\sigma$ of the subhalo’s stellar velocity dispersion.

We select subhalos based on their radial distances from the hosts and their halo-to-stellar mass ratios. 
Specifically, we include galaxies located within $3 R_\mathrm{200m}$ of the hosts at $z=0$.
While our primary goal is to apply \gnn to the Milky Way's satellites, extending the selection to more distant dwarf galaxies allows us to assess \gnn's performance across a broader subhalo population and in environments affected by tidal effects (Section~\ref{section:tidal}).

In addition, we select subhalos with halo-to-stellar mass ratios satisfying $M_\mathrm{halo} / M_\mathrm{\star} > 10^{2}$ and with the number of star particles between $N_\mathrm{\star} \in [20, 5000]$.
This roughly corresponds to a stellar mass range of approximately $M_\mathrm{\star} \in [10^5, 10^7] \, \modot$ and a halo mass range of $M_\mathrm{halo} \in [10^7, 10^{10}] \, \modot$.\footnote{Peak halo mass range of $M^\mathrm{peak}_\mathrm{halo} \in [10^8, 10^{10}] \, \modot$.}
The number of stars is chosen to encompass observations of classical and faint dwarfs.
For instance, the highest number of resolved stars observed for Milky Way dwarfs, Fornax and Sculptor, are 2483 and 1365 stars, respectively~\citep{2009AJ....137.3100W}.

Next, we center each subhalo and compute the positions and velocities of star and DM particles of each subhalo relative to the subhalo's DM barycenter.
We identify the center of each subhalo using the shrinking sphere method~\citep{2003MNRAS.338...14P} on the DM particles only.
Starting with an initial guess for the center, we iteratively compute the DM barycenter of all particles within a specified radius. 
At each step, the radius is reduced by 10\%, and the center is updated to the newly calculated barycenter.
The iterations proceed until the center converges to within 0.1\% of the previous iteration.
In rare cases, when a subhalo is located too close to its host or another subhalo, the computed center may still be offset. These cases are manually checked. 

Additionally, the stellar and DM centers are not always perfectly aligned, potentially complicating the analysis.
To quantify this misalignment, we compute the position and velocity offsets between the stellar and DM centers for all subhalos in our sample.
The median position and velocity offsets are $0.17 \, \mathrm{kpc}$ and $1.77 \, \mathrm{km/s}$, respectively, with maximum values of $0.66 \, \mathrm{kpc}$ and $9.52 \, \mathrm{km/s}$.\footnote{We note one extreme outlier with position and velocity offsets of $13.7 \, \mathrm{kpc}$ and $21.5 \, \mathrm{km/s}$, respectively, which was excluded from these statistics.}
For simplicity, we assume these centers coincide.

Finally, in this study, we do not account for observational effects.
This includes measurement uncertainties, survey selection biases, and sources of contamination, such as misidentified member stars and binaries. 
These effects will be explored in future work.

\section{Methodology}
\label{section:method}

\gnn (Graph Neural Posterior Estimator) employs simulation-based inference (SBI; see~\citealt{2020PNAS..11730055C} for a review) to overcome limitations in the likelihood construction and sampling process in Jeans modeling. 
In SBI, the likelihood function is implicitly encoded within the simulations, allowing for more realistic physical processes to be incorporated than analytic models.
Neural posterior estimation (NPE), in particular, leverages neural networks to maximize the information extracted from simulated data and map it directly to posterior distributions~\citep{cranmer_kyle_2016_198541, 2016arXiv160506376P}. 

The \gnn model consists of a graph neural network (GNN;~\citealt{4700287}) for feature extraction and a conditional normalizing flow~\citep{2015arXiv150505770J, 2019arXiv191202762P} for density estimation.
Stellar kinematic data, which consists of line-of-sight velocities and two projected coordinates, are represented as graphs.
During the forward pass, the GNN compresses the graph representation into summary features, which are then used as context by the flow to model the posterior distribution. 

Both the \gnn model and training simulations are largely the same with those introduced in N23.
Below, we summarize key assumptions and a few notable changes.

\subsection{Training Simulation \& Forward Model}

Similar to N23, we generate training samples of dwarf galaxies using a Monte Carlo simulation based on analytic models.
The simulation procedure is as follows:

For each simulated galaxy, we first adopt some parametric phase-space distribution function $f(\vec{x}, \vec{v}; \theta)$ for its tracer stars, where $\vec{x}, \vec{v}$ are the positions and velocities, respectively. 
The parameter set $\theta$ defines the DM density, velocity anisotropy, and tracer density profiles.
We can then generate the population of $N$ tracer stars by independently sampling each star from $f(\vec{x}, \vec{v}; \theta)$. 

The likelihood of each simulated galaxy, given the parameter set $\theta$, is determined by the probability of observing the set of positions and velocities, $\{\vec{x}_i, \vec{v}_i\}$, for each of the $N$ tracer stars, where $i = 1, \dots, N$:
\begin{equation}
    \label{eq:loglike_npe}
    \mathcal{L}_\mathrm{NPE} = \prod_{i=1}^N \mathcal{T}[f](\{\vec{x}_i, \vec{v}_i\}; \theta),
\end{equation}
where $\mathcal{T}$ represents some transformations on $f$ to incorporate any observational effect. 
In our forward model, $\mathcal{T}$ simply projects the 6-D positions and velocities into a 2D projected coordinates and line-of-sight velocities along a random axis.
Any additional observational effects, such as measurement uncertainties, selection biases, etc., can be folded directly into $\mathcal{T}$, which we will explore in future work. 

In the context of NPE, the likelihood informs the relationship between the observed data $\{\vec{x}_i, \vec{v}_i\}$ and the parameter $\theta$.
While the model does not explicitly learn $\mathcal{L}$, it implicitly incorporates this information during training by approximating the posterior distribution $p(\theta~|~\{\vec{x}_i, \vec{v}_i\})$.

We assume the galaxies to be spherically symmetric and in dynamical equilibrium (i.e., the distribution function $f(\vec{x}, \vec{v}; \theta)$ does not explicitly depend on time).
These are common assumptions in traditional Jeans analysis, but as previously discussed, might not be robust against realistic Milky Way dwarfs. 
Ideally, one can construct a more realistic family of distribution functions from hydrodynamic simulations, such as FIRE-2.
However, such simulations are computationally demanding and currently impractical for generating the large training sets required for our framework. 
Moreover, as we will demonstrate in Section~\ref{section:observation}, even high-resolution zoom-in simulations struggle to capture the observed population of dwarf galaxies, particularly the ultra-faints.

In addition, training on hydrodynamic simulations could make the model sensitive to specific features of the simulations, such as baryonic prescriptions or even specific code implementations.
These include differences in subgrid physics models, numerical solvers, resolution, or the treatment of star formation, feedback processes, and gas cooling.
Variations in these implementation choices across simulation codes can lead to divergent results, even when simulating similar physical systems. 
As a result, models trained on one set of simulations might overfit to these implementation-specific details, reducing their capacity to generalize to other simulations or real observational data.

For these reasons, we aim to focus only on investigating whether the additional kinematic information leveraged by \gnn can help improve the robustness of the mass modeling, while keeping the simplified assumptions in Jeans analysis. 

As in N23, we adopt the same parametric function for the DM density profile and tracer density distribution, and update the velocity anisotropy profile.
Namely, we assume the DM density profiles follow a generalized Navarro-Frenk-White profile~\citep{1997ApJ...490..493N}:
\begin{equation}
    \label{eq:gNFW}
    \rhodm^\mathrm{gNFW}(r) = \rho_0 \left(\frac{r}{r_\mathrm{dm}}\right) ^{-\gamma}\left(1 + \frac{r}{r_\mathrm{dm}}\right)^{-(3-\gamma)},
\end{equation}
where $\rho_0$ is the density normalization, $r_\mathrm{dm}$ is the DM scale radius, and $\gamma$ is the inner slope.

The tracer density distribution $\nu(r)$ follows the 3-D Plummer profile~\citep{1911MNRAS..71..460P},
\begin{equation}
    \label{eq:plummer_3-D}
    \nu(r) = \frac{3L}{4 \pi r_\star^3} \left(1 + \frac{r^2}{r_\star^2} \right)^{-5/2},
\end{equation}
where $L$ is the total luminosity and $r_\star$ is the tracer scale radius. 
The tracer mass density $\Sigmastar(R)$ is the projection of $\nu(r)$ along the line-of-sight,
\begin{equation}
    \label{eq:plummer_2D}
    \Sigmastar(R) = \frac{L}{\pi r_\star^2} \left(1 + \frac{R^2}{r_\star^2} \right)^{-2},
\end{equation}
where $R$ is the projected radius.
We assume the stellar contribution to the total mass to be negligible and ignore the luminosity term $L$, which describes the normalization of Eq.~\ref{eq:plummer_3-D} and Eq.~\ref{eq:plummer_2D}.

The velocity anisotropy profile $\beta(r)$ is defined as
\begin{equation}
    \label{eq:beta}
    \beta(r) \equiv 1 - \frac{\sigma^2_t}{\sigma^2_r},
\end{equation}
where $\sigma_r$ and $\sigma_t$ are the radial and tangential velocity dispersions. 
The velocity anisotropy ranges from $-\infty$ to $1$, where $\beta=1, 0, -\infty$ corresponds to a radial, isotropic, and tangential velocity profile, respectively. 
We assume a functional form for $\beta(r)$ following the Osipkov-Merritt (OM) profile~\citep{1979PAZh....5...77O, 1985AJ.....90.1027M},
\begin{equation}
    \label{eq:velani_OM}
    \beta^\mathrm{OM}(r) = \frac{\beta_0 + (r/r_a)^2}{1 + (r/r_a)^2},
\end{equation}
where $\beta_0$ is the normalization and $r_a$ is the anisotropy scale radius.
The scale radius $r_a$ thus determines the transition from $\beta_0$ at small radii to a radially-biased orbits at larger radii.
In contrast to the N23 model, which always set $\beta_0$ to $0$, we let $\beta_0$ be a free parameter within the range $\left[-0.5, 1\right]$.

\input{table_prior}

In total, the current model has three DM parameters ($\rho_0$, $r_s$, $\gamma$) and three stellar parameters ($\beta_0$, $r_a$, $r_\star$). 
We expand the prior range for these parameters from N23 and summarize them in Table~\ref{tab:priors}.
Additionally, we select only galaxies with 3-D velocity dispersions within $0.1-50 \, \kms$ for the final training set, which should sufficiently include the population of classical and ultra-faint dwarfs~\citep{ 2024arXiv241107424P}.\footnote{
Assuming an isotropic velocity distribution, the 3-D velocity dispersion $\sigma_\mathrm{3-D}$ is related to the line-of-sight velocity dispersion $\sigma_\mathrm{los}$ via $\braket{\sigma_\mathrm{3-D}^2} = 3\braket{\sigma_\mathrm{los}^2}$. 
For $\sigma_\mathrm{3-D}$ in the range of $(0.1-50)\,\kms$, $\sigma_\mathrm{los}$ is approximately within $(0.06-28.9) \, \kms$.}
Among Milky Way dwarfs, the highest and lowest observed line-of-sight velocity dispersions--excluding the Small and Large Magellanic Clouds (SMC and LMC)--is $12.1 \pm 0.2 \, \kms$ for Fornax~\citep{2009AJ....137.3100W, 2011AJ....141..194G, 2018ApJ...860...66M, 2019ApJ...881..118W} and $1.2^{+0.9}_{-0.6} \, \kms$ for Tucana V~\citep{2024ApJ...968...21H}.

To construct and sample distribution functions, we use the \textsc{Agama} (Action-based Galaxy Modeling Architecture; \citealt{2019MNRAS.482.1525V}) library. 
\textsc{Agama} is a C++ framework that provides tools for distribution function modeling, potential solving, and orbit integration, with a particular focus on action-based methods, making it well-suited for equilibrium dynamical models.

For each galaxy, we first sample the tracer counts from a Poisson distribution with a mean of 100, then sample that many tracers from the distribution function independently.
We then apply five random projections by selecting a random line-of-sight axis and projecting the galaxy onto the 2D plane perpendicular to this axis. 
We extract only the line-of-sight velocities \vlos and the projected coordinates $(x, y)$ for each galaxy.
We do not consider proper motions here, although they can readily include in future work and can provide additional constraints. 

\subsection{Machine learning framework}

We adopt the same machine learning setup as in N23, with minor updates to the network architecture (see Appendix~\ref{app:ml}). 
The full schematic of our method is available in Figure S1 of N23.
Each galaxy is modeled as an undirected graph consisting of a set of nodes and edges connecting them.  
Each node represents an individual tracer star, with node features given by its projected radius from the center of the galaxy, i.e., $r = \sqrt{x^2 + y^2}$, where $x$ and $y$ are the 2D projected coordinates, and its line-of-sight velocity $v_\mathrm{los}$.  
The edges are constructed by connecting each star to its $k$-nearest neighbors, with $k=20$.  

Representing galaxies as graphs is advantageous because it allows for a \textit{permutation-invariant} neural network architecture, such as a GNN.  
This approach enables the model to efficiently handle varying numbers of tracer stars, which is particularly important for realistic applications, where observed dwarf galaxies have a broad range of tracer counts.

As mentioned, the \gnn model consists of a GNN to extract summary features from the input graph representation, and a conditional flow to model the posterior distributions from these summary features.
The architecture remains largely the same with the N23 model, with a notable change in the graph convolution layers in the GNN. 
Instead of the Chebyshev convolutional layers (ChebConv), the GNN now employs graph attention layers~\citep[GATConv,][]{2017arXiv171010903V}.
These layers use the self-attention mechanism~\citep{2017arXiv170603762V} to learn the edge weighting between connected nodes based on the node features. 
This allows the model to learn the most relevant connections directly from the training data, rather than relying on some predetermined weights or features (e.g. Euclidean distance between nodes, as in \citealt{2022ApJ...937..115V, 2023ApJ...952...69D, 2023arXiv231117141C}).
We found that although both ChebConv and GATConv have similar performance on the validation dataset (as also observed in N23), GATConv are less prone to overfit and more robust against tidal effects (Section~\ref{section:tidal}).

During training, the GNN and normalizing flows are optimized simultaneously using the maximum likelihood objective of the flows (Equation~\ref{eq:loss}).
For additional training details, we refer readers to Appendix~\ref{app:ml}.
We use $5 \times 10^6$ and $5 \times 10^5$ galaxies for the training and validation set, respectively.
The increase in the number of training samples compared to N23 reflects the wider range of the prior distributions and the inclusion of $\beta_0$.
The training converges after approximately 22 hours on a NVIDIA Tesla V100.
Once trained, the model can sample the posterior almost instantaneously, \textit{regardless of the number of tracers}.

\section{Comparison with Jeans-based Methods}
\label{section:comparison}

\subsection{Simple Jeans modeling}
\label{section:jeans_gnn}

We compare the performance of \gnn and Jeans modeling on the FIRE dwarf galaxies.
While Jeans methods are generally considered fast, they can be significantly more time-consuming than \gnn when modeling the full mock dataset. 
As mentioned, once trained, \gnn can draw posterior samples almost instantaneously, whereas Jeans modeling may take several hours to an entire day, especially for galaxies with a large tracer count.
For this reason, we randomly select four galaxies from each FIRE-2 simulation, resulting in a total of 36 galaxies. 

For each galaxy, we apply the fitting procedure outlined in \citet{2008ApJ...678..614S}, which is commonly employed in the literature~\citep[e.g.][]{2015ApJ...801...74G, 2021MNRAS.507.4715C}.
Briefly, we first fit the tracer mass density profile $\Sigmastar(R)$ for $r_\star$, and then perform a joint fit for $\rhodm(r)$ and $\beta(r)$ by modeling the line-of-sight velocity distribution as an unbinned Gaussian likelihood,
\begin{equation}
    \label{eq:loglike_jeans}
    \mathcal{L}_\mathrm{Jeans} = \prod_{i=1}^{N} 
    \frac{(2\pi)^{-1/2}}{\sqrt{\sigma^2_\mathrm{los}(R_i) + \Delta_i^2}}
    \exp\left[
    -\frac{1}{2} \left(\frac{(\vec{v}_i - \braket{v})^2}{\sigma^2_\mathrm{los}(R_i) + \Delta_i^2}\right)
    \right],
\end{equation}
where $R_i$ is the projected radius, $\vec{v}_i$ and $\Delta_i$ are the line-of-sight velocity and its measurement uncertainty of star $i$, and $\braket{v}$ is the mean velocity of the tracer population.
The line-of-sight velocity dispersion profile $\sigma_\mathrm{los}(r)$ is related to $\rhodm(r)$ and $\beta(r)$ via Jeans equations (see Appendix~\ref{app:jeans}).
Since we do not account for measurement uncertainty, $\Delta_i$ is set to zero for all stars.
To ensure consistency with \gnn, the prior distribution is identical for the DM and anisotropy parameters.

For the remainder of this work, we refer to this model as ``Simple Jeans'' to distinguish it from more advanced Jeans-based methods. 
These include, for example, \gs (\citealt{2017MNRAS.471.4541R}, discussed further in Section~\ref{section:gs_gnn}), \textsc{CJAM} (Jeans Anisotropic Multi-Gaussian Expansions; ~\citealt{1994A&A...285..723E, 2008MNRAS.390...71C, 2013MNRAS.436.2598W, 2015arXiv150405533C}), and \textsc{MAMPOSSt}~\citep{2013MNRAS.429.3079M}.

For each galaxy, we also calculate the true density and anisotropy profile from the DM and star particles.
We divide the particles into non-overlapping radial bins containing equal number of particles, and calculate the mean and standard deviation of the relevant quantities within each bin. 
For $\rhodm(r)$, we use bins of $500$ DM particles, which should sufficiently resolve the density profiles.
For $\beta(r)$, we use bins of $50$ star particles, though this number may be reduced slightly when necessary to guarantee at least two bins.

\subsubsection{DM density and velocity anisotropy profiles}
\label{section:jeans_gnn_rho_beta}

\begin{figure*}
    \centering
    \includegraphics[width=\linewidth]{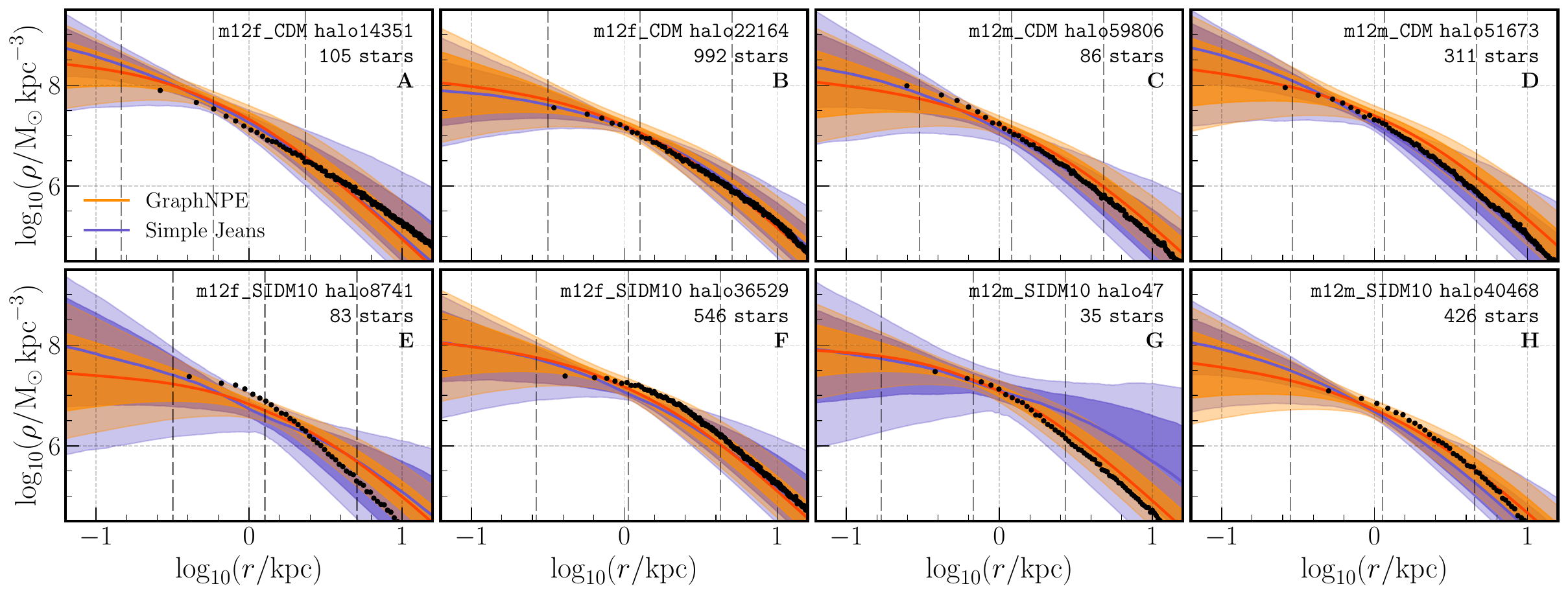}
    \caption{
    Comparison between the inferred DM density profiles $\rhodm(r)$ from \gnn (orange) and Jeans modeling (blue) for a selection of FIRE-2 galaxies. 
    Each panel shows the inferred and true profiles for an individual galaxy. 
    The solid line and shaded bands denote the median, 68\%, and 95\% confidence intervals of the inferred density profiles.
    The true density profiles, calculated directly from the DM particle data, is represented by black circles with error bars.
    The vertical dashed lines indicate $0.25, 1.0, 4.0 \, r_\mathrm{1/2}$, where $r_\mathrm{1/2}$ is the 3-D half-stellar mass radius.  
    }
    \label{fig:rho_gnn_jeans}
\end{figure*}

\begin{figure*}
    \centering
    \includegraphics[width=\linewidth]{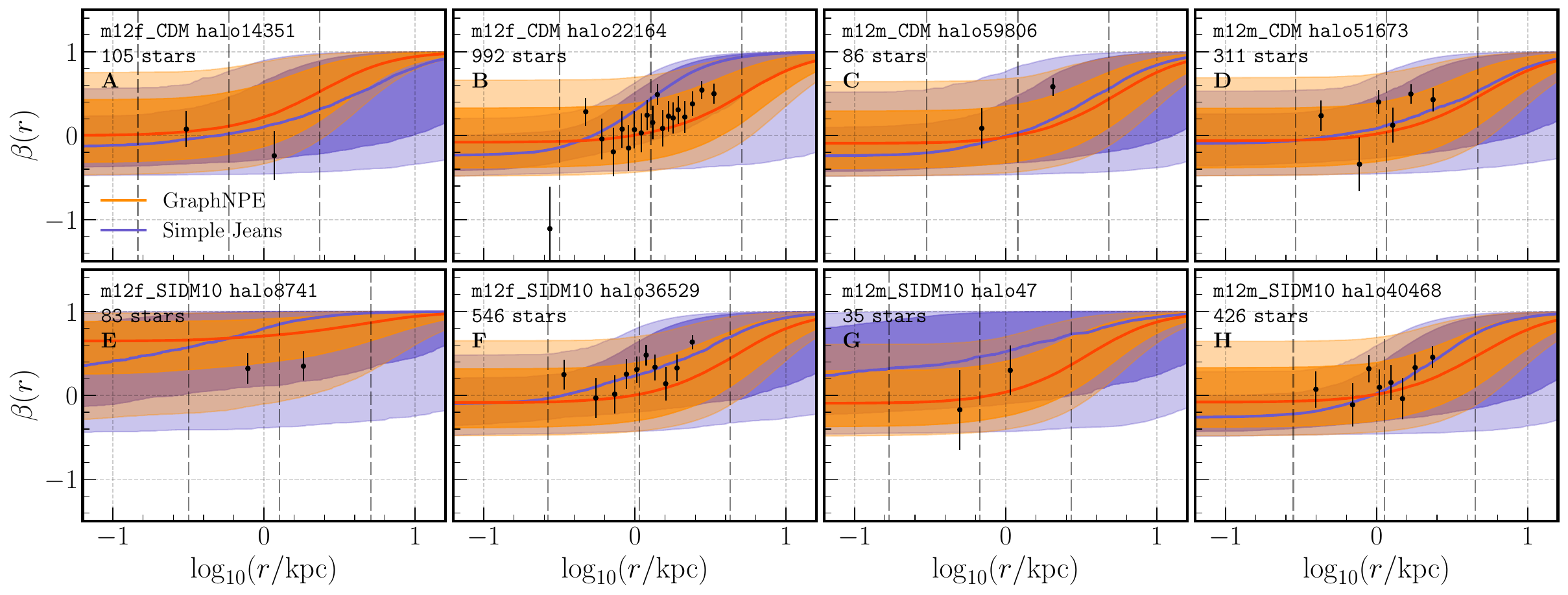}
    \caption{
    Comparison between the inferred velocity anisotropy profiles $\beta(r) = 1 - \sigma^2_t /\sigma^2_r$, where $\sigma_t$ and $\sigma_r$ are the tangential and radial velocity dispersions, from \gnn (orange) and Jeans modeling (blue) for a selection of FIRE-2 galaxies.
    Values of $\beta=0, 1 -\infty$ correspond to isotropic, radially-biased, and tangentially-biased orbits.
    Panels are the same as in Figure~\ref{fig:rho_gnn_jeans}.}
    \label{fig:beta_gnn_jeans}
\end{figure*}

Figure~\ref{fig:rho_gnn_jeans} and Figure~\ref{fig:beta_gnn_jeans} show the recovered DM density profiles $\rhodm(r)$ and velocity anisotropy profiles $\beta(r)$ for a subset of the selected galaxies, with the remaining profiles provided in Appendix~\ref{app:comparison}.
Each panel shows a single dwarf galaxy, labeled with its \rockstar halo ID at $z=0$ and a corresponding letter (e.g., A, B, C, D) for readability, and compares the profiles derived using \gnn (orange) and Jeans modeling (blue).
The posterior median profiles are shown as solid lines, with shaded bands indicating the 68\% and 95\% confidence intervals, while black circles with error bars represent the true profiles.\footnote{Note that error bars are not visible for the true density profiles.}
The dashed vertical lines indicate $(0.25, 1, 4) \, r_\mathrm{1/2}$, where $r_\mathrm{1/2}$ is the 3-D half-stellar mass radius.

Both \gnn and Jeans modeling can recover the DM density to within the 95\% confidence intervals. 
However, the velocity anisotropy profile is generally not well-constrained, with significant uncertainties persisting in most cases. 
The OM profile provides a reasonably good fit for these satellites, with the exception of \texttt{Halo~A}, which may also be affected by the low number of tracer stars.

In general, \gnn provides tighter constraints on the profiles compared to Jeans modeling.
This is particularly evident when the tracer count is low, as seen in \texttt{Halo~E} and \texttt{Halo~G}.
Conversely, when the tracer star count is high, the performance of \gnn and Jeans modeling becomes more similar.
In the case of \texttt{Halo~B}, the posterior distributions of the recovered density profiles $\rhodm(r)$ show nearly perfect overlap, with Jeans modeling offering slightly tighter constraints.
However, \gnn outperforms Jeans modeling in constraining the anisotropy profile $\beta(r)$.
We observe similar performance trend across all profiles, as detailed in Appendix~\ref{app:comparison}.

The trend in the performance differences between \gnn and Jeans is consistent with findings in N23. 
The improvement at low tracer counts likely results from \gnn's ability to account for the full distribution function $f(\vec{x}, \vec{v})$. 
By training on Monte Carlo simulations with the likelihood described in Equation~\ref{eq:loglike_npe}, which explicitly incorporates the full $f(\vec{x}, \vec{v})$, \gnn effectively performs amortized inference on this likelihood.
In contrast, the Jeans likelihood in Equation~\ref{eq:loglike_jeans} depends explicitly only on the velocity dispersion, which is the second-order moment of $f(\vec{x}, \vec{v})$.
This also explains the similarity in the recovered density profiles at high tracer counts. 
As the tracer count increases, the line-of-sight velocity dispersion profiles can be better constrained to larger radii.
This helps break the degeneracy between the density profile $\rhodm(r)$ and the anisotropy profile $\beta(r)$, allowing Jeans modeling to recover more accurate density profiles that closely align with those from \gnn~\citep[e.g.][]{2017MNRAS.471.4541R, 2021MNRAS.501..978R, 2021MNRAS.507.4715C}.

\begin{figure*}
    \centering
    \includegraphics[width=\linewidth]{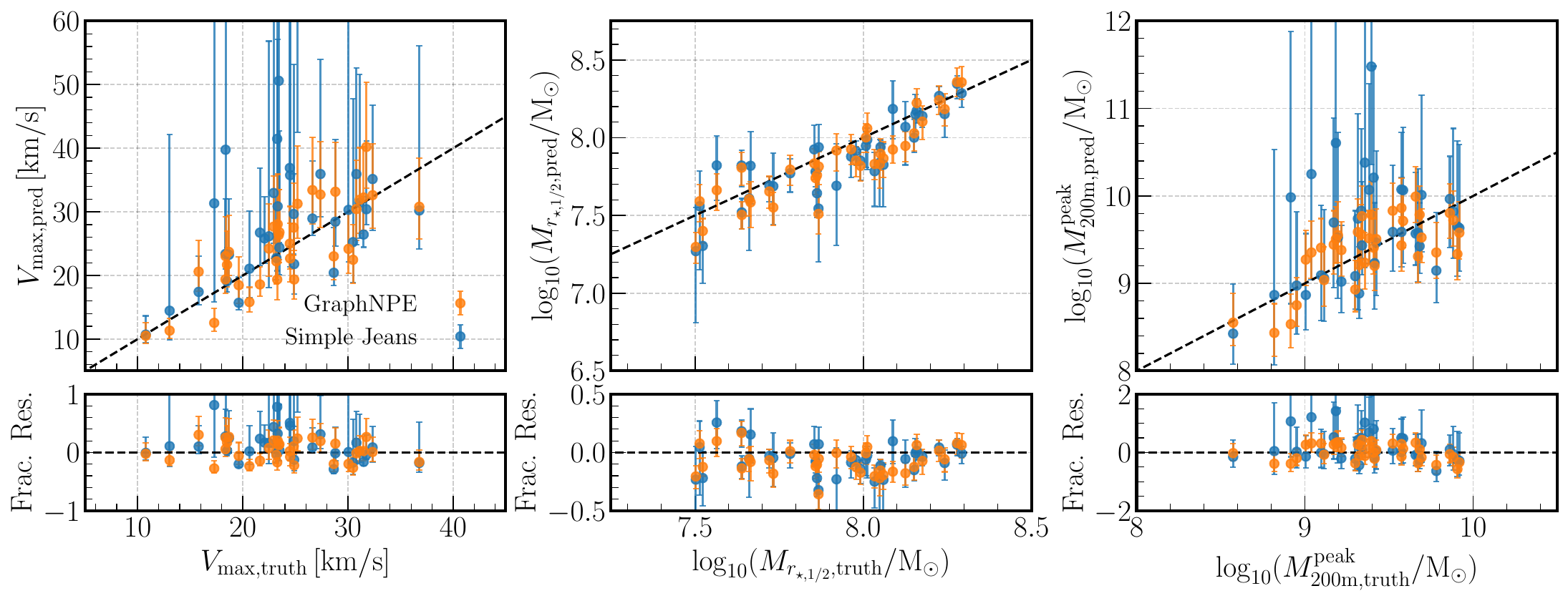}
    \caption{
    Comparison between the inferred mass and structural parameters from \gnn (orange) and Jeans modeling (blue).
    From left to right, the columns show the peak circular velocity \vmax, the half-stellar radius mass \mhl, and the virial mass \mvirp.
    \vmax and \mvirp are calculated by extrapolated the DM density profiles to $r_\mathrm{max}$ and $r_\mathrm{200m}$, respectively.
    The top and bottom rows show the median and fractional residual of the predicted values against the true values, respectively, with the error bars denoting the 64th percentile ranges.
    The black dashed line represents the one-to-one correlation.
    Since the virial mass \mvirp might not be well-defined for tidally stripped halos, the recovered \mvirp values are compared to the peak virial mass achieved throughout the halos' evolutionary history,
    }
    \label{fig:jeans_summary}
\end{figure*}

\begin{figure*}
    \centering
    \includegraphics[width=\linewidth]{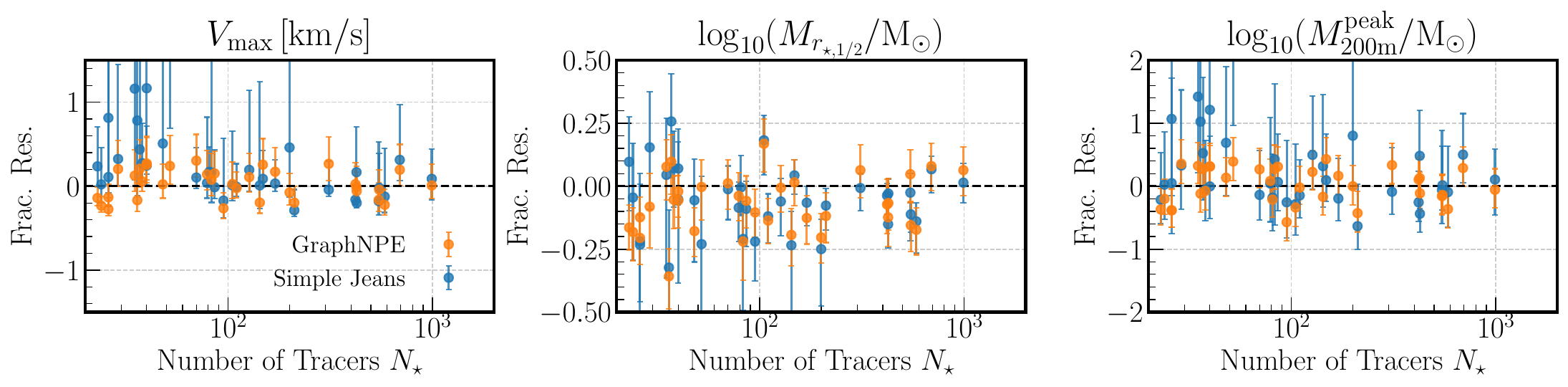}
    \caption{
    The fractional residual of the predicted mass and structural parameters as a function of the tracer counts $N_\star$.
    From left to right, the panels show the peak circular velocity \vmax, the half-stellar radius mass \mhl, and the virial mass \mvirp.
    The \gnn and Simple Jeans results are shown as orange and blue data points, respectively. 
    }
    \label{fig:jeans_summary_nstar}
\end{figure*}

\subsubsection{Mass and structural parameters}
\label{section:jeans_gnn_summary}

We perform a detailed comparison between the performance of \gnn and Jeans modeling by recovering key mass and structural parameters for all selected galaxies. 
Specifically, for each galaxy, we calculate:
\begin{enumerate}
\item \textbf{The peak circular velocity:} $\vmax \equiv V_\mathrm{circ}(r_\mathrm{max})$, where $r_\mathrm{max}$ is the radius at which the circular velocity, $V_\mathrm{circ}(r) = \sqrt{G M(r) / r}$, reaches its maximum value.
\vmax provides a robust mass estimator that is relatively insensitive to the halo boundary~\citep[e.g.,][]{1999ApJ...522...82K, 2008ApJ...673..226P}. 

\item \textbf{The half-stellar radius total mass:} $\mhl \equiv M(r_{\star, 1/2})$, where $\rhl$ is the 3-D radius enclosed half of the stellar mass. 
Previous studies have shown that the mass profile is maximally constrained near $\rhl$ for a single tracer population with only line-of-sight velocity measurements, since \mhl is approximately proportional to the velocity dispersion $\braket{\sigmalos^2}$~\citep[e.g.][]{2009ApJ...704.1274W, 2010MNRAS.406.1220W, 2017MNRAS.469.2335C, 2018MNRAS.481.5073E}.

\item \textbf{The peak virial mass:} The virial mass is defined as $M_\mathrm{200m} \equiv M(r_\mathrm{200m})$, where $r_\mathrm{200m}$ is the radius at which the mean halo density is 200 times the \textit{mean matter density} of the Universe.
However, since many Milky Way dwarf galaxies, as well as those in our test samples, have likely experienced some degree of tidal stripping, the outer regions of their DM halos have already been significantly stripped, and the present-day \mvir is not well-defined.
Therefore, we compare our mass estimates to the peak virial mass of the halos throughout their evolutionary history, \mvirp.
Unless stated otherwise, the virial mass will refer to this peak value throughout our analysis.

\end{enumerate}

Note that although the inner slope of the DM density profiles are of considerable interest, we do not include it in our analysis for the following reasons. 
Since the simulated profiles do not strictly follow gNFW, we estimated the slope using the innermost radial bins.
However, this approach yielded noisy and unreliable estimates due to several factors: (1) strong dependence on the binning scheme, (2) limited numbers of DM particles in the innermost regions, particularly for lower-mass subhalos, introducing significant Poisson noise, and (3) finite resolution effects that prevent probing sufficiently deep into the inner profiles.
Given these limitations, we concluded that reliable constraints on the inner slope would require either higher-resolution simulations or alternative fitting approaches.

\input{table_jeans_gnn}

Since the model outputs the parameterized gNFW density profiles in Equation~\ref{eq:gNFW}, for each profile, we calculate the predicted peak circular velocity, half-stellar radius total mass, and peak virial mass by integrating the profiles to their respective radii, i.e. $r_\mathrm{max}$, $\rhl$, $r_\mathrm{200m}$.

We emphasize that this is the first study to evaluate the ability of mass modeling methods to recover \vmax and \mvirp in mock galaxies within realistic environments.
\citet{2020MNRAS.498..144G}, which applied an older version of \gs on dwarf galaxies in the APOSTLE simulations, is the most similar to this work.
Their study examined how well \gs can constrain \mhl and compared this with other mass estimators (see their Figures 3 and 4) and explored the impact of tidal stripping on mass profile recovery (their Figure 9).
However, they did not investigate the recovery of \vmax or \mvirp, which we analyze for the first time in this context.
The effects of tidal stripping on these quantities are further explored in Section~\ref{section:tidal}.

Figure~\ref{fig:jeans_summary} displays the predicted and true values for \vmax, \mhl, and \mvirp (left to right columns) as derived from \gnn (orange) and Jeans modeling (blue). 
The predicted values correspond to the median of the posterior, with error bars representing the 68\% confidence interval for each parameter.
The fractional residual (bottom panel) is defined as $\Delta\vmax = (V_\mathrm{max, pred} - V_\mathrm{max, truth}) / V_\mathrm{max, truth}$ for \vmax and $\Delta M = \log (M_\mathrm{pred} / M_\mathrm{truth})$ for \mhl and \mvirp.

To further quantify the performance of each method, we compute standard performance metrics, including the absolute error (AE), squared error (SE), and negative log-likelihood (NLL).
Table~\ref{tab:jeans_gnn} presents the mean and the standard error for each metric across all selected galaxies.
The AE and SE metrics are computed between the true values and predicted values (which we take to be the median of the posteriors) with the SE metric more sensitive to outliers.
The NLL metric, on the other hand, takes into account the overall shape of the posterior.
If the posterior were perfectly Gaussian, the NLL would be reduced to a chi-square statistic, with an additional normalization term that accounts for the standard deviation.
However, since the posteriors of \vmax and \mvirp exhibit long tails, as hinted from the asymmetric error bars in Figure~\ref{fig:jeans_summary}, the NLL provides a more robust assessment.
To compute the NLL, we first fit the posterior using Gaussian Kernel Density Estimation\footnote{We use a KDE bandwidth of 1 for \vmax and 0.1 for \mhl and \mvir.} (KDE) and then evaluate it at the true value. 

Both \gnn and Jeans modeling recover \vmax and \mvirp within the 95\% confidence intervals.
However, it is evident from Figure~\ref{fig:jeans_summary} and Table~\ref{tab:jeans_gnn} that \gnn provides more accurate and tighter constraints compared to Jeans modeling.
Specifically, Jeans tends to overestimate \vmax and \mvirp, while also producing significantly wider 68\% confidence intervals. 
As shown in Table~\ref{tab:jeans_gnn}, Jeans performs worse according to all three performance metrics: AE, SE, and NLL.

On the other hand, both \gnn and Simple Jeans achieve similarly accurate performance in recovering \mhl.
This is expected, as \mhl strongly correlates with the line-of-sight velocity dispersion, making it less susceptible to the mass-anisotropy degeneracy and uncertainties in modeling $\beta(r)$. 
This relationship has been established by various mass estimators~\citep{2009ApJ...704.1274W, 2010MNRAS.406.1220W, 2017MNRAS.469.2335C, 2018MNRAS.481.5073E}.
Similarly, mass modeling methods have demonstrated comparable accuracy in estimating \mhl, even in realistic galactic environments~\citep{2013A&A...558A..35B, 2020MNRAS.498..144G}, and additionally shown that \mhl is less sensitive to assumptions on DM profile~\citep{2013A&A...558A..35B}. 
From Table~\ref{tab:jeans_gnn}, we observe that the AE and SE metrics for \gnn and Simple Jeans are similar, with \gnn achieving a marginally better NLL; however, the performance differences remain well within one standard deviation for both methods.

Lastly, we examine the fractional residuals of the mass and structural parameters as a function of the tracer counts $N_\star$ in Figure~\ref{fig:jeans_summary_nstar}, comparing \gnn and Simple Jeans results.
As expected, \gnn demonstrates consistent performance even at low tracer counts ($N_\star < 100$), while Simple Jeans shows degraded performance in this regime.
This is especially true for \vmax and \mvirp recovery.
At high tracer counts ($N_\star > 100$), the residuals become more similar, though \gnn still slightly outperforms Simple Jeans.
We present additional results for \gnn across the full test dataset in Appendix~\ref{app:nstar}.

\subsection{Higher-moment methods}
\label{section:gs_gnn}

\begin{figure*}
    \centering
    \includegraphics[width=\linewidth]{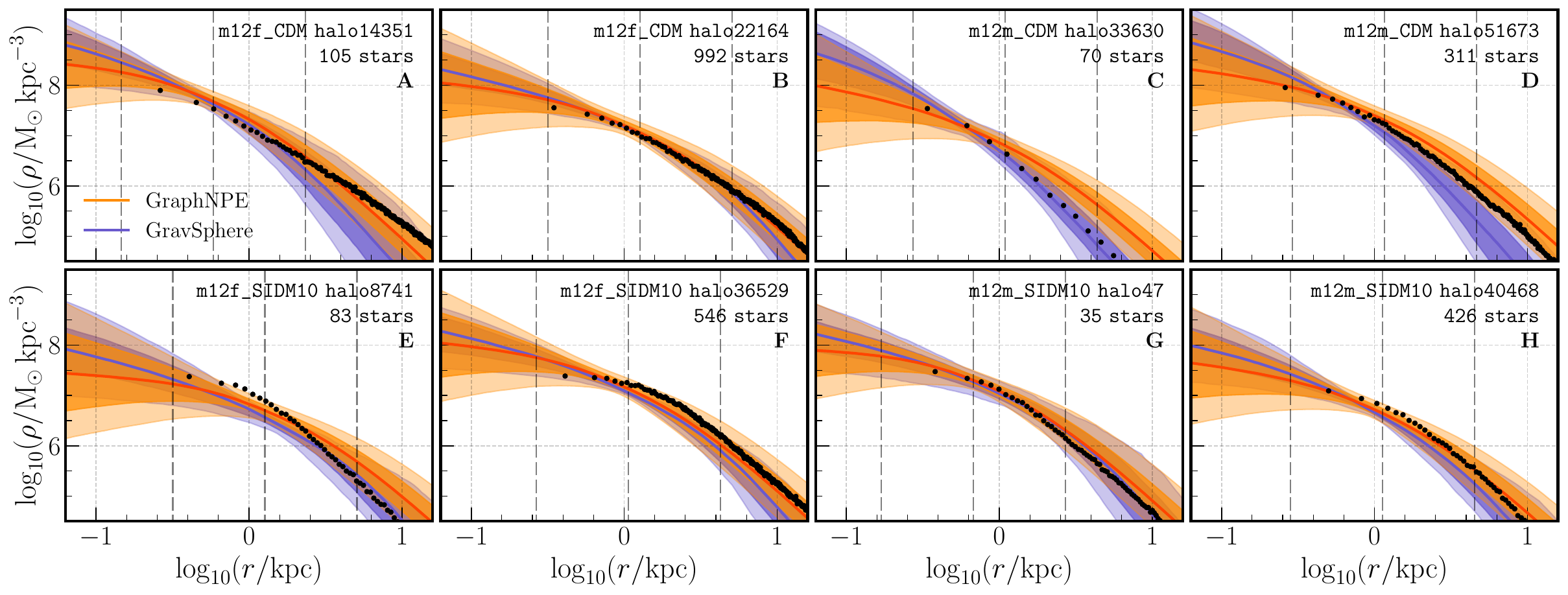}
    \caption{
    Comparison between the inferred DM density profiles $\rhodm(r)$ from \gnn (orange) and \gs (blue) for a selection of FIRE-2 galaxies. 
    Panels are the same as in Figure~\ref{fig:rho_gnn_jeans}.
    }
    \label{fig:rho_gnn_gs}
\end{figure*}
\begin{figure*}
    \centering
    \includegraphics[width=\linewidth]{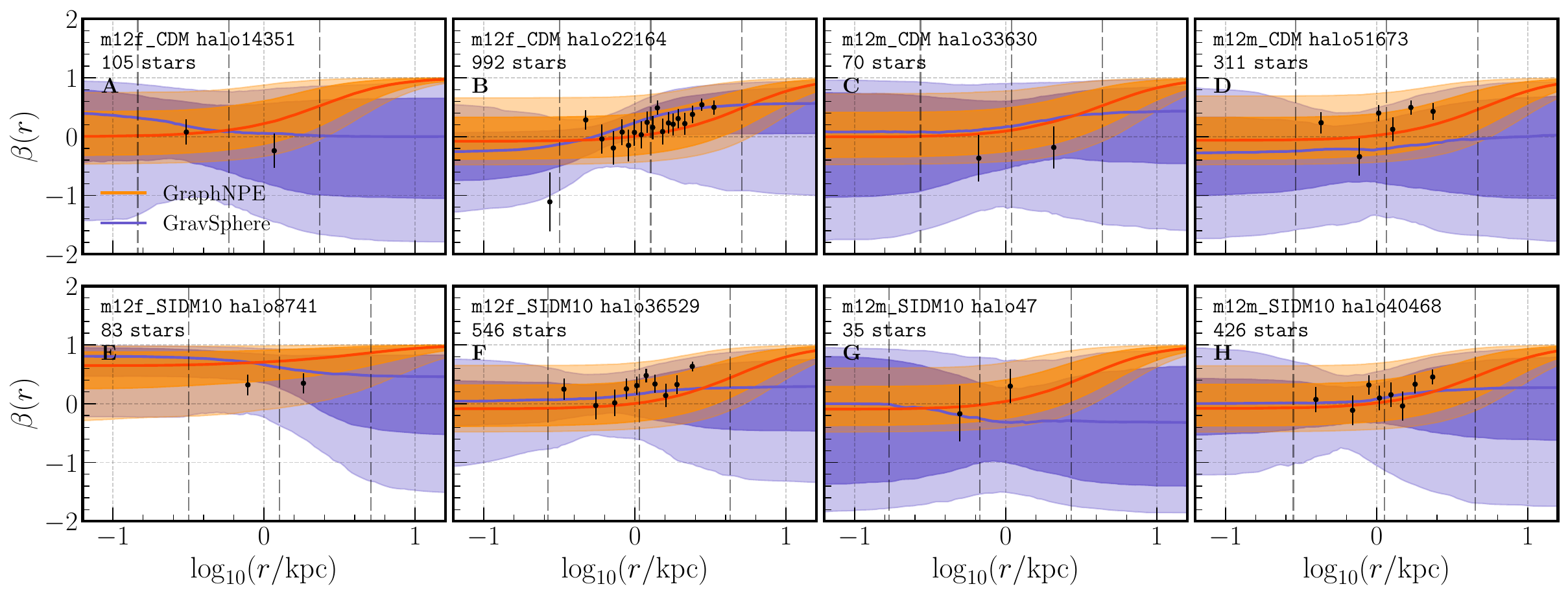}
    \caption{
    Comparison between the inferred velocity anisotropy profiles $\beta(r)$ from \gnn (orange) and \gs (blue) for a selection of FIRE-2 galaxies. 
    Panels are the same as in Figure~\ref{fig:rho_gnn_jeans}.}
    \label{fig:beta_gnn_gs}
\end{figure*}
\begin{figure*}
    \centering
    \includegraphics[width=\linewidth]{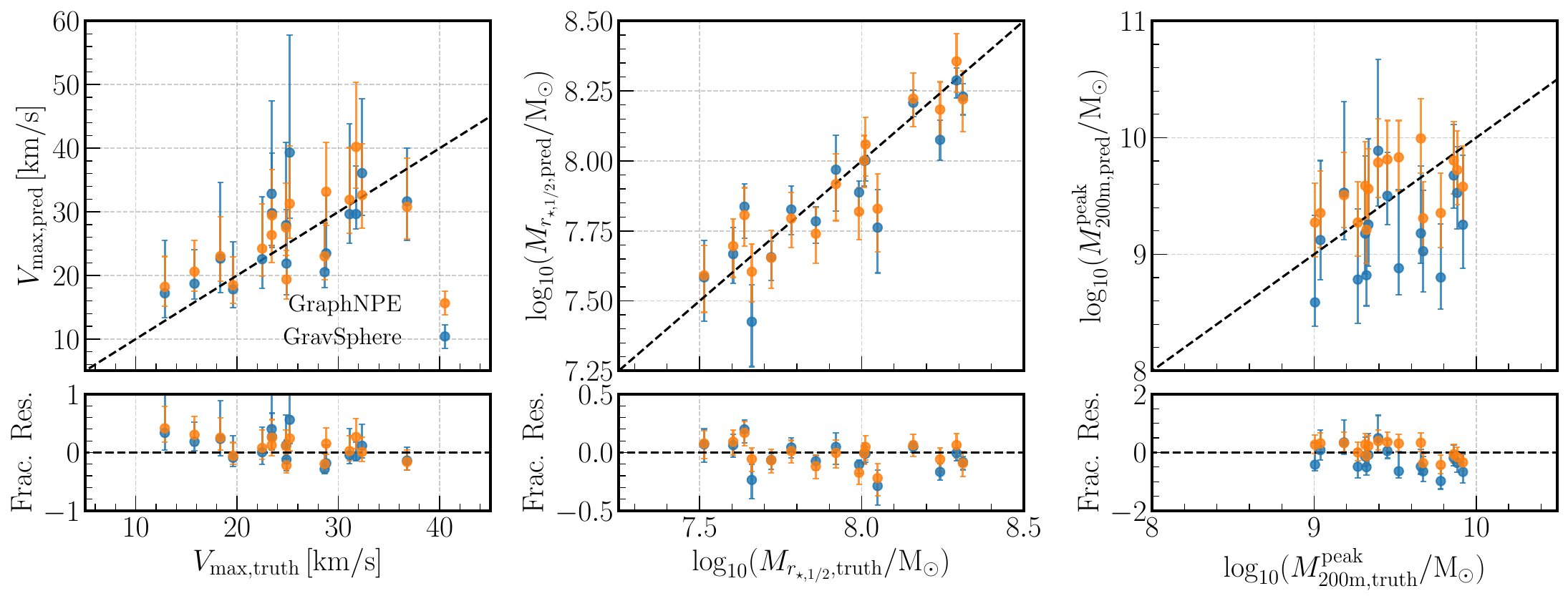}
    \caption{
    Comparison between the inferred mass and structural parameters from \gnn (orange) and \gs (blue).
    From left to right, the columns show the peak circular velocity \vmax, the enclosed mass \mhl, and the virial mass \mvir.
    Panels are the same as in Figure~\ref{fig:jeans_summary}.
    }
    \label{fig:gs_summary}
\end{figure*}
\begin{figure*}
    \centering
    \includegraphics[width=\linewidth]{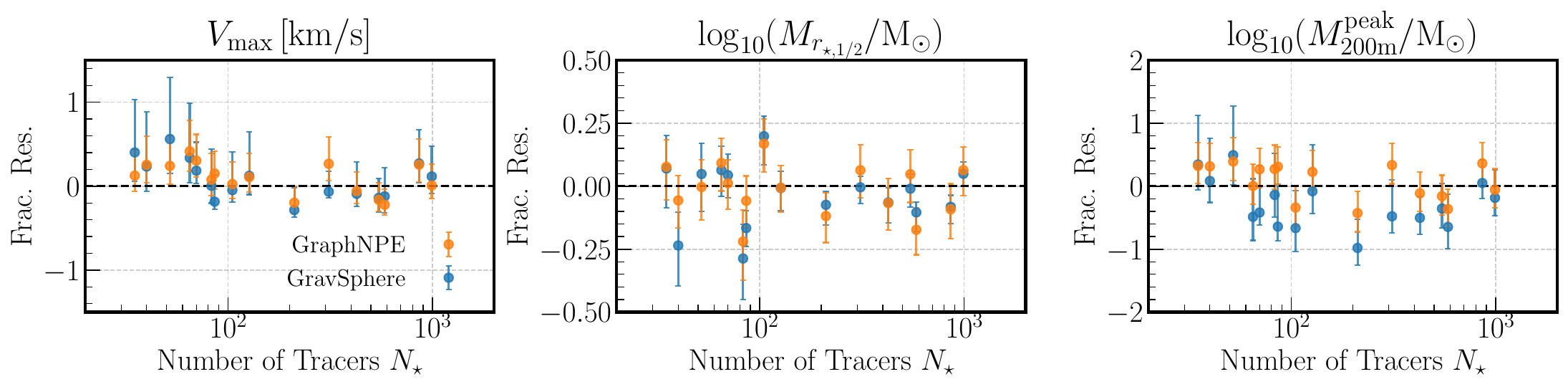}
    \caption{
    The fractional residual of the predicted mass and structural parameters as a function of the tracer counts $N_\star$.
    From left to right, the panels show the peak circular velocity \vmax, the half-stellar radius mass \mhl, and the virial mass \mvirp.
    The \gnn and \gs results are shown as orange and blue data points, respectively. 
    }
    \label{fig:gs_summary_nstar}
\end{figure*}

Having demonstrated that \gnn produces tighter and more accurate constraints on the dark matter density and velocity anisotropy profiles compared to Simple Jeans, we now turn to a comparison with a more advanced Jeans-based method. 
Specifically, we evaluate the performance of \gnn against \gs, a non-parametric mass modeling approach that incorporates higher-order velocity moments~\citep{2017MNRAS.471.4541R}.
In addition to the velocity dispersion, \gs computes the ``Virial Shape Parameters''~\citep[VSPs;][]{1990AJ.....99.1548M, 2014MNRAS.441.1584R}:
\begin{equation}
    \label{eq:vsp1}
    v_{s1} = \frac{2}{5} \int_0^\infty G M (5 - 2 \beta) \nu \sigma^2_r r \diff r = \int_0^\infty \Sigmastar \braket{\vlos^4}R \diff R
\end{equation}
and 
\begin{equation}
    \label{eq:vsp2}
    v_{s2} = \frac{4}{35} \int_0^\infty G M (7 - 6 \beta) \nu \sigma^2_r r^3 \diff r = \int_0^\infty \Sigmastar \braket{\vlos^4} R^3 \diff R.
\end{equation}
Here $r$ and $R$ denote the 3-D and projected radius, respectively. 
VSPs are particularly advantageous as they depend only on $\braket{\vlos^4}$ and $\beta(r)$ (where the anisotropy term arises from the integration over the projected radius), rather than directly involving its fourth-order moment counterpart (see e.g. \citealt{1990AJ.....99.1548M, 2014MNRAS.441.1584R})
This imposes two additional constraints on the velocity anisotropy, helping to break the mass-anisotropy degeneracy.

\gs has been described and extensively tested in \citet{2017MNRAS.471.4541R, 2018MNRAS.481..860R, 2020MNRAS.498..144G, 2021MNRAS.505.5686C, 2024MNRAS.535.1015D}.
Additionally, \citet{2021MNRAS.501..978R} provides a comprehensive comparison between an earlier version of \gs and other mass modeling techniques.
In this paper, we use the public version of \gs as described in \citet{2021MNRAS.505.5686C} which bins the data using the \textsc{binulator} method.\footnote{Code available at \url{https://github.com/justinread/gravsphere}.}
Below, we highlight the key considerations for this comparison:

\gs assumes the following form for the velocity anisotropy \citep{2007A&A...471..419B},
\begin{equation}
    \label{eq:velani_gs}
    \beta(r) = \beta_0 + (\beta_\infty - \beta_0) \frac{1}{1 + \left(r/r_a\right)^\eta},
\end{equation}
where $\beta_0$ and $\beta_\infty$ are the asymptotic anisotropies at small and large radii, $r_a$ is the anisotropy transition radius, and $\eta$ controls the sharpness of the transition.
The OM anisotropy described in Equation~\ref{eq:velani_OM} is a special case of this parameterization, where $\beta_\infty=1$ and $\eta=2$.
In \gs, the priors are defined on the symmetrized version of $\beta(r)$, defined as
\begin{equation}
    \tilde{\beta}(r) = \frac{\sigma_r^2 - \sigma_t^2}{\sigma_r^2 + \sigma_t^2} = \frac{\beta}{2 - \beta}.
\end{equation}
This reparameterization ensures that $\tilde{\beta}$ remains finite, as it is constrained to $[-1,1]$, whereas $\beta$ can diverge to $-\infty$.
The priors of \gs are uniform over $\tilde{\beta}_0 \in [-0.5, 1]$, $\tilde{\beta}_\infty \in [-0.5, 1]$, $\eta \in [0, 3]$, with a fixed $r_a = 1 \, \mathrm{kpc}$.
For comparison, the priors of \gnn in Table~\ref{tab:priors} roughly translate to a range of $\tilde{\beta}_0 \in [-0.2, 1]$ and $r_a \in [2 \times 10^{-3}, 10^3] \, \mathrm{kpc}$, with $\tilde{\beta}_\infty = 1$ and $\eta = 2$ fixed by the definition of the OM profile.

Additionally, \gs uses the \textsc{coreNFWtides} model for the DM density profile~\citep{2021MNRAS.505.5686C} and a ``three Plummer'' model for the tracer density profile~\citep{2016MNRAS.459.3349R, 2017MNRAS.471.4541R}.
For a more detailed description of the model, we refer readers to Section 4.1 in~\citet{2021MNRAS.505.5686C}.
Here, we summarize the asymptotic behavior of the \textsc{coreNFWtides} profile. 
The DM density follows $\rhodm^\mathrm{cNFWt} \propto r^{-\gamma}$, where $\gamma$ transitions from $\gamma_\mathrm{in}$ at small radii to $\gamma_\mathrm{out}$ at large radii.
Both the inner and outer slopes, $\gamma_\mathrm{in}$ and $\gamma_\mathrm{out}$, are free parameters, with prior distributions $[0, 1]$ and $[3.01,5]$ respectively.\footnote{The parameters $\gamma_\mathrm{in}$ and $\gamma_\mathrm{out}$ are equivalent to $n$ and $\delta$ in $\rhodm^\mathrm{cNFWt}$. We use this notation for consistency with the gNFW profile.}
In comparison, the gNFW profile in \gnn follows $\rhodm^\mathrm{gNFW} \propto r^{-\gamma}$ at small radii, where $\gamma$ can vary between $[-1,2]$ (see Table~\ref{tab:priors}), and always transition to $\rhodm^\mathrm{gNFW} \propto r^{-3}$ at large radii.

Lastly, unlike \gnn, \gs partitions the data into radial bins using the \textsc{binulator} code, described in Section 4.1.1 of \citet{2021MNRAS.505.5686C}.
\textsc{binulator} automatically adjusts the bins to contain an equal number of stars and fits a generalized Gaussian probability to each bin to robustly estimate its mean, variance, and kurtosis.
This improves upon the binning routines of the previous version of \gs and enhances performance in the low tracer count limit.
In this work, \gs uses 10 stars per radial bin for the light profile and 20 stars per radial bin for the line-of-sight velocity dispersion profile. 
The stars are assumed to have negligible velocity uncertainties.
The model simultaneously fits the mass, velocity anisotropy, and tracer density profiles by minimizing the chi-square statistic between the observable quantities--namely $\Sigmastar$ and $\sigmalos$ within each bin, and the VSPs.

Due to differences in modeling choices and prior distributions, it is difficult to directly compare between \gnn and \gs. 
Thus, unlike in Section~\ref{section:jeans_gnn}, here we do not seek to establish which method can provide a better constraint on the density profile.
Instead, our goal is to establish a baseline performance for \gnn and to highlight its key strengths and limitations relative to advanced Jeans-based methods like \gs.
To this end, we apply \gs to the same sample of FIRE-2 dwarf galaxies presented in Section~\ref{section:jeans_gnn}. 
Due to the higher computational cost of \gs as compared to Simple Jeans, we limit our analysis to four simulations: \mf, \mm, \mfsidmt, and \mmsidmt, comprising a total of 16 dwarf galaxies.

\subsubsection{Result}

\input{table_gs_gnn}

Figure~\ref{fig:rho_gnn_gs} shows the inferred DM density profiles $\rhodm(r)$ for a selection of FIRE-2 galaxies (with the rest shown in Appendix~\ref{app:comparison}).
Each panel shows an individual galaxy and follows the same format as Figure~\ref{fig:rho_gnn_jeans}: the median, 68\%, and 95\% confidence intervals of the inferred $\rhodm(r)$ from \gnn and \gs are displayed in orange and blue, respectively, while the true profile is shown as black circles with error bars. 
Note that the galaxy samples are selected to highlight the differences between \gnn and \gs, and thus do not necessarily overlap with the sample in Figure~\ref{fig:rho_gnn_jeans}.

In general, the performance between \gs and \gnn are comparable.
Both methods effectively constrain the density profiles within their respective 95\% confidence intervals.
Interestingly, we do not observe the significant performance degradation, as seen with Simple Jeans in Section~\ref{section:jeans_gnn}.
It may seem surprising that \gs performs so well even when there are few data points, as earlier versions of the code certainly struggled in this regime~\citep[e.g.][]{2021MNRAS.501..978R}. 
However, here we use the improved version of \gs introduced in \citet{2021MNRAS.505.5686C} that uses the \textsc{binulator} to bin the data. 
This was designed to work well even for very few stellar velocity data points, making \gs suitable also for modeling ultra-faint dwarfs.

The inner profile appears to be more tightly constrained by \gs.
However, we note that this is likely due to differences in the prior distribution on the inner slope $\gamma$, with $\gamma^\mathrm{GS}_\mathrm{in} \in [0, 1]$ and $\gamma^\mathrm{gNFW} \in [-1, 2]$.
The upper-95\% confidence intervals of the density profiles between the two methods tend to be in agreement, though the profiles from \gs are slightly steeper in some cases.
For the lower-95\%, \gnn allows the density to flatten more significantly and even decline as $r$ decreases, thus favoring more core-like structures.
The decline in the central density is unphysical; however, by allowing $\gamma$ to go negative, \gnn avoids running into the prior edge for profiles with pronounced cores.

We find that \gs predicts steeper outer density profiles than \gnn, likely also due to differences in the prior distribution of $\gamma$. 
Specifically, \gs allows the outer slope $\gamma_\mathrm{out}^\mathrm{GS}$ to vary between $[3.01, 5]$, whereas \gnn assumes a fixed outer behavior $\rhodm^\mathrm{gNFW} \propto r^{-3}$.
As a result, \gs struggles with shallower outer profiles (e.g. \texttt{Halo~A} and \texttt{Halo~D}), while \gnn has difficulty with steeper ones (e.g. \texttt{Halo~C} and \texttt{Halo~E}).

Figure~\ref{fig:beta_gnn_gs} presents the velocity anisotropy profiles $\beta(r)$. 
Within the $(0.25, 4) \, r_{\star, 1/2}$ region, where kinematic tracers are well-sampled, the inferred anisotropy profiles from both \gs and \gnn are well constrained. 
Outside this range, the weaker constraints from \gs arise from its more flexible prior and the lack of tracer data, whereas both \gnn and Simple Jeans assumes the OM profile, which imposes stricter assumptions on $\beta(r)$ but may be less accurate in some cases.
For example, this is evident in \texttt{Halo~A} and \texttt{Halo~E}, where \gs provides a better fit compared to the OM profile.

As in Section~\ref{section:jeans_gnn_summary}, we compute the same mass and structural parameters, i.e. \vmax, \mhl, and \mvirp, for each galaxy, with the results shown in Figure~\ref{fig:gs_summary}. 
Compared to Simple Jeans, the performance of \gs closely aligns with that of \gnn. 
Notably, despite differences in modeling choices and prior distributions, the predicted values of \vmax, \mhl, and \mvirp from \gs and \gnn are remarkably consistent, with their 68\% confidence intervals largely overlapping.
However, we note that \gs consistently underestimates \mvirp, which is somewhat expected given its preference for steeper outer slopes (Figure~\ref{fig:rho_gnn_gs}).
Additionally, we evaluate the AE, SE, and NLL performance metrics and present the mean and standard error in Table~\ref{tab:gs_gnn}. 
The performance metrics are comparable for \gnn and \gs across all mass and structural parameters, with \gnn slightly outperforming \gs.

Lastly, Figure~\ref{fig:gs_summary_nstar} shows the fractional residuals of the mass and structural parameters as a function of $N_\star$.
Unlike the comparison with Simple Jeans, it is more difficult to draw definitive conclusions about \gnn versus \gs performance due to the limited sample size.
However, the results suggest that \gnn may also provide a modest advantage over \gs at low tracer counts, at least in the case of \vmax.

\section{Tidal Effects}
\label{section:tidal}

\begin{figure*}
    \centering
    \includegraphics[width=\linewidth]{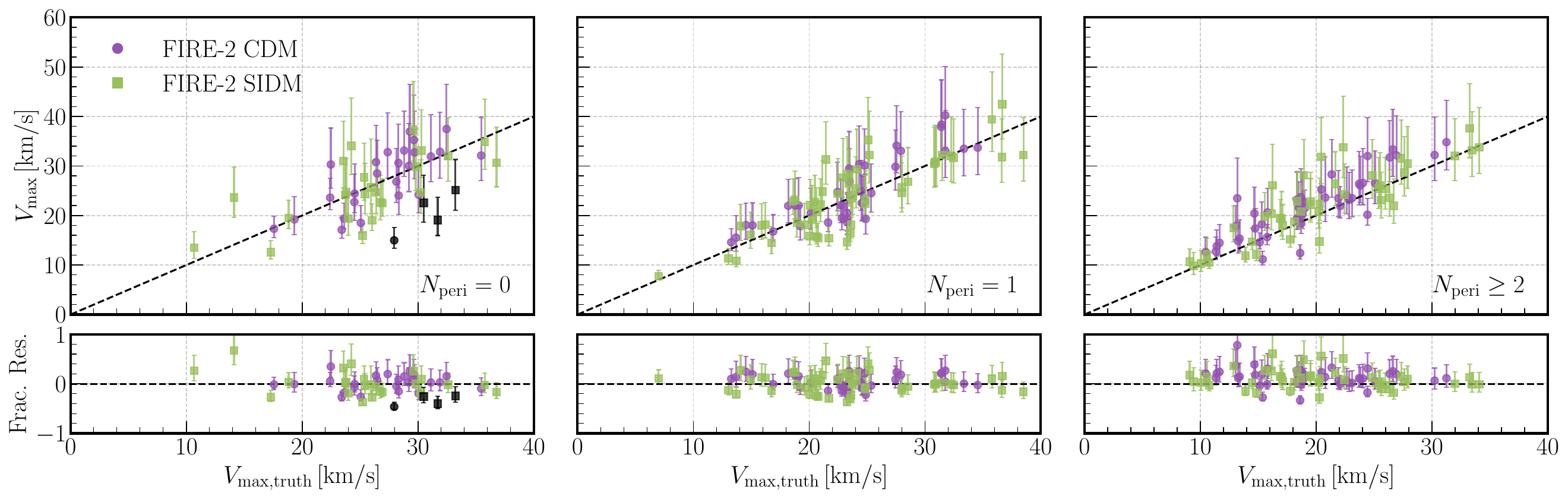}
    \caption{    
    Recovery of the peak circular velocity \vmax of dwarf galaxies in FIRE-2 dataset.  
    The top row shows the predicted \vmax (the median of the posteriors) versus the true \vmax, while the bottom row displays the residuals.  
    Error bars represent the 68\% confidence intervals.
    CDM galaxies are shown as purple circles, while SIDM galaxies are represented by green squares; outliers for both are highlighted in black.
    The black dashed line represents the one-to-one correlation.
    Galaxies are grouped by the number of pericentric passages \nperi , with each column corresponding to a different grouping.  
    }
    \label{fig:vmax-nperi}
\end{figure*}
\begin{figure*}
    \centering
    \includegraphics[width=\linewidth]{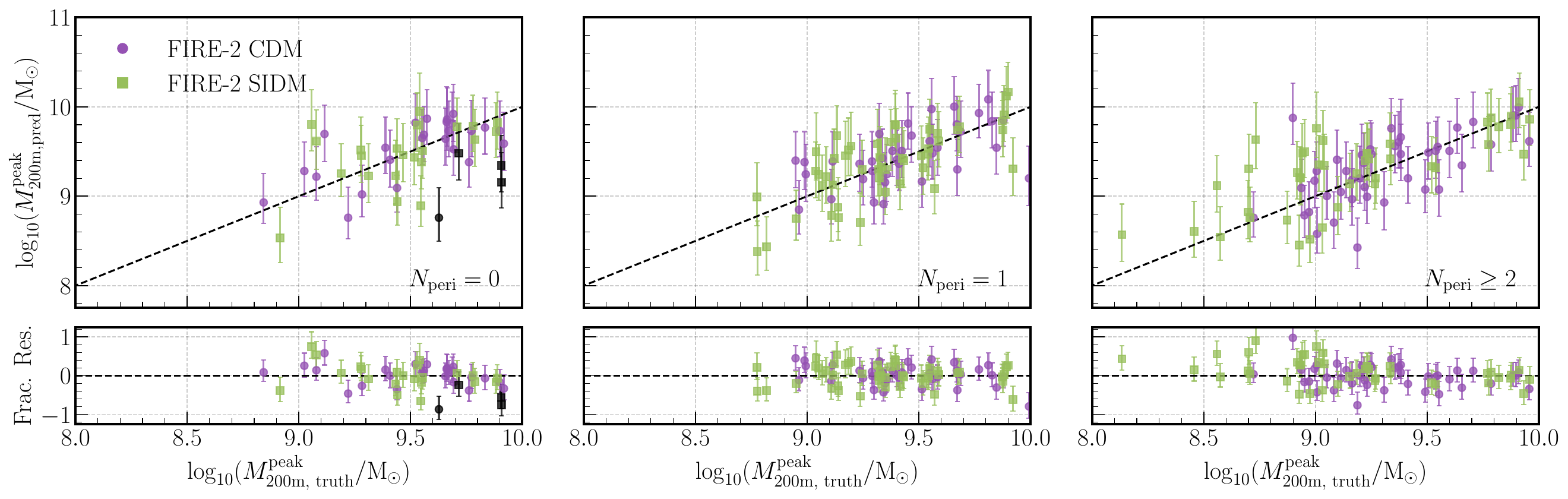}
    \caption{    
    Recovery of the peak virial mass \mvirp of dwarf galaxies in the FIRE-2 dataset.
    Panels are the same as Figure~\ref{fig:vmax-nperi}.
    }
    \label{fig:m200-nperi}
\end{figure*}

We now evaluate \gnn's performance on the full FIRE-2 mock dataset.
In this section, we assess the robustness of \gnn across varying degree of tidal effects, which has been extensively studied in dwarf galaxies in the Local Group~\citep[e.g.][]{2006MNRAS.367..387R, 2015MNRAS.454.2401B, 2022ApJ...940..136P, 2024MNRAS.535.1015D} and cosmological simulations~\citep[e.g.][]{2011ApJ...726...98K, 2016ApJ...818..193T, 2017MNRAS.472.3378F, 2018MNRAS.476.3816F, 2023ApJ...949...44S, 2024arXiv241009143S}.
Here, we do not focus on satellites that are massively disrupted, but rather on those where tidal signatures, such as tidal tails or velocity gradients, are not clearly detectable.
In FIRE-2, \citet{2023ApJ...949...44S} shows that this is common, with more than half of satellites exhibiting significant tidal features that remain undetected when analyzed using mock Dark Energy Survey (DES;~\citealt{2015AJ....150..150F, 2018ApJS..239...18A, 2021ApJS..255...20A}) observations.

\subsection{Quantifying tidal effects with orbital parameters}

To quantify the degree of tidal effects, we track the orbital history of each galaxy over the simulation snapshots and compute the following parameters: (1) the number of pericentric passages \nperi, (2) the last pericentric distance \rperi, (3) the time since the last pericentric passage \tperi, and (4) the current distance to host \rcurr.
Satellites that are more tidally disrupted are expected to have lower pericentric distances \rperi and to have completed more orbits around their host galaxies (higher \nperi).
The latter condition can result from either a shorter orbital period or an early infall time; however, we do not examine these factors separately.
The time since the last pericentric passage \tperi provides insight into how long the satellite has had to relax or recover from its most recent tidal interaction, which could influence the persistence of tidal features.
Lastly, the current distance \rcurr indirectly measures the ongoing tidal forces and also a directly observable quantity.
For, a detailed exploration of the orbital dynamics and histories of satellites in Milky Way-mass halos within FIRE-2 simulations, we refer readers to \citet{2023MNRAS.518.1427S}.

To account for variations in the mass and size of the host halos, \rperi and \rcurr are expressed in units of the virial radius, $R_\mathrm{200m}$, of the host halo at the time of the pericentric passage or the current epoch, respectively. 

It is important to note that the above orbital parameters do not directly probe tidal effects.
For instance, the density within the half-light radius of a satellite relative to its host can influence its susceptibility to tidal effects, with more compact satellites being more resilient.
As a result, tidal effects are commonly quantified using density-based metrics~\citep[e.g.,][]{2022ApJ...940..136P, 2023ApJ...949...44S} or mass loss indicators, such as the ratio of current to peak values of \vmax, \mvir, and $M_\star$~\citep[e.g.,][]{2015MNRAS.447.1112B, 2018MNRAS.476.3816F}.
However, while density metrics and mass loss indicators provide a more direct measure of tidal effects, they are not readily available in observations, making them difficult to apply consistently. 
Orbital parameters, though more indirect, are more easily inferred from observations.

\subsection{Example: result with \nperi}
\label{section:tidal_nperi}
\input{table_tidal}

We assess the performance of \gnn as a function of the orbital parameters discussed above, with a primary focus on \vmax and \mvir.

Figures~\ref{fig:vmax-nperi} and \ref{fig:m200-nperi} present the predicted versus true values (top rows) and residuals (bottom rows) of \vmax and \mvir, respectively, across three different bins of the number of pericentric passages $\nperi= 0$, $1$, and $\geq 2$ (left to right columns).
Similar figures for other orbital parameters, including \rperi, \tperi, and \rcurr, are provided in Appendix~\ref{app:tidal}.
As before, the predicted values and their error bars are the median and 68\% confidence intervals of the posterior distributions.
Additionally, we distinguish CDM and SIDM galaxies with purple circles and green squares, respectively, with outliers denoted in black.

Figures~\ref{fig:vmax-nperi} and \ref{fig:m200-nperi} clearly illustrate the impact of tides on the predictions of the present-day \vmax and the peak \mvirp.
As the number of pericentric passages increase, tidal effects accumulate, and \gnn tends to overestimate the \vmax and \mvirp. 
Despite this, we find a good alignment between the predicted and true values in all three bins.
In particular, we find that \gnn can recover \vmax and \mvirp within the 68\% confidence interval for approximately 64\% and 71\% of galaxies in the full samples, respectively.
Although not shown in the figures, \gnn captures approximately 93 \% and 95\% of galaxies within the 95\% confidence interval for \vmax and \mvirp, respectively. 
Interestingly, this is similar to \citet{2020MNRAS.498..144G}, which showed that \gs recover the enclosed mass distributions to within 68\% and 95\% confidence for 60\% and 90\% of their samples, respectively.

To further investigate the effects of tides, we calculate the fraction of galaxies within the 68\% and 95\% confidence intervals, which we now refer to as the ``coverage fractions'', across different \nperi bins.
Table~\ref{tab:tidal} presents the mean and associated error of the coverage fractions for \vmax and \mvirp for three \nperi bins.
The errors are estimated using bootstrapping methods, where we resample the dataset with replacement multiple times to generate a distribution of coverage fractions, using the standard deviation of this distribution as the uncertainty estimate.

As \nperi increases, the coverage fraction decreases, which is expected due to \gnn's tendency to overestimate \vmax and \mvirp for more tidally disrupted galaxies.
The recovery is overall quite strong: even for galaxies with multiple pericentric passages, we recover their true \vmax and \mvirp values within the 68\% confidence interval for $57.2 \pm 5.1\%$ and $71.2 \pm 5.0\%$ of the galaxies, and within the 95\% confidence interval for $89.6 \pm 3.4\%$ and $93.0 \pm 2.8\%$, respectively.
We note that the first bin has a lower coverage fraction, which we attribute to a few outlier data points, highlighted as black data points in Figures~\ref{fig:vmax-nperi} and \ref{fig:m200-nperi}.
These outliers will be discussed in more detail below; here, we note that removing them increases the 68\% and 95\% coverage fractions in the first \nperi bin to $69.6\%$ and $92.0\%$ for \vmax and to $71.6\%$ and $91.9\%$ for \mvirp.

As noted, a few outlier galaxies in the $\nperi=0$ bin exhibit significantly underestimated \vmax and \mvirp values by \gnn. 
These galaxies, highlighted as black data points in Figures~\ref{fig:vmax-nperi} and \ref{fig:m200-nperi}, are located at the high-mass end, with true values of $\vmax \gtrsim 28 , \kms$ and $\log_{10} \mvirp \gtrsim 9.5$, making them less likely to be strongly affected by tidal effects.
Upon closer examination, we find that these galaxies are all nearing their first pericentric passage and are likely experiencing significant tidal shocks due to their proximity to the host galaxy. 
This example underscores a key limitation of \nperi: while it effectively tracks cumulative tidal effects, it does not capture instances where a galaxy is undergoing strong, transient tidal interactions.
Therefore, to gain a more comprehensive understanding of the model’s performance and the complexity of tidal interactions, it is crucial to consider additional orbital parameters, as discussed in Section~\ref{section:tidal_all}.

\gnn's tendency to overestimate \vmax and \mvirp can be explained as follows.
As the satellite passes through its pericenter, the energy injection from tidal shock can ``puff up'' the stellar profiles, extending the stellar velocity dispersion profiles to larger radii while simultaneously lowering the overall velocity dispersion~\citep[e.g.,][]{2006MNRAS.367..387R, 2008ApJ...673..226P}.
Additionally, \citet{2006MNRAS.367..387R} have shown that the tidal tails can lead to a rise in the \textit{projected} velocity dispersion profiles beyond the tidal stripping radius, making the satellites appear dynamically hotter in projection. 
In the context of Jeans modeling, this increased velocity dispersion at large radii can result in an overestimation of the enclosed mass in these regions. 
Although \gnn is trained on the full distribution function, as defined in Equation~\ref{eq:loglike_npe}, the velocity dispersion is still expected to play a dominant role in determining the mass profiles.

The presence of unbound stars can artificially inflate the velocity dispersion, further leading to an overestimation of the mass. 
The member star assignment procedure described in Section~\ref{section:method} mitigates this issue but is unlikely to completely eliminate it. 
While assignments based on the halo potential, such as using \vmax~\citep{2020MNRAS.491.1471S}, can return a more pristine sample of member stars, we opt to use our procedure since such information is not directly accessible in observational data.

In a recent work, \citet{2024arXiv241103192C} demonstrates that tidal stripping tends to isotropize the anisotropy profile.
Since \gnn assumes the OM anisotropy profile, which inherently tends towards radially-biased orbits at large radii, it can overestimate $\beta$ at these radii for disrupted galaxies.  
This, in turn, leads to an overestimation in the 3-D velocity dispersion and, consequently, the mass.

In the case of \vmax, because we infer present-day values, highly disrupted galaxies often have their outer halos stripped, causing an overestimation of their current \vmax.

To briefly summarize, we emphasize that, despite these challenges, \gnn's overall performance remains strong. 
In particular, \gnn recovers \vmax and \mvirp within the 68\% and 95\% confidence intervals for a large fraction of galaxies. 
This accuracy matches that of the previous version of \gs in \citet{2020MNRAS.498..144G} for enclosed mass distributions, while here we specifically demonstrate similar reliability for \vmax and \mvirp.
The strong performance on \mvirp suggests that (1) present-day stellar kinematics encode sufficient information to infer the peak halo mass and (2) \gnn is capable of effectively extracting this information. 
The former is somewhat unsurprising, as past studies have shown that \mvirp can be recovered if the tidal radius is sufficiently far from the half-stellar mass radius of satellites~\citep[e.g.][]{2006MNRAS.367..387R, 2018MNRAS.481.5073E, 2018MNRAS.481..860R, 2019MNRAS.487.5799R}.
This result is further highlighted by the fact that the simple Jeans model in Section~\ref{section:jeans_gnn} appears to overestimate the peak mass more consistently, as shown from the samples in Figure~\ref{fig:jeans_summary}.

\subsection{Result with all orbital parameters}
\label{section:tidal_all}

\begin{figure*}
    \centering
    \includegraphics[width=\linewidth]{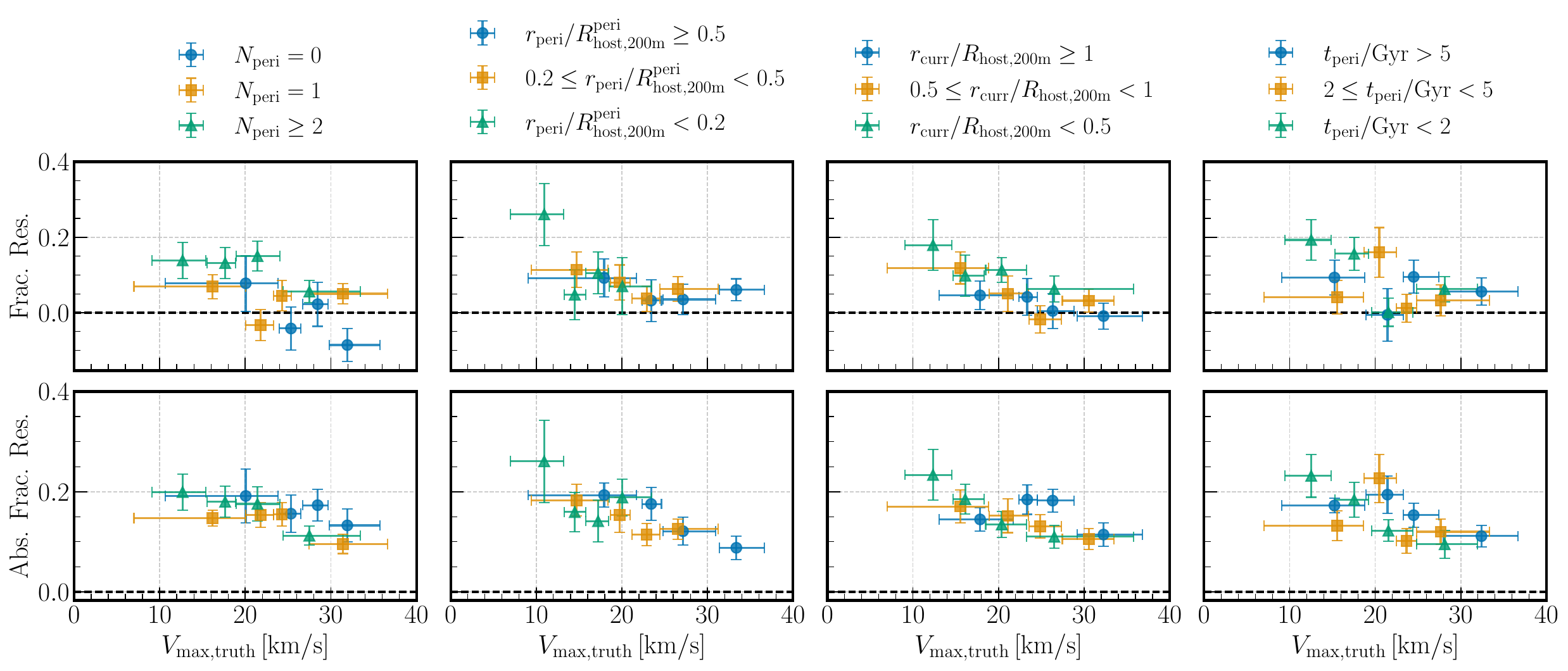}
    \caption{The net (top rows) and absolute (bottom rows) fractional residuals of \vmax across different bins of the orbital parameters.
    From left to right, the panels show bins of the number of pericentric passages \nperi, the last pericentric distance \rperi in units of the host $R_\mathrm{200m}$ at the time, the current distance \rcurr in units of the host current $R_\mathrm{200m}$, and the time since last pericentric passage \tperi.
    }
    \label{fig:vmax-residual}
\end{figure*}

\begin{figure*}
    \centering
    \includegraphics[width=\linewidth]{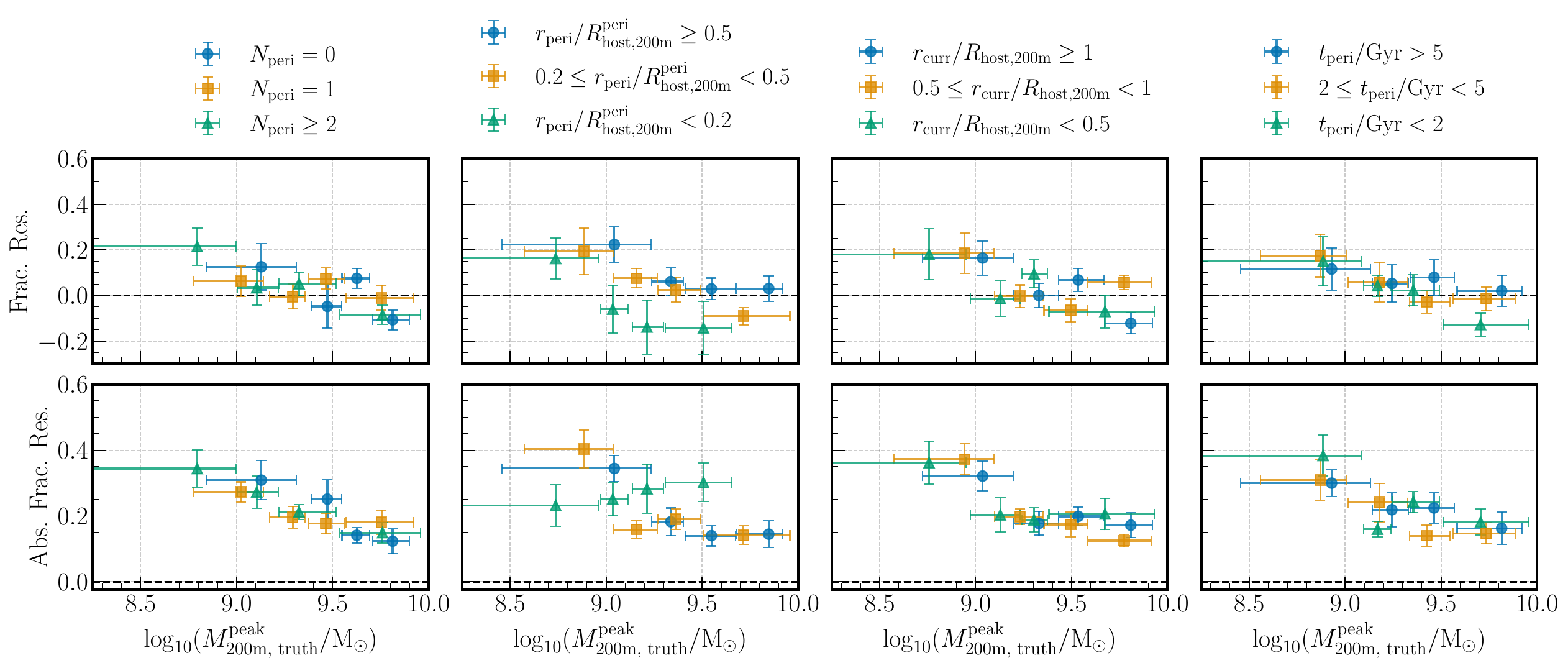}
    \caption{The net (top rows) and absolute (bottom rows) residuals of \mvirp across different bins of the orbital parameters.
    Panels are the same as Figure~\ref{fig:vmax-residual}.}
    \label{fig:m200-residual}
\end{figure*}

We now examine the performance of \gnn for the remaining orbital parameters: namely, the last pericentric distance \rperi, the current distance \rcurr, and the time since the last pericenter \tperi.
As before, Table~\ref{tab:tidal} presents the 68\% and 95\% coverage fractions, now calculated across three separate bins for each of the above parameters, resulting in a total of nine bins.

The coverage fractions generally align with their respective confidence intervals. 
In the case of \vmax, the trends in \rperi, \rcurr, and \tperi are consistent with that of \nperi: more tidally disrupted bins tend to have lower coverage due to \textsc{GraphNPE's} bias towards higher \vmax values. 
On the other hand, the trends for \mvirp are much less pronounced. 
With the exception of \rperi, the coverage fractions remain roughly the same across all bins of orbital parameters. 

Interestingly, \gnn appears to be conservative in estimating \mvirp, with overly large confidence intervals that cover more samples than expected.
In some cases (e.g., $\tperi > 5 \, \mathrm{Gyr}$, $\nperi=1$), \gnn can recover \mvirp for close to 100\% of the populations.
The broad confidence intervals might arise from extrapolating the DM density profiles when computing \mvirp, which introduces additional uncertainties. 
This issue could be mitigated with a more accurate model for the profile tails or a more extended tracer population.

Among the four orbital parameters, the pericentric distance \rperi shows the clearest trend, with the coverage fraction dropping significantly in the most disrupted bin, specifically $\rperi < 0.2 R^\mathrm{peri}_\mathrm{host, 200m}$.
This is consistent with findings from prior works~\citep[e.g.][]{2023ApJ...949...44S, 2024arXiv241009143S, 2024MNRAS.527.5868M}.
Nevertheless, even under these conditions, we still recover $\vmax$ within the 68\% confidence interval for about 50\% of the sample and within the 95\% confidence interval for about 90\%. 
For $\mvirp$, approximately 60\% of the sample falls within the 68\% confidence interval, while 90\% falls within the 95\% confidence interval.

To better quantify the accuracy of \gnn on predicting \vmax and \mvirp , we compute their mean residuals across different bins of the four orbital parameters.
We first note an important consideration.
It is evident from Figures~\ref{fig:vmax-nperi} and \ref{fig:m200-nperi} that there are distinct populations of galaxies within the three \nperi bins. 
Specifically, more massive galaxies--both in their peak and present-day masses--are more likely to occupy lower \nperi bin.
Galaxies with high \nperi likely fell into their host halos at earlier times, resulting in generally lower values of \mvirp, as they had less time to accrete mass before being incorporated into their host.
Additionally, these galaxies are expected to have experienced more stripping, leading to lower values of present-day \vmax.
It is therefore crucial to disentangle tidal effects from those of galaxy populations. 
More massive galaxies tend to host more stars, and the tighter constraints by \gnn observed in the lower \nperi bin could arise either from the absence of significant tidal effects or from the larger number of stars.
Therefore, for each bin of the tidal parameter, we further subdivide the galaxies into three equal-sized bins based on \vmax and \mvirp. 
Each sub-bin contains at least 10 galaxies to ensure statistical robustness in our analysis.

Figures~\ref{fig:vmax-residual} and \ref{fig:m200-residual} present the fractional residuals for \vmax and \mvirp, respectively, across different bins of \nperi, \rperi, \rcurr, and \tperi (left to right columns). 
The top row shows the net (signed) residuals, while the bottom row shows the absolute (unsigned) residuals.
In each panel, the color and marker indicate the bins of the corresponding tidal parameter, with blue circles, yellow squares, and green triangles representing bins in order from least to most likely to be tidally disrupted.
The $y$-position of each data point represents the mean residuals of galaxies in the bin, with the vertical error bars showing the standard errors in estimating the mean.
The $x$-position corresponds to the mean of the true \vmax or \mvirp values for all galaxies in the bin, while the horizontal error bars indicate the range of true \vmax or \mvirp values within the bin.
We do not include the outliers discussed previously in the calculation.

The fractional residual of \vmax exhibits similar trends across the panels of Figure~\ref{fig:vmax-residual}.
Within the same tidal bin, \gnn overestimates \vmax for galaxies at low \vmax, as expected since these galaxies are more susceptible to tides.
As \vmax increases, both the net residual and absolute residual steadily decrease. 
Beyond $\vmax \gtrsim 20 \, \kms$, the net residual approaches zero, while the absolute residual flattens out at approximately $10-20\%$.
When comparing different tidal bins, we find that galaxies in the more disrupted bins tend to have lower values of \vmax, consistent with the findings in Figure~\ref{fig:vmax-nperi}.
At similar \vmax values, the mean residuals tend to be larger in more disrupted bins, although there is significant scatter in both \vmax and residuals within each bin.
For $\vmax$ in the range $10-20 \, \kms$, the absolute residuals are typically around $20\%$, peaking at approximately $25 \pm 10\%$ for galaxies with $\vmax \approx 11 \, \kms$ in the $\rperi < 0.2 \, R_\mathrm{host, 200m}^\mathrm{peri}$ bin. 
At $\vmax \gtrsim 20 \, \kms$, both the net and absolute residuals flatten across all tidal bins, as also previously noted, resulting in no significant performance differences.

We observe a similar trend in the residual of \mvirp in Figure~\ref{fig:m200-residual} for bins of \nperi, \rperi, and \rcurr. 
Specifically, \gnn overestimates the true values at low \mvirp, with both the net and absolute residuals decreasing in magnitude as \mvirp increases. 
At higher masses, the residuals flatten across all tidal bins, similar to the behavior seen for \vmax.
This transition occurs around $\mvirp \sim 10^9 \, \modot$, consistent with the expectation for halos with $\vmax \sim 20\kms$, based on $\mvir-\vmax$ relation (see e.g. \citealt{2001MNRAS.321..559B, 2016MNRAS.462..893R}).
Quantitatively, the absolute residual decreases from $\sim 0.4$ dex at the low-mass end to  $\sim 0.2$ dex at higher \mvirp.
In the case of \tperi, although we find a clear correlation between the true \mvirp and the residuals, there are no significant performance differences across the \tperi bins, indicating that time since pericenter does not strongly impact \gnn’s ability to recover \mvirp.

\section{Comparison between CDM and SIDM}
\label{section:cdm_sidm}

\begin{figure}
    \centering
    \includegraphics[width=0.8\linewidth]{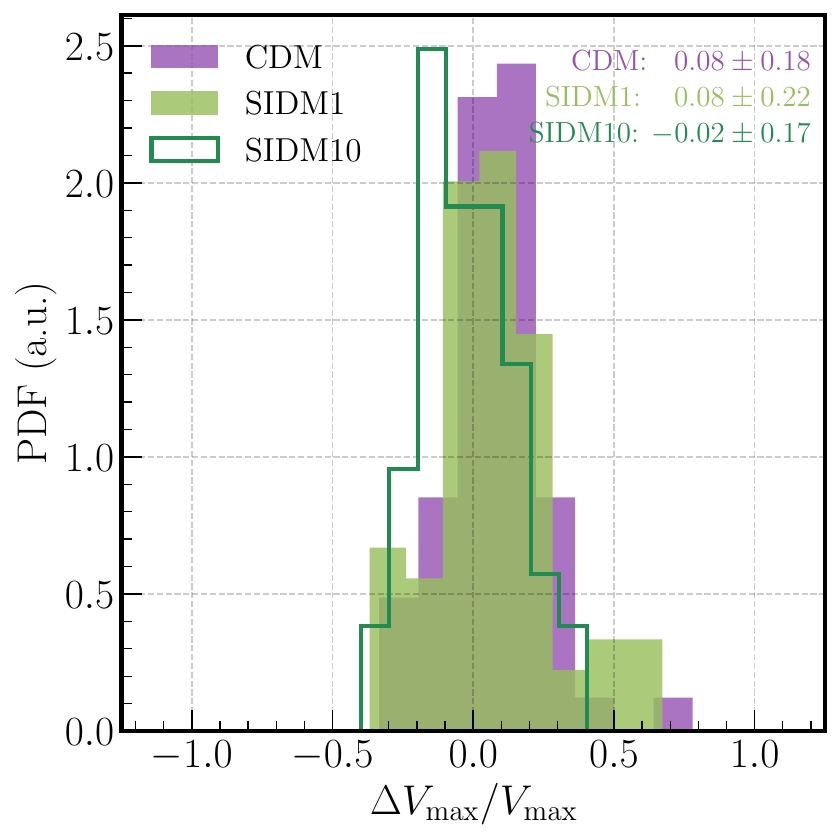}
    \includegraphics[width=0.8\linewidth]{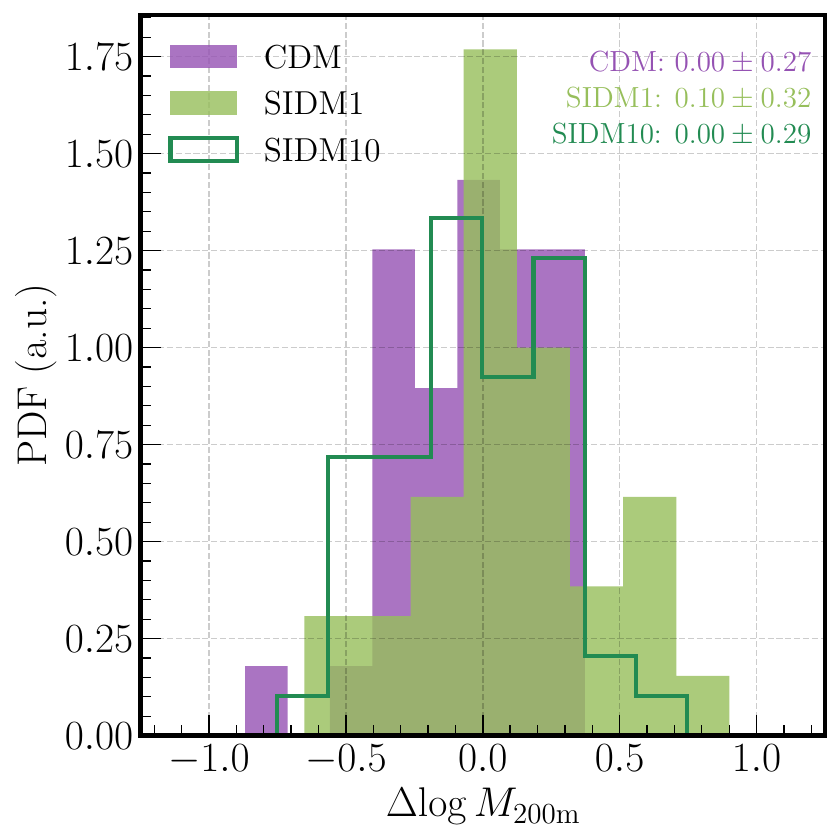}
    \caption{
    Normalized distributions of fractional residuals in \vmax (top) and \mvirp (bottom) for dwarf galaxies in the FIRE-2 CDM (purple), SIDM1 (green), and SIDM10 (dark green) simulations.
    The mean and standard deviation of the distribution for each simulation are indicated in the legend.
    The CDM, SIDM1, and SIDM10 simulations have 50, 72, and 53 mock galaxies, respectively. 
    }
    \label{fig:res_vmax_m200_hist}
\end{figure}

\input{table_sidm_cdm}

We briefly compare the performance of \gnn between the FIRE-2 CDM and SIDM dwarf galaxies. 
Prior works have shown that SIDM subhalos are more prone to tidal disruption effects than their CDM counterparts, due to their lower concentration and evaporation driven by ram pressure stripping~\citep{2000PhRvL..84.3760S, 2018PhR...730....1T, 2017MNRAS.472.2945R, 2022MNRAS.516.2389V}.
Here, we only consider the global trend across the entire population, irrespective of the orbital parameters discussed in Section~\ref{section:tidal}.

To minimize the environmental impacts of the host galaxies, we only consider \texttt{m12f\_CDM} and \texttt{m12m\_CDM} from the CDM simulations, comparing them against their SIDM counterparts.
We also compare the SIDM simulations with different cross-sections: $\sigma/m=1 \, \mathrm{cm^2 /g}$ and $\sigma/m = 10 \, \mathrm{cm^2 /g}$, which we simply denote as SIDM1 and SIDM10, respectively. 
The CDM, SIDM1, and SIDM10 simulations consist of 50, 72, and 53 mock galaxies, respectively. 

We remind readers of an important caveat: the SIDM simulations are run with a modified version of the FIRE-2 physics model that ignore the thermal-to-kinetic energy conversion during shock expansion from massive star mass loss (Section~\ref{section:simulation}).
As a result, the SIDM galaxies exhibit lower star formation rates and stellar masses, with a higher number of surviving galaxies due to increased survival probabilities. 
While we do not expect this modification to affect performance on individual CDM vs. SIDM galaxies, it introduces a systematic difference (independent of SIDM physics) in the galaxy populations.

Figure~\ref{fig:res_vmax_m200_hist} shows the normalized distribution of fractional residuals for \vmax (top) and \mvirp (bottom), with CDM and SIDM simulations represented in purple and green, respectively. 
The mean and standard deviation of each distribution are indicated in the figure. 

In the top panel, the mean residuals suggest that \gnn tends to overestimate \vmax for CDM and SIDM1, consistent with the trends discussed in Section~\ref{section:tidal}. 
This is further evidenced by the extended tails of the CDM and SIDM1 distributions. 
The model performs best on the SIDM10 samples, where the \vmax residual distribution is more centered around zero.
The residual distributions of \mvirp in the bottom panel are more comparable between the simulations. 
The SIDM1 distribution peaks the highest at zero, but also exhibits some outliers toward positive residuals, contributing to a larger overall spread.

Table~\ref{tab:sidm_cdm} presents the 68\% and 95\% coverage fractions for both \vmax and \mvirp across the CDM and SIDM samples.
The reported means and associated errors are derived from 1000 bootstrap resamples. 
The model shows comparable performance on the CDM and SIDM10 simulations. 
In contrast, SIDM1 exhibits lower overall coverage, likely driven by the long tail of high residuals shown in Figure~\ref{section:cdm_sidm}.

To conclude, despite these differences, the distributions remain broadly comparable. 
However, given the limited sample size, the large spread in the distributions, and the caveats mentioned above, it is difficult to draw definitive conclusions. 
For both \vmax and \mvir, the mean residuals between the simulations remain consistent within $1\sigma$ with each other. 
We leave a more detailed analysis to future work.

\section{Discussion}
\label{section:discussion}

\subsection{FIRE-2 Test Samples vs. Observational Samples}
\label{section:observation}

\begin{figure}
    \centering
    \includegraphics[width=0.8\linewidth]{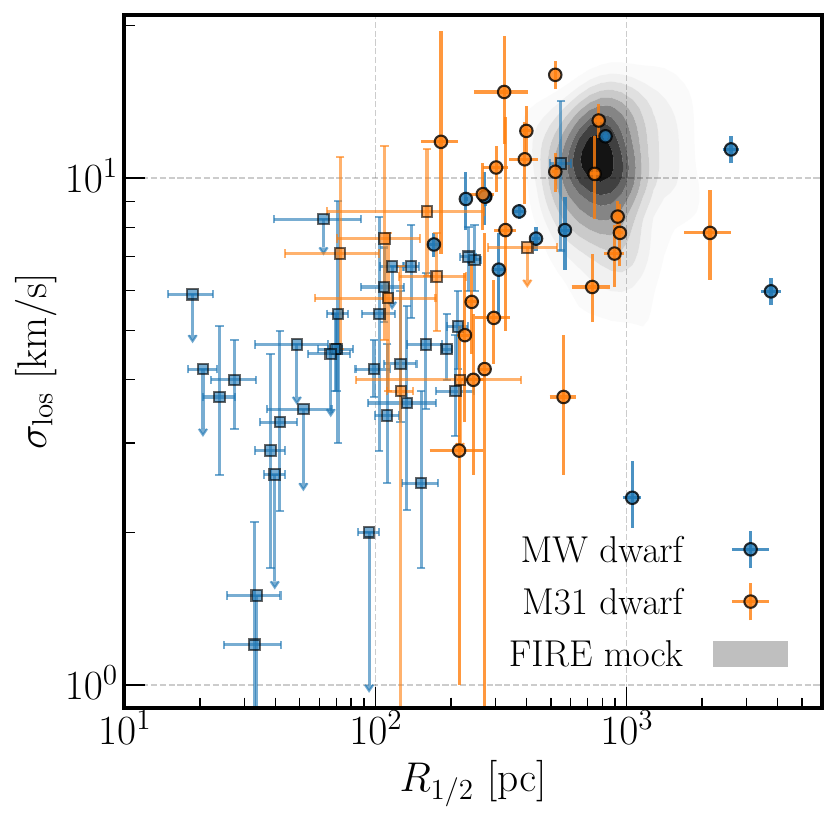}
    \caption{The line-of-sight velocity dispersion $\sigmalos$ versus the projected half-stellar mass radius $r_{\star, 1/2}$ for Milky Way dwarfs (blue, excluding SMC and LMC), M31 dwarfs (orange), and dwarfs in our FIRE-2 mock dataset (gray contours).
    The contours are shown at every 10\% intervals.
    Upper limits are denoted by downward arrows. 
    Ultra-faint dwarfs $(M_V > -7.7)$ are denoted as squares.
    }
    \label{fig:vdisp_rhalf}
\end{figure}

\begin{figure}
    \centering
    \includegraphics[width=0.8\linewidth]{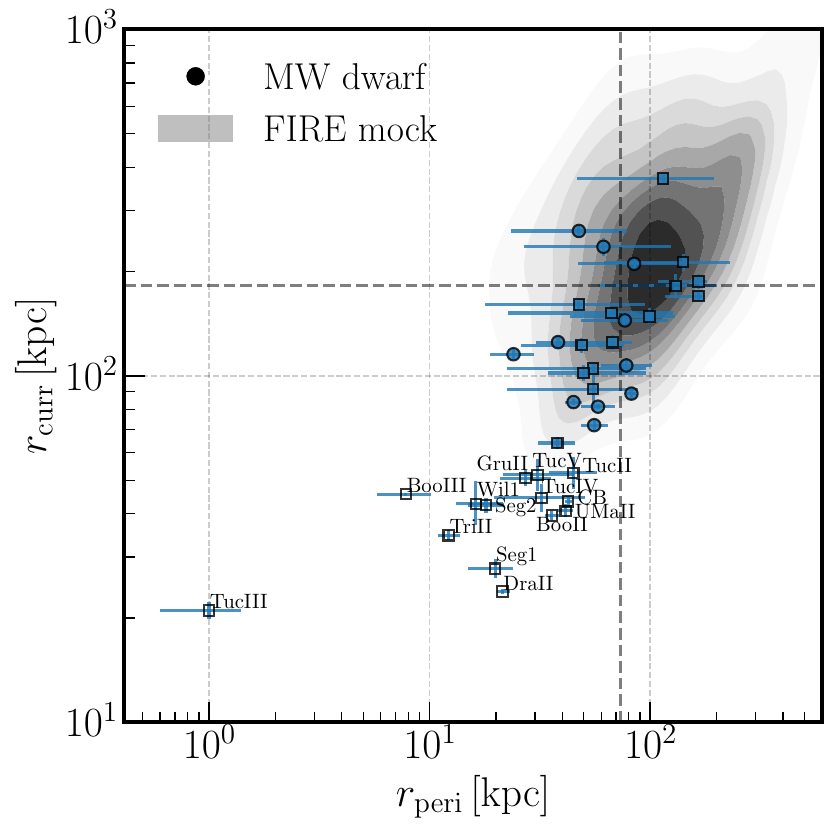}
    \caption{
    The pericentric distance \rperi versus current distance \rcurr for Milky Way dwarfs and our FIRE-2 mock dataset (gray contours).
    The contours are shown at every 10\% intervals.
    Ultra-faint dwarfs $(M_V > -7.7)$ are denoted as squares.
    For clarity, dwarf galaxies within and outside of the 90\% contour are denoted as filled and unfilled data points. 
    }
    \label{fig:orbit_mw_dwarf}
\end{figure}

Our ultimate goal is applying \gnn to real observational datasets of dwarf galaxies in the Local Group.
Therefore, it is crucial not only to evaluate the performance of \gnn on more realistic simulations, such as FIRE-2, but to also determine how well the dwarf galaxies in our test datasets match observations.
Here, we compare the properties of FIRE-2 dwarf galaxies with those of the Milky Way to identify key discrepancies and what constitutes a reliable dwarf target for our mass modeling method in future analysis.

We emphasize that here we do not seek to compare the FIRE-2 physics model directly with observations. 
Rather, we aim to assess how the ranges of properties in existing observational samples compare to those of the FIRE-2 dwarf galaxies included in our test samples, and highlight which real dwarf galaxies are considered ``out-of-distribution''. 
This is important, as it allows us to identify the regimes where the results in Section~\ref{section:tidal} may not fully generalize, providing critical context for interpreting the results of this study.

Previous works have shown that satellites in FIRE-2 simulations broadly match the observed satellite population in the Local Group. 
\citet{2016ApJ...827L..23W} and \citet{2019MNRAS.487.1380G} demonstrated that the simulated satellite stellar mass functions and line-of-sight velocity dispersions are consistent with those of the Milky Way and M31, down to approximately $M_\mathrm{\star} \sim 10^5 \, \modot$ and $\sigmalos \sim 5 \, \kms$.
This agreement excludes ultra-faint dwarfs, as FIRE-2 does not yet resolve this regime due to limitations in mass and spatial resolutions.

Figure~\ref{fig:vdisp_rhalf} shows the line-of-sight velocity dispersion $\sigmalos$ versus the projected half-stellar mass radius $R_{\star, 1/2}$.
Data for Milky Way (blue) and M31 (orange) dwarfs are from the Local Volume catalog presented in~\citet{2024arXiv241107424P}.
The gray contours represent the distribution of $\sigmalos$ and $R_\mathrm{1/2}$ of dwarf galaxies from our FIRE-2 mock catalog.
Ultra-faint dwarfs, defined as $M_\mathrm{V} > -7.7$ \citep{2019ARA&A..57..375S}, are marked as squares to highlight that they are not expected to be resolved in FIRE-2.

As expected, the dwarf galaxies in our mock samples predominantly occupy regions of the parameter space corresponding to higher $\sigmalos$ compared to the Milky Way and M31 samples. 
As discussed above, this discrepancy arises because FIRE-2 simulations do not resolve ultra-faint dwarfs.
Additionally, the FIRE-2 dwarfs generally exhibit larger half-stellar mass radii $R_{\star, 1/2}$ compared to the observed Milky Way and M31 samples, consistent with the findings of \citet{2024ApJ...966..131S}.
Of the 43 Milky Way dwarfs shown, only three are within this 90\% contour of the simulation, namely Bootes III, Fornax, and Sextans. 

Despite differences between the FIRE-2 mock samples and observations, it is important to note that the majority of the observed data samples remain well within the prior distribution.
From Table~\ref{tab:priors}, the Plummer scale radius can range from $r_\mathrm{\star} \in [20, 20000]\,\mathrm{pc}$,\footnote{For a Plummer profile, the projected half-stellar mass radius is equal to the scale radius, i.e., $R_{\star, 1/2} = r_\mathrm{\star}$.} while the velocity dispersion can range from $\sigma_\mathrm{3-D} \in [0.1, 50] \, \kms$ (or $\sigmalos \in [0.06, 28.9] \, \kms$, assuming isotropy).
The three galaxies with the smaller half-stellar mass radii are Triangulum II, Segue 1, and Willman 1 with values of $20.56^{+2.74}_{-2.61}\,\mathrm{pc}$, $23.90^{+3.74}_{-3.40}\,\mathrm{pc}$, and $27.43^{+5.89}_{-5.35}\,\mathrm{pc}$, respectively (see \citealt{2024arXiv241107424P} and references therein).
While these technically fall within the prior range, Triangulum II is very close to the lower boundary. This proximity may artificially truncate the posterior distribution, preventing a proper exploration of smaller values and potentially biasing parameter inferences.
To better accommodate such observations, we plan to extend the prior limits in future analyses.

More importantly, as shown in N23, the performance of \gnn is largely scale-free. 
In our idealized simulations, we observed no significant differences in performance across samples with varying DM and stellar profiles (see Figure 1 of N23). 
This robustness likely extends to more realistic simulations, such as FIRE. 
That said, environmental effects--particularly tidal effects--could introduce scale-dependent differences in performance, as less massive and less compact galaxies are likely to be more disrupted.
Comparing the FIRE-2 populations to observed dwarf galaxies is not straightforward. The FIRE-2 mock samples are generally more massive yet less compact than observed galaxies, leading to competing effects that may partially offset each other. 
Additionally, \citet{2023ApJ...949...44S} showed that FIRE-2 satellites may be more prone to tidal effects than observations, with many satellites disrupting with relatively high densities. 

Therefore, we now compare the orbital properties of our mock galaxies with Milky Way dwarfs to highlight out-of-distribution real dwarf galaxies. 
A prior work from \citet{2020MNRAS.491.1471S} has shown that the radial distribution of FIRE-2 satellites are consistent with Local Group and Milky Way-analog in the SAGA survey~\citep{2017ApJ...847....4G, 2021ApJ...907...85M, 2024ApJ...976..117M} for $M_\star \gtrsim 10^5 \, \modot$.

Figure~\ref{fig:orbit_mw_dwarf} shows the pericentric distance \rperi versus the current distance \rcurr of the Milky Way dwarfs (circles) and the FIRE-2 simulations (gray contours, shown at every 10\% interval.
For clarity, we represent dwarfs within and outside of the 90\% contour with filled and unfilled data points, respectively.
Similarly, ultra-faint dwarfs $(M_\mathrm{V} > -7.7)$ are marked as squares and not expected to be resolved by FIRE-2.
The vertical and horizontal dashed lines correspond to $0.2$ and $0.5$ times $R_\mathrm{200m}$ of the Milky Way, assumed to be $365 \pm 76 \, \mathrm{kpc}$~\citep{2020MNRAS.496.3929D}.
Orbital properties are taken from \citet{2022ApJ...940..136P}, which uses mixture models to determine the systemic proper motions of the Milky Way satellites and subsequently fits their orbits, accounting for both the Milky Way and LMC potentials.
About half of the Milky Way satellites are within the 90\% contour of the simulations, which typically occupy the high \rperi and \rcurr regions. 
This difference highlights a numerical limitation: simulated satellites too close to their hosts are often disrupted and become much more challenging to track with \rockstar~\citep{2013ApJ...762..109B, 2024ApJ...970..178M}.

There are 13 satellites within the 90\% contours that have a mean $\rperi < 0.2 R_\mathrm{200m}$ and $\rcurr < 0.5 R_\mathrm{200m}$ of the Milky Way. These satellites belong to the more disrupted tidal bins in Section~\ref{section:tidal}, so we expect \gnn to perform worse for these cases. 

The 14 satellites outside of the 90\% contour are annotated in Figure~\ref{fig:orbit_mw_dwarf}.
Note that all of these satellites are necessarily disrupted--some may have large densities and thus are more resilient to tidal effects. 
However, for those that are disrupted, the degree of tidal effects may not be fully captured in the FIRE-2 simulations due to numerical limitations.
As such, the results presented in Section~\ref{section:tidal} may require extrapolation to account for these effects.

Additionally, we list satellites identified as potentially disrupted in \citet{2022ApJ...940..136P} and other studies.
Tucana III and Antlia III has been identified as being tidally disrupted based on priors studies (see \citealt{2015ApJ...813..109D, 2018ApJ...862..114S, 2018ApJ...866...22L} for Tucana III and \citealt{2021ApJ...921...32J, 2022ApJ...926...78V} for Antlia III).
\citet{2022ApJ...940..136P} also identified additional satellites potentially undergoing tidal disruption due to their low pericentric distances and low densities relative to the Milky Way, including Bootes I, Bootes III, Crater II, Grus II, Segue 2, and Tucana IV.
In a forthcoming work, we will provide a more detailed discussion of the tidal effects on these satellites and their implications for \gnn. 
For now, we direct readers to the discussion in \citet{2022ApJ...940..136P} and the references therein.

\subsection{Comparison with distribution function fitting}
\label{section:df}

As briefly discussed in Section~\ref{section:intro}, our method is a form of massively accelerated distribution function (DF) fitting~\citep{2002MNRAS.330..778W, 2018MNRAS.480..927P, 2021MNRAS.501..978R}.
Traditional DF methods rely on the assumption of steady-state dynamics, where the phase-space distribution is expressed as a function of conserved integrals of motion  $I(\{\vec{x}_i, \vec{v}_i\})$.
Under this assumption, the likelihood is given by,
\begin{equation}
    \label{eq:loglike_df}
    \mathcal{L}_\mathrm{DF} = \prod_{i=1}^N f(I(\{\vec{x}_i, \vec{v}_i\}); \theta).
\end{equation}
The integral of motion $I(\{\vec{x}_i, \vec{v}_i\})$ is chosen to be the energy-angular momentum $\{E, L\}$, or the action-angle variable 
$J$, as in~\citet{2021MNRAS.501..978R}.
The likelihood of a given phase-space coordinate, $\{\vec{x}_i, \vec{v}_i\}$, is calculated by first computing $I$ and evaluating the DF at $f(I)$.
Conversely, to sample $\{\vec{x}_i, \vec{v}_i\}$ from $f(I)$, \textsc{Agama} can generate samples in the space of $I$ and then transform them back to phase-space coordinates using an improved version of the torus mapping approach in~\citet{2016MNRAS.456.1982B}.

We expect the performance of \gnn to be comparable to the \textsc{Agama} DF method~\citet{2021MNRAS.501..978R}.
However, we note a few key advantages:

First, as with other NPE methods, \gnn is amortized, i.e. once trained, the model can be applied to any observation without re-training. 
In contrast, likelihood-based methods, including DF methods, must be rerun for each new dataset.  
This significantly limits the parameter space that can be explored with these models, particularly when extending to non-spherical cases.  
For systems with 1000 tracers, the \textsc{Agama} DF method in~\citet{2021MNRAS.501..978R} takes approximately $\mathcal{O}(10^3)$ minutes per run, comparable to the total training time of \gnn.  
However, once trained, \gnn performs inference in milliseconds, regardless of the tracer counts, making it far more efficient for large-scale applications.
While computational cost may not be a major concern for real observations, given that the current number of observed dwarf galaxies does not yet pose a computational barrier~\citep{2024arXiv241107424P}, it becomes critical when testing and exploring the modeling parameter space on mock simulations.
For instance, the detailed evaluations of mass modeling and tidal effects in Section~\ref{section:tidal} would be impractical with DF methods on the full FIRE-2 mock dataset, due to their high computational cost.

Second, to evaluate the likelihood, traditional DF methods must map an observed phase-space coordinate $\{\vec{x}_i, \vec{v}_i\}$ to the integrals of motion $I$.
For datasets with incomplete phase-space information (e.g., missing proper motions or line-of-sight distances), these methods require marginalization over the missing coordinates through Monte Carlo sampling and evaluation of complex multidimensional integrals. 
This process is not only computationally intensive, but can also introduce numerical noise.
In contrast, the NPE likelihood in Equation~\ref{eq:loglike_npe} is never explicitly evaluated, but instead learned implicitly through the training process.
Consequently, generating the training set requires only sampling from $f(I)$, mapping $I \rightarrow \{\vec{x}_i, \vec{v}_i\}$, and discarding the corresponding components of $\{\vec{x}_i, \vec{v}_i\}$ that are missing in the data.
This approach thus entirely circumvents the need for complex marginalization integrals required in traditional DF methods.

Finally, it is unclear how traditional DF methods can be extended to model non-equilibrium dynamics. 
While \gnn in this study also assumes pseudo-dynamical equilibrium--via the use of orbital integrals of motion for generating the training set--this assumption is not intrinsic to the method.
By construction, \gnn can incorporate more general dynamical states simply by expanding the training data to include systems undergoing tidal disruption, mergers, or other out-of-equilibrium processes.
This flexibility makes simulation-based approaches such as \gnn a promising path forward for extending mass modeling beyond equilibrium assumptions~\citep[e.g.][]{2015NatCo...6.7599U, 2025arXiv250113148W}, an avenue we leave for future work.

\section{Conclusion}
\label{section:conclusion}

Dwarf galaxies and their mass density profiles hold the key to understanding structure formation and the particle nature of DM.
Accurate mass modeling of these galaxies is essential for constraining DM properties and distinguishing between different theoretical scenarios, such as CDM and SIDM.
As new spectroscopic surveys such as MUSE-Faint~\citep{2020A&A...635A.107Z, 2021A&A...651A..80Z, 2021arXiv211209374Z, 2023A&A...678A..38J} and DELVE (DECam Local Volume Exploration survey;~\citealt{2022ApJS..261...38D, 2025ApJ...979..176T}) continue to expand kinematic measurements of dwarf galaxy populations, robust modeling techniques that maximize the information extracted from stellar kinematics become increasingly valuable. 

In this work, we assess the performance of \gnn, an SBI framework first introduced in \cite{2023PhRvD.107d3015N}, on simulated dwarf galaxies in Milky Way-like galactic environments.
The framework employs GNNs and normalizing flows to infer DM density profiles from line-of-sight stellar velocities. 
We train \gnn on a Monte Carlo simulation of dwarf galaxies, where each system is generated by sampling from a spherical equilibrium DF model.
Unlike traditional Jeans-based approaches, \gnn can leverage the full phase-space distribution of tracers, incorporating higher-order velocity moments and spatial correlations to maximize the information extracted from stellar kinematics.

We apply our framework to 96 dwarf galaxies from the five CDM simulations of the \latte~\citep{2016ApJ...827L..23W, 2023ApJS..265...44W} suite of FIRE-2 simulations~\citep{2018MNRAS.480..800H}, along with 61 satellites from the SIDM simulations with a self-interaction cross-section of $\sigma/m = 1 \, \mathrm{cm^2/g}$ and 46 from those with $\sigma/m = 10 \, \mathrm{cm^2/g}$~\citep{2021MNRAS.507..720S, 2022MNRAS.516.2389V, 2024ApJ...974..223A}.

In Sections~\ref{section:jeans_gnn} and \ref{section:gs_gnn}, we compared \gnn to two Jeans-based methods: a Simple Jeans model using an unbinned Gaussian likelihood~\citep{2008ApJ...678..614S}, and \gs, a higher-order moment method incorporating virial shape parameters~\citep{2017MNRAS.471.4541R}.
We randomly selected 36 and 16 FIRE-2 galaxies for these comparisons, respectively.
For each galaxy, we assess the inferred DM density $\rho(r)$ and velocity anisotropy profiles $\beta(r)$, as well as three mass and structural parameters: the peak circular velocity \vmax, the half-stellar radius mass \mhl, and the peak virial mass \mvirp.
We summarize our findings below:
\begin{itemize}
    \item  
    Compared to Simple Jeans, we find that \gnn produces tighter and more accurate constraints on the DM density and anisotropy profiles. 
    This is particularly evident when tracer counts are low, as \gnn continues to provide robust constraints, whereas Simple Jeans struggles to reliably recover the density and anisotropy profiles. 
    When tracer counts are high, however, the performances of \gnn and Simple Jeans become more comparable, as the additional line-of-sight velocities can help constrain the anisotropy and mitigate the mass-anisotropy degeneracy.
    
    \item When we compare \gnn to the \gs higher-order Jeans model, we find that the performance is overall similar, with discrepancies primarily arising from differences in modeling choices and prior distributions.
    Notably, the differences in prior distributions results in \gs (\gnn) favoring steeper (shallower) inner and outer density profiles.

    \item To assess the accuracy of different methods in recovering \vmax, \mhl, and \mvir, we compute the absolute errors and square errors between the predicted and true values of these parameters, as well as the negative log-likelihood of the predicted posterior distributions. 
    We summarize these metrics in Table~\ref{tab:jeans_gnn} for \gnn versus Jeans and Table~\ref{tab:gs_gnn} for \gnn versus \gs.
    Overall, \gnn significantly outperforms the Jeans model across all metrics. 
    While \gnn also achieves slightly better recovery of mass and structural parameters compared to \gs, the improvements are relatively modest.
    We discuss key advantages of \gnn against traditional modeling methods such as \gs and DF fitting in Section~\ref{section:df}.
\end{itemize}

In Section~\ref{section:tidal}, we analyzed how well \gnn can recover \vmax and \mvirp across different levels of tidal effects n in the full FIRE-2 mock dataset.
Our focus is not on satellites that are heavily disrupted, but rather on those where tidal signatures, such as tidal tails or velocity gradients, are subtle or difficult to detect, a scenario found to be common in \cite{2023ApJ...949...44S}.
To quantify tidal effects, we consider four orbital parameters: the number of pericentric passages \nperi, the last pericentric distance \rperi, the current distance from the host \rcurr, and the time since the last pericentric passage \tperi.
We present results on the impact of tidal effects on the recovery of \vmax and \mvirp, with Section~\ref{section:tidal_nperi} providing a detailed example using three \nperi bins and Section~\ref{section:tidal_all} extending the analysis to the remaining orbital parameters.
Our conclusion is as follows:
\begin{itemize}
    \item 
    Overall, the recovery is strong for both \vmax and \mvirp. 
    We observe that \gnn systematically overestimates both parameters for the more tidally disrupted bins.
    Despite this, for the full FIRE-2 dataset, \gnn recovers \vmax within the 68\% and 95\% confidence intervals for 64\% and 93\% of galaxies, respectively, while for \mvirp, the corresponding fractions are 71\% and 95\%.
   
    \item The overestimation is primarily observed in lower-mass galaxies ($\vmax < 20,\kms$ and $\mvirp < 10^9 \modot$) that have undergone significant tidal disruption. 
    In contrast, \gnn remains robust for more massive galaxies, even when they experience strong tidal effects.
    The mean absolute residuals in \vmax increase from $10\%$ in the least disrupted galaxies to $20\%$ in the most disrupted ones, while for \mvirp, they rise from $0.2$ to $0.4$ dex.
    
    \item 
    For \vmax, the coverage fraction, defined as the proportion of true values that fall within the expected confidence interval, shows a clear decreasing trend in the more disrupted bins for all orbital parameters.
    In contrast, the trends for \mvirp are much less pronounced (see Table~\ref{tab:tidal}).

    \item 
    Of the four orbital parameters, \rperi shows the clearest trends, with the 68\% coverage drops significantly for the most disrupted bin ($\rperi < 0.2 R^\mathrm{peri}_\mathrm{host, 200m}$), consistent with past findings~\citep[e.g.][]{2023ApJ...949...44S, 2024arXiv241009143S, 2024MNRAS.527.5868M}.
    However, even in this case, \gnn recovers \vmax within the 68\% and 95\% confidence interval for approximately 50\% and 90\% of the samples, respectively. 
    Similarly, for \mvirp, around 60\% of the sample falls within the 68\% confidence interval, while just under 90\% falls within the 95\% confidence interval.

    \item
    We find that \gnn tends to be slightly conservative when predicting \mvirp, as indicated by coverage fractions that exceed their expected confidence intervals, typically by a few percents. 
    For example, across the entire population, the coverage fraction is 71\% at the 68\% confidence level (see Table~\ref{tab:tidal}).
    This weaker constraint likely comes from the additional uncertainties introduced when extrapolating the DM density profiles to $r_\mathrm{200m}$. 
    These uncertainties could be mitigated by employing more accurate and flexible tail models for the DM density profiles or by extending the tracer populations to larger radii. 
\end{itemize}

Lastly, we briefly compare the performance of \gnn across the CDM and SIDM simulations with different cross-sections by examining their residuals. 
Overall, the model performs similarly for CDM and the high-interaction SIDM scenario ($\sigma/m = 10 \, \text{cm}^2/\text{g}$), while showing lower coverage for the low-interaction SIDM case ($\sigma/m = 1 \, \text{cm}^2/\text{g}$). 
Despite some differences in distributions, the mean residuals for both \vmax and \mvirp remain consistent within $1\sigma$ across simulations. 
These findings demonstrate the robustness of \gnn while highlighting the need for further testing with larger datasets to draw more definitive conclusions.

It is important to note that \gnn is trained on idealized dynamical models assuming spherical equilibrium, rather than on galaxies in complex hydrodynamical environments like those in FIRE.
A common challenge for machine learning methods is their ability to generalize to out-of-distribution data, where deviations from the training distribution can degrade performance. 
Despite this, \gnn robustly recovers the density and velocity anisotropy profiles, as well as key mass and structural parameters, even when applied to galaxies in realistic hydrodynamical simulations.

We further highlight the novelty and importance of evaluating how well mass modeling methods can recover \vmax and \mvirp of dwarf galaxy simulations under realistic galactic environments. 
A previous study by \cite{2020MNRAS.498..144G} conducted a comprehensive test of \gs on APOSTLE dwarf galaxies, demonstrating that its recovery of \mhl is comparable to that of mass estimators while also examining the impact of tidal effects.
Our work additionally examines \vmax and \mvirp, both of which are crucial for constraining cosmological models~\citep[e.g.][]{1995ApJ...447L..25B, 2001MNRAS.321..559B, 2003MNRAS.345..923V, 2016MNRAS.462..893R, 2019MNRAS.487.5799R, 2021arXiv210609050K}.
This evaluation is particularly timely, as discrepancies between leading mass modeling approaches, such as \textsc{CJAM}~\citep{2013MNRAS.436.2598W} and \gs, suggest that different methods can yield significantly different mass estimates when applied to the same dataset~\citep{2021arXiv211209374Z}, underscoring the need for rigorous validation.

In ongoing work, we aim to apply \gnn to observed dwarf galaxies in the Milky Way and the Local Group. 
We discuss key differences between galaxies in our mock FIRE-2 dataset and real observations in Section~\ref{section:discussion}.
Before applying GraphNPE to observational data, several considerations need to be addressed: incorporating measurement uncertainties, accounting for selection biases, and masking contamination from foreground and binary stars. 
The first two can be readily handled by integrating them into the training or inference process (see e.g. \citealt{2023ApJ...952L..10W}), while contamination requires careful data processing and member star selection.

An important direction for future work is extending \gnn to ultra-faint dwarfs, which represent some of the most DM-dominated systems in the universe. 
Although here we focus on classical dwarfs due to the FIRE-2 resolution limit, there is a compelling case that \gnn should perform well on the ultra-faints given their extremely low stellar tracer counts, a regime where \gnn has shown advantages over traditional methods. 
However, to properly validate \gnn for ultra-faint applications, we require higher-resolution cosmological simulations to stress-test and better quantify its performance on systems with low velocity dispersions and sparse kinematic data. 
As next-generation surveys continue to discover ultra-faint systems, establishing the reliability of mass modeling approaches in these data-limited regimes will be crucial for maximizing their scientific potential.

Furthermore, while this study applies \gnn to spherical equilibrium models, its framework is not inherently restricted to such cases, unlike traditional DF methods.
\gnn can be extended to out-of-equilibrium systems~\citep[e.g.][]{2015NatCo...6.7599U, 2025arXiv250113148W} by expanding the training set to include galaxies undergoing tidal disruption, mergers, and other dynamical perturbations, which we will pursue in future work. 
This will allow \gnn to recover mass profiles in systems where equilibrium assumptions break down, further broadening its applicability to realistic astrophysical environments.

To conclude, in this work, we establish \gnn as a robust and efficient DF-based mapping method for inferring DM density profiles in dwarf galaxies, providing a promising avenue for constraining DM models. 
By leveraging machine learning, \gnn enables rapid, amortized inference, making it particularly well-suited for large-scale applications such as systematically exploring the mass modeling parameter space and investigating the effects of tidal disruption across mock simulations.
More broadly, our work highlights the power of SBI and graph-based learning in astrophysics, paving the way for more adaptable and data-driven approaches to mass modeling in the era of increasingly precise kinematic surveys.

\section*{Acknowledgments}
We thank Jenna Samuel, Nondh Panithanpaisal, Nora Shipp, Stephanie O'Neil, Arpit Arora, Xiaowei Ou for helpful discussions.

TN, SM, and LN is supported by the National Science Foundation under Cooperative Agreement PHY-2019786 (The NSF AI Institute for Artificial Intelligence and Fundamental Interactions, \url{http://iaifi.org/}).
TN is also supported by a CIERA Postdoctoral Fellowship.
LN is also supported by the Sloan Fellowship, the NSF CAREER award 2337864, NSF award 2307788.
JIR would like to acknowledge support from STFC grants ST/Y002865/1 and ST/Y002857/1.
CAFG was supported by NSF through grants AST-2108230 and AST-2307327; by NASA through grants 21-ATP21-0036 and 23-ATP23-0008; and by STScI through grant JWST-AR-03252.001-A.
AW received support from NSF, via CAREER award AST-2045928 and grant AST-2107772.
TS gratefully acknowledges the support of the NSF-Simons AI-Institute for the Sky (SkAI) via grants NSF AST-2421845 and Simons Foundation MPS-AI-00010513 and support by NASA grants 22-ROMAN22-0055 and 22-ROMAN22-0013.
This material is based upon work supported by the U.S. Department of Energy, Office of Science, Office of High Energy Physics of U.S. Department of Energy under grant Contract Number  DE-SC0012567.

This work used Bridges-2~\citep{10.1145/3437359.3465593} at Pittsburgh Supercomputing Center through allocation phy210068p from the Advanced Cyberinfrastructure Coordination Ecosystem: Services \& Support (ACCESS) program, which is supported by National Science Foundation grants \#2138259, \#2138286, \#2138307, \#2137603, and \#2138296. 
Other numerical calculations were run on the Northwestern computer cluster Quest, the Caltech computer cluster Wheeler, Frontera allocation FTA-Hopkins/AST20016 supported by the NSF and TACC, XSEDE/ACCESS allocations ACI-1548562, TGAST140023, and TG-AST140064 also supported by the NSF and TACC, and NASA HEC allocations SMD-16-7561, SMD-17-1204, and SMD-16-7592. 
The computations in this work were, in part, run at facilities supported by the Scientific Computing Core at the Flatiron Institute, a division of the Simons Foundation.
The data used in this work were, in part, hosted on equipment supported by the Scientific Computing Core at the Flatiron Institute, a division of the Simons Foundation.

\section*{Software}
This research makes use of the following packages:
\textsc{Agama}~\citep{2019MNRAS.482.1525V},
\gs~\citep{2018MNRAS.481..860R, 2021MNRAS.505.5686C},
\textsc{IPython}~\citep{PER-GRA:2007}, 
\textsc{Jupyter}~\citep{2016ppap.book...87K},
\textsc{Matplotlib}~\citep{2007CSE.....9...90H},
\textsc{NumPy}~\citep{harris2020array},
\textsc{PyTorch}~\citep{2019arXiv191201703P}, 
\textsc{PyTorch Geometric}~\citep{2019arXiv190302428F}, 
\textsc{PyTorch Lightning}~\citep{william_falcon_2020_3828935},
\textsc{SciPy}~\citep{2020SciPy-NMeth},
\textsc{zuko}~\citep{2023zndo...7625672R}

\section*{Data Availability}

The CDM FIRE-2 simulations are publicly available through the FIRE project website (\url{http://fire.northwestern.edu/data/}). 
The SIDM simulation data used in this study are not publicly available but can be provided upon reasonable request to the corresponding author.
Additional simulation data, including the specific analysis outputs and derived data products generated for this study, are available upon reasonable request to the corresponding author.
The trained model weights are available at upon reasonable request to the corresponding author.



\bibliographystyle{mnras}
\bibliography{references} 

\appendix
\input{appendix}

\bsp	
\label{lastpage}
\end{document}

%% file: table_simulation.tex
\begin{table}
\centering
\begin{tabularx}{1\linewidth}{lIIII}
\hline
simulation & $M_\mathrm{200m}$ &  $R_\mathrm{200m}$ & $\sigma / m$  & $N_\mathrm{gal}$  \\
& $[\mathrm{M_\odot}]$ & $[\mathrm{kpc}]$ & $[\mathrm{cm^{2}/g}]$ \\ 
\hline
\texttt{m12c\_CDM} & $1.35 \times10^{12}$ & 351 & 0 & 23 \\ 
\texttt{m12b\_CDM} & $1.43 \times10^{12}$ & 358 & 0 & 16 \\
\texttt{m12f\_CDM} & $1.71 \times10^{12}$ & 380 & 0 & 24 \\
\texttt{m12i\_CDM} & $1.18 \times10^{12}$ & 336 & 0 & 12 \\
\texttt{m12m\_CDM} & $1.58 \times10^{12}$ & 371 & 0 & 36 \\ 
\hline
\texttt{m12f\_SIDM1} & $1.40 \times 10^{12}$ & 352 & 1 & 30\\ 
\texttt{m12m\_SIDM1} & $1.24 \times 10^{12}$  & 337 & 1 & 42 \\ 
\texttt{m12f\_SIDM10} & $1.35 \times 10^{12}$ & 346 & 10 & 19 \\ 
\texttt{m12m\_SIDM10} & $1.20 \times 10^{12}$ & 333 & 10 & 34\\ 
\hline
\end{tabularx}
\caption{Simulation specifications of the FIRE-2 simulations used in this work. Values for virial mass and radius are taken from~\citet{2023ApJS..265...44W}.}
\label{tab:simulations}
\end{table}

%% file: table_prior.tex
\begin{table*}
\begin{center}
\begin{tabular}{llll}
\hline
Parameter & Name & Equation & Prior\Tstrut\Bstrut	\\   
\hline
$\log_{10}(\rho_0 / (\mathrm{M_\odot \, kpc^{-3}} ))$ & DM density normalization & gNFW (Eq.~\ref{eq:gNFW}) & $ [3, 10]$ \\
$\log_{10}(r_\mathrm{dm} / \mathrm{kpc})$ & DM scale radius & gNFW (Eq.~\ref{eq:gNFW}) & $ [-2, 2]$ \\  
$\gamma$ & DM inner slope & gNFW (Eq.~\ref{eq:gNFW}) & $ [-1, 2]$\\ 
\hline
$r_\star / r_\mathrm{dm}$ & Stellar scale radius & Plummer (Eq.~\ref{eq:plummer_2D}) & $ [0.2, 1]$\\ 
\hline
$r_a / r_\star$ & Velocity anisotropy scale radius & OM (Eq.~\ref{eq:velani_OM}) & $ [0.1, 10]$\\ 
$\beta_0 $ & Velocity anisotropy normalization & OM (Eq.~\ref{eq:velani_OM}) & $ [-0.5, 1]$  \\ 
\hline
\end{tabular}
\end{center}
\caption{Prior ranges of the training simulations before the velocity dispersion cut.}
\label{tab:priors}
\end{table*}

%% file: table_jeans_gnn.tex
\begin{table}
\begin{center}
\begin{tabular}{c|ccc}
\hline
\multirow{2}{*}{} & \multicolumn{3}{c}{$\vmax \, (\mathrm{km / s})$} \\
& AE$^{1}$ & SE$^{2}$ & NLL$^{3}$ \\ 
\hline
Simple Jeans 
& $7.73 \pm 2.07$
& $210.01 \pm 138.03$
& $4.42 \pm 0.17$
\\ 
\gnn
& $3.60 \pm 0.38$
& $18.15 \pm 3.02$
& $3.51 \pm 0.31$
\\
\hline \hline
& \multicolumn{3}{c}{$\log_{10} (\mhl$ / \modot)} \\
& AE  & SE & NLL \\ 
\hline
Simple Jeans
& $0.11 \pm 0.01$
& $0.02 \pm 0.00$
& $-0.46 \pm 0.09$
\\
\gnn 
& $0.11 \pm 0.01$
& $0.02 \pm 0.00$
& $-0.59 \pm 0.08$
\\
\hline \hline
& \multicolumn{3}{c}{$\log_{10} (\mvirp$ / \modot)} \\
& AE  & SE & NLL \\ 
\hline
Simple Jeans 
& $0.41 \pm 0.08$
& $0.38 \pm 0.14$
& $1.07 \pm 0.07$
\\ 
\gnn 
& $0.24 \pm 0.02$
& $0.08 \pm 0.01$
& $0.33 \pm 0.03$
\\
\hline
\end{tabular}
\end{center}
$^1$ {\footnotesize Absolute Error(AE). Lower is better.}
\\
$^2$ {\footnotesize Squared Error (SE). Lower is better.}
\\
$^3$ {\footnotesize Negative Log-Likelihood (NLL). Lower is better. To estimate the NLL of each parameter, the posterior was first fitted using Gaussian Kernel Density Estimate (KDE) and evaluated at the target point.}
\caption{Performance of \gnn and Jeans modeling for the maximum circular velocity \vmax, the half-light mass \mhl, and the virial mass \mvir over the selected sample of galaxies for multiple metrics.}
\label{tab:jeans_gnn}
\end{table}

%% file: table_gs_gnn.tex
\begin{table}
\begin{center}
\begin{tabular}{c|ccc}
\hline
\multirow{2}{*}{} & \multicolumn{3}{c}{$\vmax \, (\mathrm{km / s})$} \\
& AE & SE & NLL \\ 
\hline
\gs
& $4.71 \pm 0.87$
& $33.60 \pm 12.65$
& $3.59 \pm 0.06$
\\ 
\gnn 
& $4.15 \pm 0.58$
& $22.23 \pm 4.74$
& $3.33 \pm 0.09$
\\
\hline \hline
& \multicolumn{3}{c}{$\log_{10} (\mhl$ / \modot)} \\
& AE  & SE & NLL \\ 
\hline
\gs
& $0.09 \pm 0.02$
& $0.02 \pm 0.01$
& $-0.78 \pm 0.24$
\\
\gnn 
& $0.08 \pm 0.02$
& $0.01 \pm 0.00$
& $-0.90 \pm 0.14$
\\
\hline \hline
& \multicolumn{3}{c}{$\log_{10} (\mvirp$ / \modot)} \\
& AE  & SE & NLL \\ 
\hline
\gs
& $0.41 \pm 0.06$
& $0.23 \pm 0.06$
& $0.79 \pm 0.08$
\\ 
\gnn 
& $0.27 \pm 0.03$
& $0.09 \pm 0.01$
& $0.36 \pm 0.03$
\\
\hline
\end{tabular}
\end{center}
\caption{Performance of \gnn and \gs for the maximum circular velocity \vmax, the half-light mass \mhl, and the virial mass \mvir over the selected sample of galaxies for multiple metrics.}
\label{tab:gs_gnn}
\end{table}

%% file: table_tidal.tex
\begin{table*}
\begin{center}
\begin{tabular}{lcccc}
\hline
 & \multicolumn{2}{c}{\vmax} & \multicolumn{2}{c}{\mvirp}\\ 
 & 68\% conf. & 95\% conf. & 68\% conf. & 95\% conf.\\
\hline
$\nperi=0$ & $64.0\pm 6.8 \%$ & $90.7\pm 3.9 \%$ & $67.7\pm 6.4 \%$ & $90.7\pm 4.0 \%$ \\
$\nperi=1$ & $69.4\pm 4.8 \%$ & $96.6\pm 1.9 \%$ & $71.6\pm 4.8 \%$ & $98.9\pm 1.1 \%$ \\
$\nperi\geq 2$ & $57.2\pm 5.1 \%$ & $89.6\pm 3.4 \%$ & $71.2\pm 5.0 \%$ & $93.0\pm 2.8 \%$ \\
\hline
$\rperi / R_\mathrm{host, 200m}^{\mathrm{peri}} \geq 0.5$ & $70.2\pm 5.7 \%$ & $93.7\pm 3.1 \%$ & $75.1\pm 5.6 \%$ & $98.4\pm 1.6 \%$\\
$0.2 \leq \rperi / R_\mathrm{host, 200m}^{\mathrm{peri}} < 0.5$ & $63.5\pm 5.4 \%$ & $94.8\pm 2.6 \%$ & $72.9\pm 4.9 \%$ & $96.2\pm 2.2 \%$\\
$\rperi / R_\mathrm{host, 200m}^{\mathrm{peri}} < 0.2$ & $49.5\pm 8.3 \%$ & $88.1\pm 5.4 \%$ & 
$61.4\pm 8.4 \%$ & $91.0\pm 5.0 \%$ \\
\hline
$\rcurr / R_\mathrm{host, 200m} \geq 1$ & $67.3\pm 5.2 \%$ & $92.2\pm 2.9 \%$ & $68.2\pm 5.1 \%$ & $95.4\pm 2.2 \%$ \\
$0.5 \leq \rcurr / R_\mathrm{host, 200m} < 1$ & $67.2\pm 5.2 \%$ & $93.2\pm 2.8 \%$ & $71.6\pm 4.9 \%$ & $96.4\pm 2.0 \%$ \\
$\rcurr / R_\mathrm{host, 200m} < 0.5$ & $54.9\pm 6.9 \%$ & $94.3\pm 3.2 \%$ & $75.4\pm 5.9 \%$ & $94.3\pm 3.1 \%$ \\
\hline
$\tperi / \mathrm{Gyr} > 5$ & $69.5\pm 6.7 \%$ & $96.0\pm 2.9 \%$ & $67.3\pm 6.7 \%$ & $100.0\pm 0.0 \%$ \\
$2 \leq \tperi / \mathrm{Gyr} < 5$ & $61.2\pm 6.2 \%$ & $93.3\pm 3.3 \%$& $77.7\pm 5.4 \%$ & $91.4\pm 3.8 \%$ \\
$\tperi / \mathrm{Gyr} < 2$ & $61.0\pm 6.0 \%$ & $91.1\pm 3.5 \%$ & $68.8\pm 5.5 \%$ & $97.1\pm 2.1 \%$ \\
\hline
\end{tabular}
\end{center}
\caption{Summary of coverage fractions for the 68\% and 95\% confidence intervals for different bins of orbital parameters. The mean and associated errors are reported for each case.}
\label{tab:tidal}
\end{table*}

%% file: table_sidm_cdm.tex
\begin{table*}
\begin{center}
\begin{tabular}{lcccc}
\hline
 & \multicolumn{2}{c}{\vmax} & \multicolumn{2}{c}{\mvir} \\ 
 & 68\% conf. & 95\% conf. & 68\% conf. & 95\% conf. \\
\hline
CDM & $64.5 \pm 6.4\%$ & $93.3 \pm 3.3\%$ & $71.9 \pm 7.6\%$ & $97.2 \pm 2.7\%$ \\
SIDM1 & $62.4 \pm 5.9\%$ & $88.3 \pm 4.0\%$ & $65.4 \pm 5.6\%$ & $89.8 \pm 3.7\%$ \\
SIDM10 & $71.0 \pm 6.6\%$ & $94.3 \pm 3.2\%$ & $71.0 \pm 6.5\%$ & $97.9 \pm 2.0\%$ \\
\hline
\end{tabular}
\end{center}
\caption{Summary of coverage fractions for the 68\% and 95\% confidence intervals for the CDM and SIDM samples. The mean and associated errors are reported for each case.}
\label{tab:sidm_cdm}
\end{table*}

%% file: appendix.tex
\appendix
\makeatletter

\section{Jeans dynamical modeling}
\label{app:jeans}
Let $f(\Vec{x}, \Vec{v})$ be the phase-space distribution function that describes the positions and velocities of tracer stars in a self-gravitating system (i.e. the dwarf galaxy).
Following the derivation from \cite{1980MNRAS.190..873B, 2008gady.book.....B}, we assume the system follows the collisionless Boltzmann equations:
\begin{equation}
    \label{eq:boltzmann}
    \frac{\partial f}{\partial t} + \Vec{v} \frac{\partial f}{\partial \Vec{x}} - \frac{\partial \Phi}{\partial \Vec{x}} \cdot \frac{\partial f}{\partial \Vec{v}} = 0,
\end{equation}
where $\Phi$ is the gravitational potential of the system.
For dwarf galaxies, the potential is dominated by their DM components.
We can thus ignore contributions of the tracer populations and write the potential in terms of the DM enclosed mass or density profiles. 

We will now solve Eq.~\ref{eq:boltzmann} and derive a relation between $\Phi$ and the stellar velocity dispersions. 
Multiplying by a velocity component $v_i$ and integrating over all velocities, Eq.~\ref{eq:boltzmann} becomes:
\begin{equation}
    \frac{\partial}{\partial t}(\nu \braket{v_j}) + \frac{\partial}{\partial x_i} (\nu \braket{v_i v_j}) + \nu \frac{\partial \Phi}{\partial x_j} = 0,
\end{equation}
where $\nu = \int d^3\Vec{v} f(\Vec{x}, \Vec{v})$ is the number density of the tracer stars, and $\braket{}$ denotes the average over the velocities.
Assuming a steady-state solution, we can ignore the first term with the explicit time derivative. 
Then, assuming spherical symmetry and working in a spherical coordinate system centered on the dwarf galaxy $(r, \theta, \phi)$, we may rewrite the equation as:
\begin{equation}
    \label{eq:jeans}
    \frac{1}{\nu} \left[\frac{\partial}{\partial r} (\nu \sigma_r^2) + \frac{2 \beta(r)}{r} (\nu \sigma_r^2)\right] =  -\frac{GM(r)}{r^2},
\end{equation}
where $\sigma_i = \sqrt{\braket{v_i^2} - \braket{v_i}^2} $ denotes the velocity dispersion of the system.  
This is commonly known as the ``spherical Jeans equations''.
Here, we rewrite the gravitational potential as a function of the enclosed mass $M(r)$ using the Poisson equation, i.e., $\Phi = -G M(r) / r$.
We have also introduced the velocity anisotropy term:
\begin{equation}
    \beta(r) = 1 - \frac{\sigma^2_\theta + \sigma^2_\phi}{2\sigma_r^2},
\end{equation}
which effectively captures how the distribution of stellar velocities deviates from isotropy.
The velocity anisotropy $\beta$ ranges from $(-\infty, 1]$, where $\beta=1, 0, -\infty$ indicates a radial, isotropic, and tangential velocity profile, respectively. 

We integrate the spherical Jeans equations (Eq.~\ref{eq:jeans}) to get the radial velocity dispersion profile $\sigma_r(r)$,
\begin{equation}
    \label{eq:veldisp_3d}
    \sigma_r^2(r) = \frac{1}{\nu(r) g(r)} \int_r^\infty g(\Tilde{r}) \frac{G M(\Tilde{r}) \nu(\Tilde{r})} {\Tilde{r}^2} \mathrm{d}\Tilde{r},
\end{equation}
where the function $g(r)$ is
\begin{equation}
    \label{eq:gint}
    g(r) = \exp \left( 2 \int \frac{\beta(\tilde{r})}{\tilde{r}} \mathrm{d}\Tilde{r} \right).
\end{equation}
The radial velocity dispersion profile $\sigma_r(r)$ is a function of the enclosed mass profile $M(r)$ and thus the density profile $\rho(r)$.
From Eq.~\ref{eq:veldisp_3d} and Eq.~\ref{eq:gint}, we see that different combinations of the velocity anisotropy $\beta(r)$ and the enclosed mass profile $M(r)$ can result in the same radial velocity dispersion profile $\sigma_r(r)$.
This manifests as the mass-anisotropy degeneracy, as mentioned, and represents a fundamental limitation of the spherical Jeans equations. 
In practice, because only the projected radii and line-of-sight velocities are available, we project Eq.~\ref{eq:veldisp_3d} using the Abel transformation.
This gives the line-of-sight velocity dispersion profile $\sigma_{\mathrm{los}}(R)$:
\begin{equation}
    \label{eq:veldisp_los}
    \sigmalos^2(R) = \frac{2}{\Sigma_\star(R)} \int_R^\infty \left( 1 - \beta(r) \frac{R^2}{r^2} \right) \frac{\nu(r) \sigma_r^2(r) r}{\sqrt{r - R^2}} \mathrm{d} r,
\end{equation}
where $\Sigma_\star (R)$ is the surface mass density of the tracer stars at the projected radius $R$.
It is simply the projection of $\nu(r)$ along the line-of-sight. 
Note that in Eq.~\ref{eq:veldisp_3d}, we again see the $\beta(r)$ degenerates with $\sigma_r(r)$ and thus $M(r)$.

In the framework of Jeans dynamical modeling, the density profiles $\rho(r)$, velocity anisotropy profiles $\beta(r)$, and the tracer mass density $\Sigma_\star(R)$ are assumed some functional forms with free parameters and fit simultaneously. 
To fit the free parameters, one typically construct the likelihood of the line-of-sight velocity dispersion profile $\sigma_\mathrm{los}(R)$ and apply maximum likelihood estimation (MLE) with Bayesian sampling techniques, e.g. Markov Chain Monte Carlo (MCMC, ~\citealt{2022AnRSA...9..557J, 2019arXiv190912313S}), Nested Sampling~\citep{2004AIPC..735..395S, 2022NRvMP...2...39A}.
The likelihood can be either binned (e.g.,~\citealt{2007PhRvD..75h3526S, 2011MNRAS.418.1526C}) or unbinned (e.g.,~\citealt{2008ApJ...678..614S}).
Traditionally, the Gaussian unbinned likelihood from~\cite{2008ApJ...678..614S} has been commonly used (e.g.,~\citealt{2015ApJ...801...74G}):
\begin{equation}
    \mathcal{L} = \prod_i^{N_\mathrm{star}} 
    \frac{(2\pi)^{-1/2}}{\sqrt{\sigma^2_\mathrm{los}(R_i) + \Delta_i^2}}
    \exp\left[
    -\frac{1}{2} \left(\frac{(v_i - \braket{v})^2}{\sigma^2_\mathrm{LOS}(R_i) + \Delta_i^2}\right)
    \right],
\end{equation}
where $R_i$ is the projected radius, and $v_i$ and $\Delta_i$ are the line-of-sight velocity and its measurement uncertainty of star $i$.
$\braket{v}$ is the mean velocity of the tracer population, and $\sigma^2_\mathrm{los}$ is the intrinsic line-of-sight velocity dispersion given by Eq.~\ref{eq:veldisp_los}.

\section{Additional details on machine learning architecture and training}
\label{app:ml}

As noted, the machine learning architecture is similar to that in N23.
The model consists of a GNN embedding network and a normalizing flow for density estimation. 

During the forward pass, node features are first projected onto a 64-dimensional latent space using a multi-layer perceptron (MLP). The graph is then processed through 3 GATConv layers, each with a hidden size of 128, two attention heads, and a Leaky ReLU activation~\citep{2015arXiv150201852H}. 
Next, we average the node features and pass the result through 4 MLP layers, each with a hidden size of 128 and a Gaussian Error Linear Unit (GELU;~\citealt{2016arXiv160608415H}) activation. 
Each MLP layer is followed by batch normalization and dropout with a rate of 0.4.
The resulting summary feature is a 128-dimensional vector that is invariant to the ordering of the nodes (i.e. permutation-invariant).
Finally, the summary feature is used as the conditioning features for the normalizing flows, which consists of 6 Neural Spline Flows (NSF;~\cite{2019arXiv190604032D}) transformations with 8 knots.
Note that this is another difference from the architecture in N23, which used 4 Masked Autoregressive Transformations (MAF;~\citealt{2015arXiv150203509G, 2017arXiv170507057P}).
We find that while NSFs have similar performance on the validation dataset as MAFs, they better generalize to the FIRE dataset.

During training, we minimize the negative log-likelihood given by the flows. 
Let the embedding network, which includes the projection MLP layer, the 4 GAT layers, and the output 4 MLP layers, be $g_\phi(\mathcal{G}_i)$ where $\phi$ is the trainable parameters and $\mathcal{G}_i$ is the input graph.
The optimization objective is simply:
\begin{equation}
    \label{eq:loss}
    \mathcal{L} = -\log p_\varphi(\theta | g_\phi(\mathcal{G}_i)),
\end{equation}
where $\varphi$ is the trainable parameters of the flows and $\theta$ is the parameters of interest, i.e. the DM and stellar parameters.

We train the feature extractor and the flows simultaneously using the AdamW~\citep{adamw2019, kingma2014adam} gradient descent optimizer with a cosine annealing learning rate scheduler~\citep{2016arXiv160803983L}.
The optimizer has a peak learning rate $5 \times 10^{-4}$ and weight decay coefficient $0.01$.
The scheduler has 40,000 warm-up steps and 80,000 decay steps.
The training batch size is 64. 
As noted, we use $5 \times 10^6$ and $5 \times 10^5$ galaxies for the training and validation set, respectively.
The training converges after approximately 50 epochs or about 22 hours on a NVIDIA Tesla V100.

\section{Additional results}

\subsection{Comparison with Jeans-based methods}
\label{app:comparison}

\begin{figure*}
    \centering
    \includegraphics[width=\linewidth]{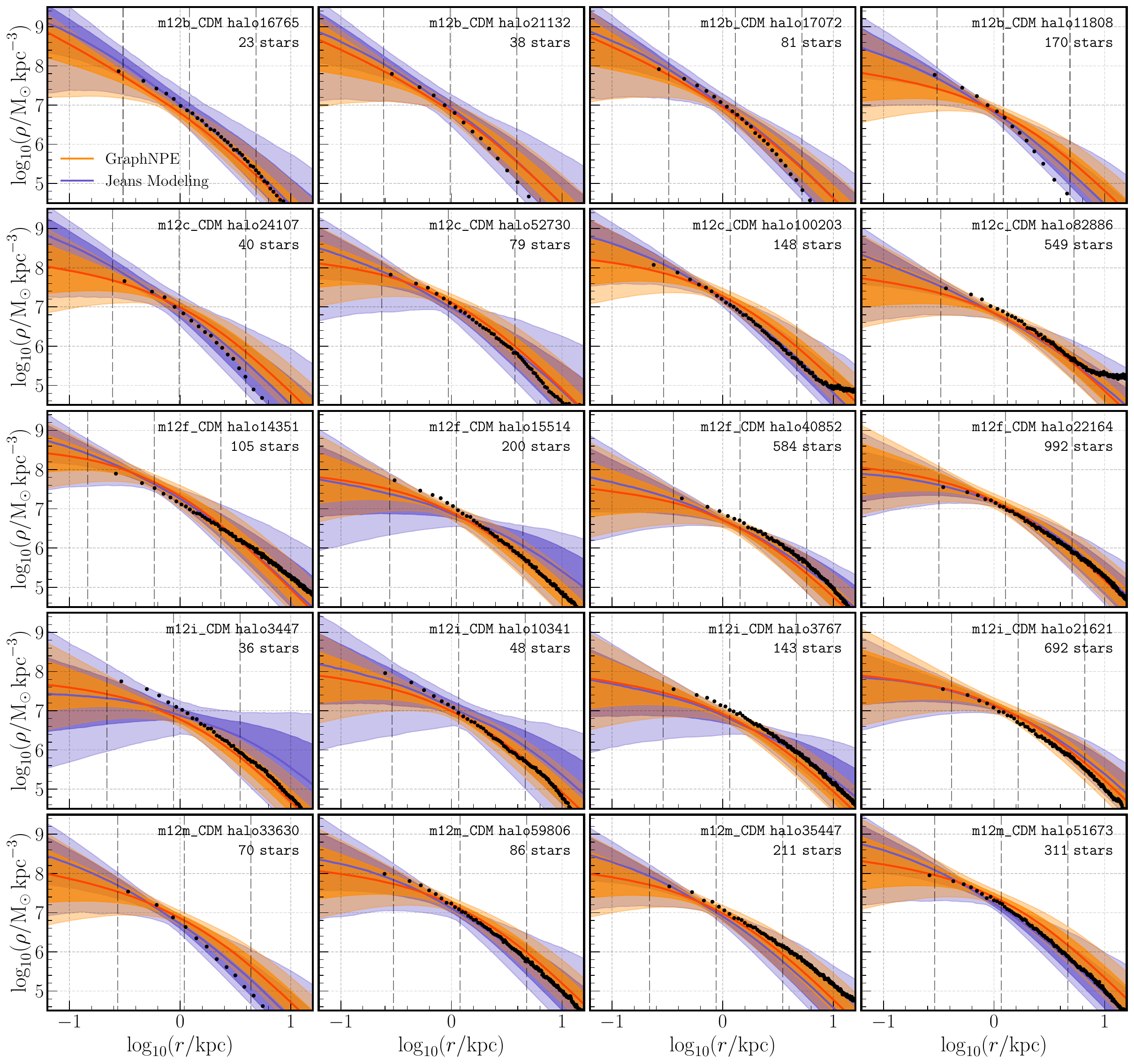}
    \caption{
    Comparison between the inferred DM density profiles $\rhodm(r)$ from \gnn (orange) and Jeans modeling (blue) for the CDM FIRE galaxies. 
    Panels are the same as in Figure~\ref{fig:rho_gnn_jeans}.
    }
    \label{fig:rho_comp_cdm_app}
\end{figure*}

\begin{figure*}
    \centering
    \includegraphics[width=\linewidth]{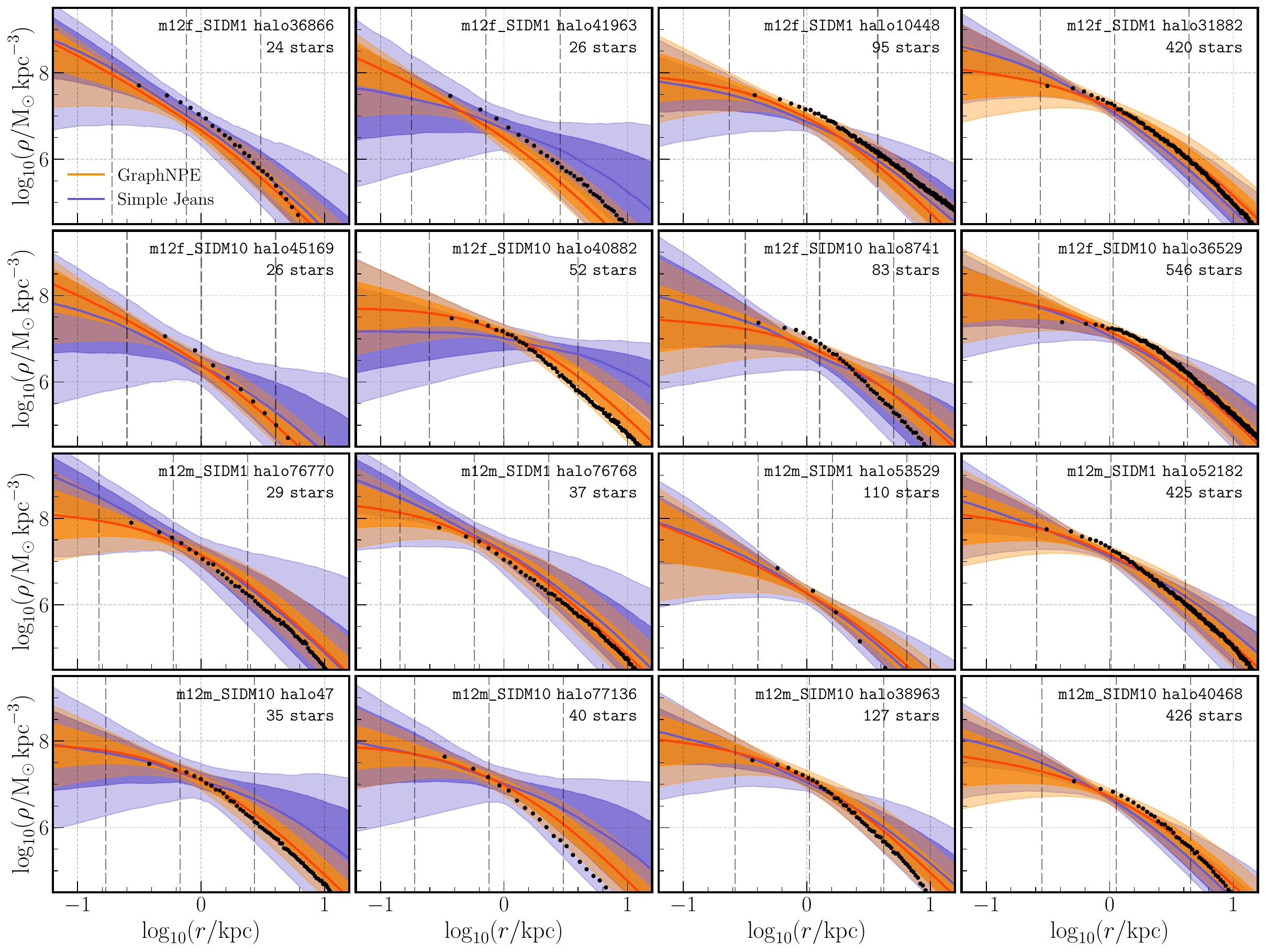}
    \caption{
    Comparison between the inferred DM density profiles $\rhodm(r)$ from \gnn (orange) and Jeans modeling (blue) for the SIDM FIRE galaxies. 
    Panels are the same as in Figure~\ref{fig:rho_gnn_jeans}.}
    \label{fig:rho_comp_sidm_app}
\end{figure*}

\begin{figure*}
    \centering
    \includegraphics[width=\linewidth]{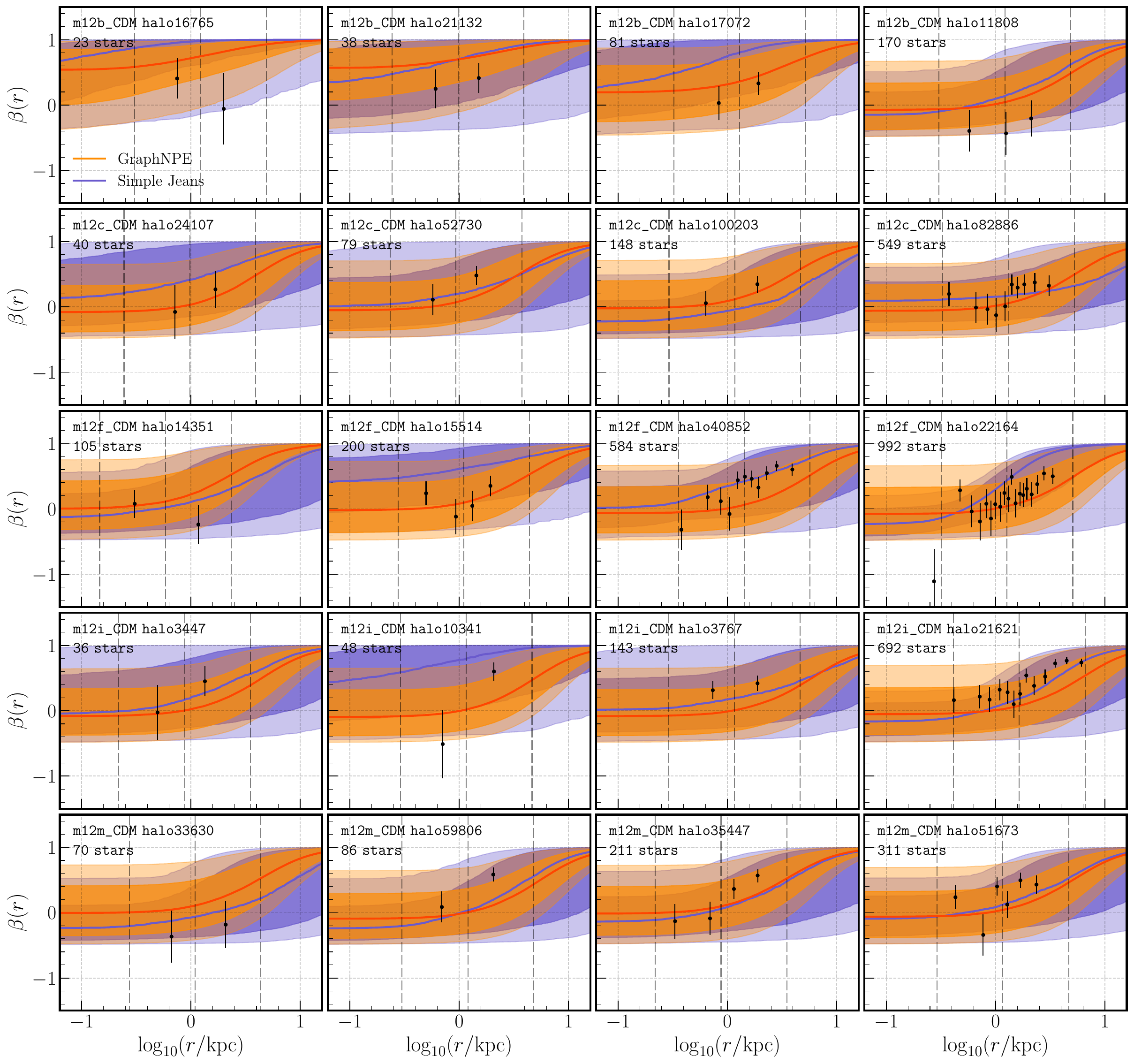}
    \caption{
    Comparison between the inferred velocity anisotropy profiles $\beta(r)$ from \gnn (orange) and Jeans modeling (blue) for the CDM FIRE galaxies. 
    Panels are the same as in Figure~\ref{fig:beta_gnn_jeans}.}
    \label{fig:beta_comp_cdm_app}
\end{figure*}

\begin{figure*}
    \centering
    \includegraphics[width=\linewidth]{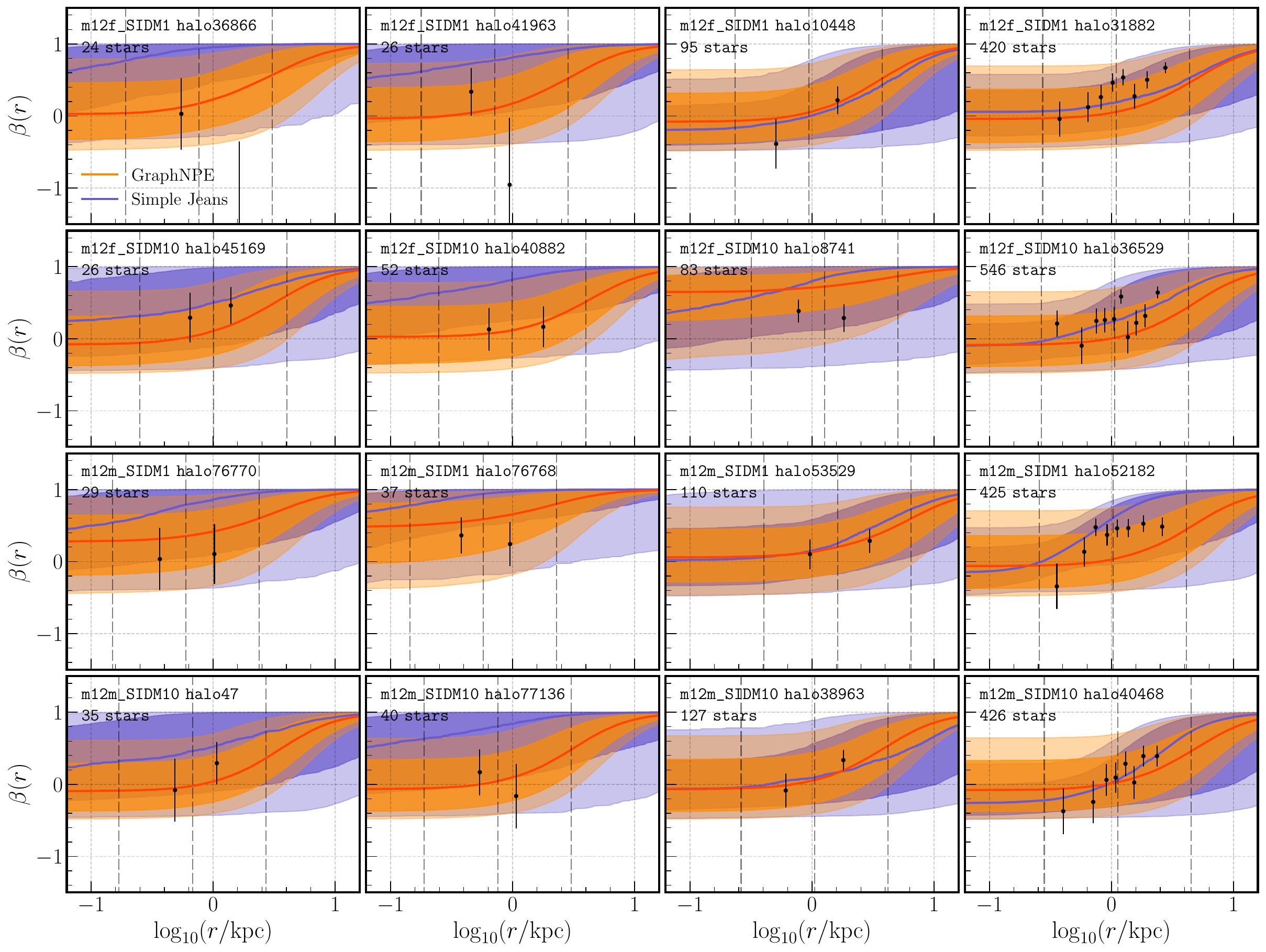}
    \caption{
    Comparison between the inferred DM density profiles $\rhodm(r)$ from \gnn (orange) and \gs (blue) for the SIDM FIRE galaxies. 
    Panels are the same as in Figure~\ref{fig:beta_gnn_jeans}.}
    \label{fig:beta_comp_sidm_app}
\end{figure*}

\begin{figure*}
    \centering
    \includegraphics[width=\linewidth]{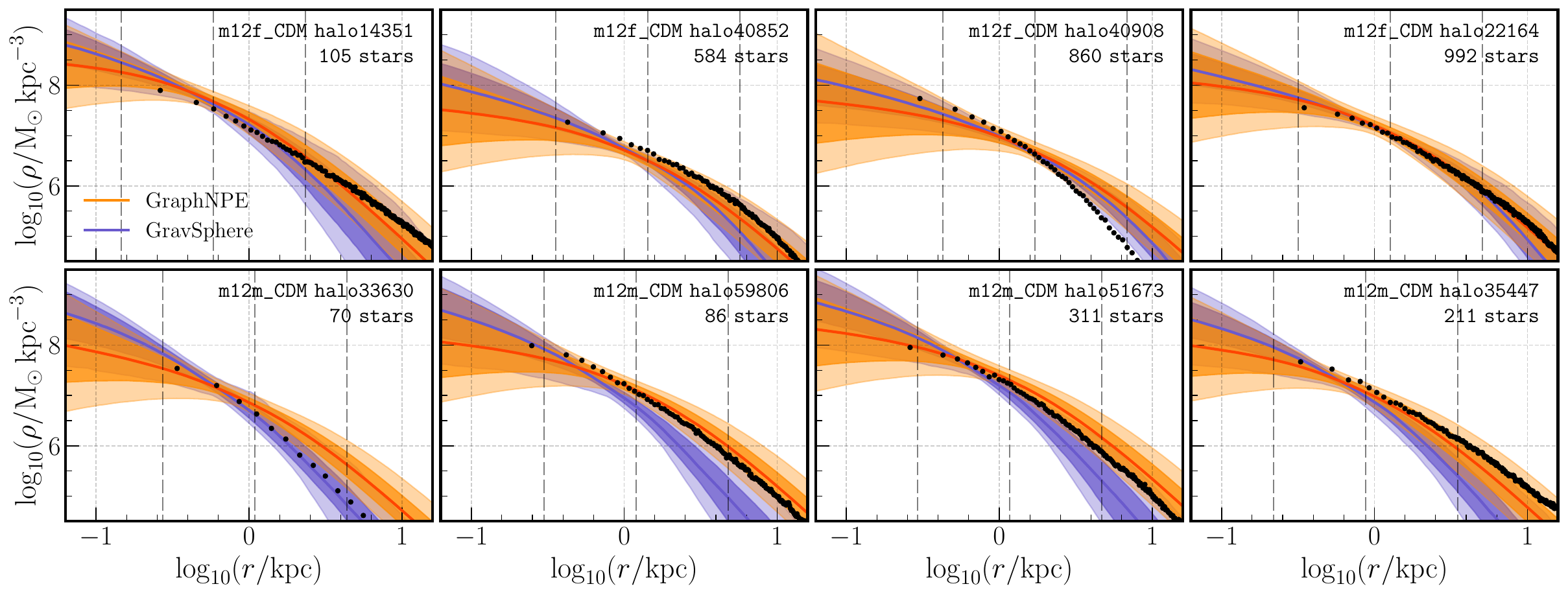}
    \caption{
    Comparison between the inferred DM density profiles $\rhodm(r)$ from \gnn (orange) and \gs (blue) for the CDM FIRE galaxies. 
    Panels are the same as in Figure~\ref{fig:rho_gnn_jeans}.
    }
    \label{fig:rho_comp_gs_cdm_app}
\end{figure*}

\begin{figure*}
    \centering
    \includegraphics[width=\linewidth]{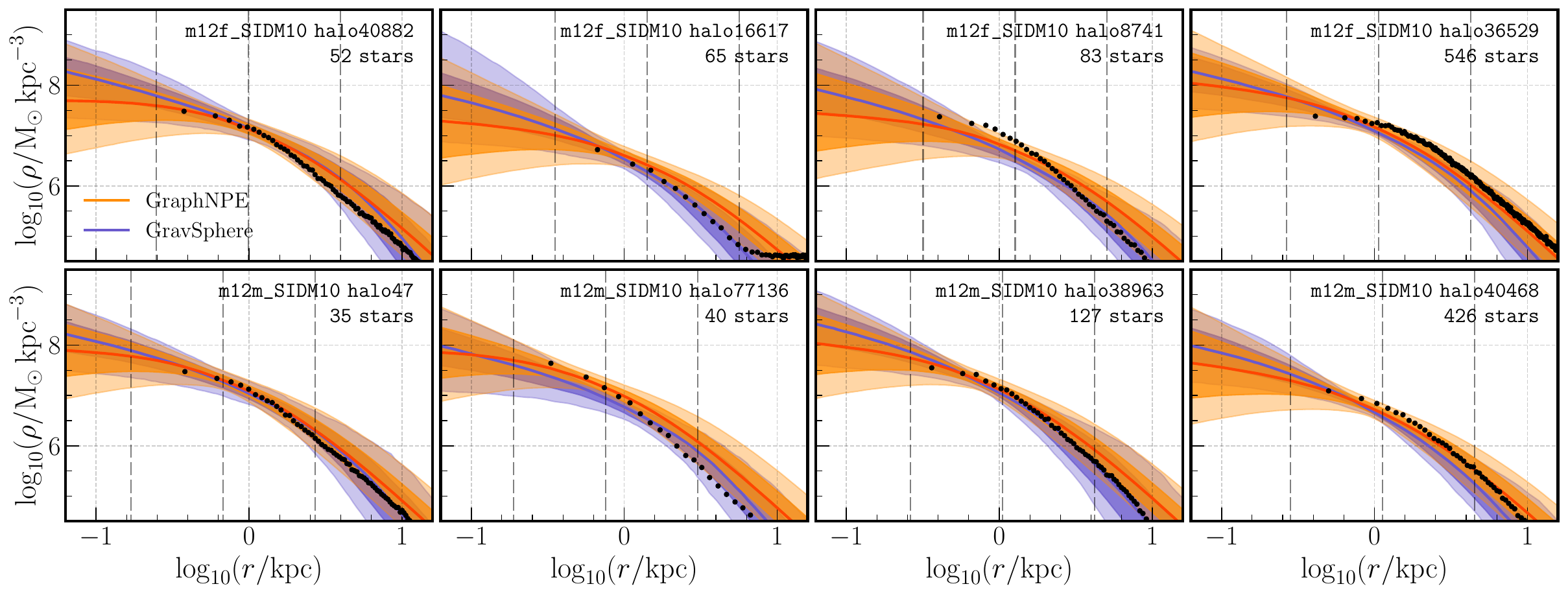}
    \caption{
    Comparison between the inferred DM density profiles $\rhodm(r)$ from \gnn (orange) and \gs (blue) for the SIDM FIRE galaxies. 
    Panels are the same as in Figure~\ref{fig:rho_gnn_jeans}.}
    \label{fig:rho_comp_gs_sidm_app}
\end{figure*}

Figures~\ref{fig:rho_comp_cdm_app} and \ref{fig:rho_comp_sidm_app} compare the inferred DM density profiles $\rho(r)$ for \gnn and Jeans modeling across the full set of selected FIRE galaxies in the CDM and SIDM simulations, respectively. 
Figures~\ref{fig:beta_comp_cdm_app} and \ref{fig:beta_comp_sidm_app} show the velocity anisotropy profiles $\beta(r)$ for the same galaxies. 
Similarly, Figures~\ref{fig:rho_comp_gs_cdm_app} and \ref{fig:rho_comp_gs_sidm_app} compare the inferred $\rho(r)$ profiles for \gnn and \gs in the CDM and SIDM simulations, respectively, while Figures~\ref{fig:beta_comp_cdm_app} and \ref{fig:beta_comp_sidm_app} present the corresponding $\beta(r)$ profiles.


\begin{figure*}
    \centering
    \includegraphics[width=\linewidth]{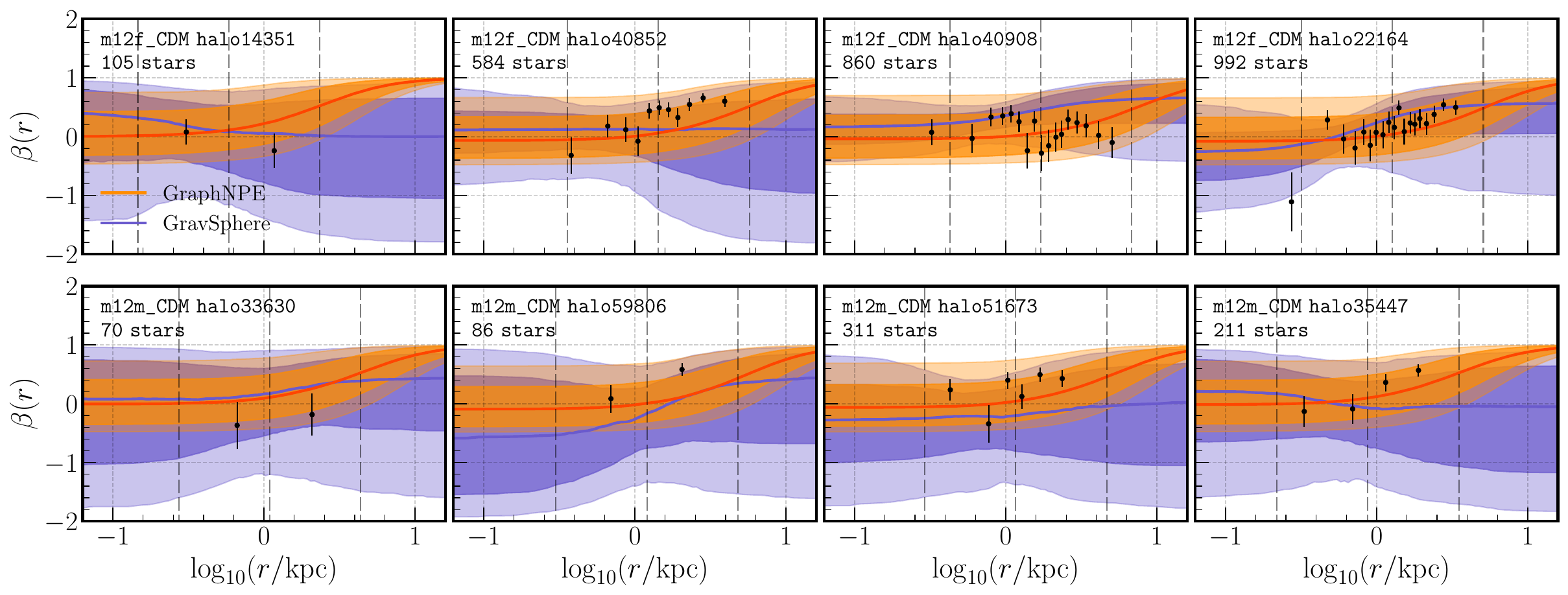}
    \caption{
    Comparison between the inferred velocity anisotropy profiles $\beta(r)$ from \gnn (orange) and \gs (blue) for the CDM FIRE galaxies. 
    Panels are the same as in Figure~\ref{fig:beta_gnn_jeans}.}
    \label{fig:beta_comp_gs_cdm_app}
\end{figure*}
\begin{figure*}
    \centering
    \includegraphics[width=\linewidth]{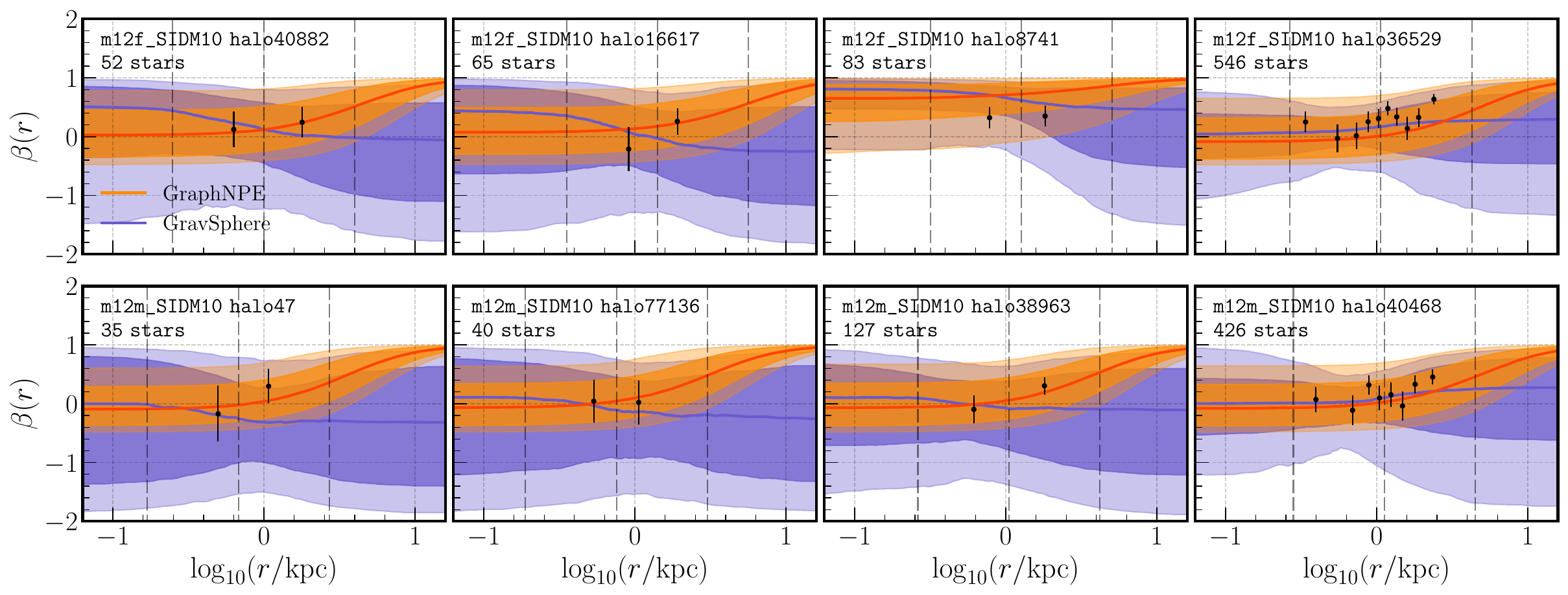}
    \caption{
    Comparison between the inferred velocity anisotropy profiles $\beta(r)$ from \gnn (orange) and \gs (blue) for the SIDM FIRE galaxies. 
    Panels are the same as in Figure~\ref{fig:beta_gnn_jeans}.}
    \label{fig:beta_comp_gs_sidm_app}
\end{figure*}

\subsection{Performance as a function of tracer counts}
\label{app:nstar}

\begin{figure*}
    \centering
    \includegraphics[width=0.4\linewidth]{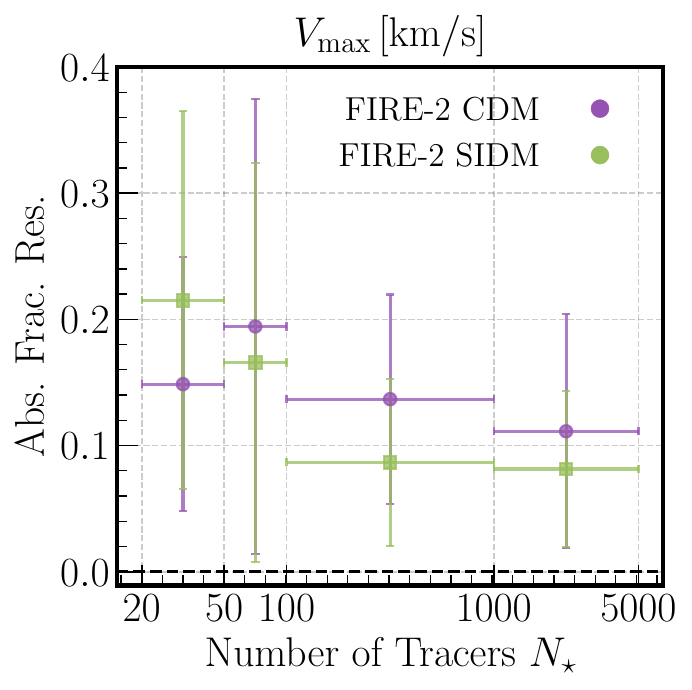}
    \includegraphics[width=0.4\linewidth]{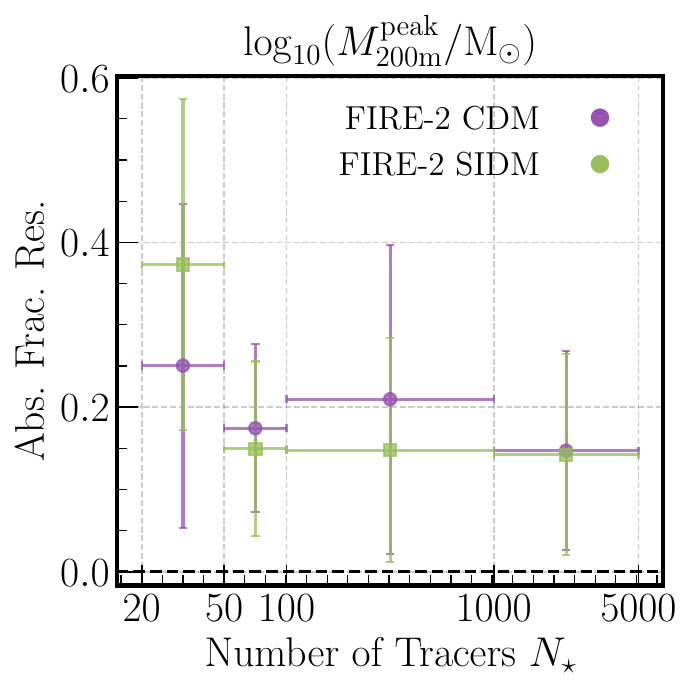}
    \caption{The absolute residuals of \vmax (left) and \mvirp (right) as a function of the tracer counts. FIRE-2 CDM galaxies are shown as purple data points, while SIDM galaxies are shown as green data points. 
    The vertical error bars denote the standard deviation of the absolute residuals in each bin.
    The horizontal error bars denote the bin width. 
    }
    \label{fig:vmax-m200-nstar}
\end{figure*}

We examine the performance of \gnn as a function of tracer counts, $N_\star$.
Figure~\ref{fig:vmax-m200-nstar} presents the absolute residuals of \vmax (left panel) and \mvirp (right panel) across four $N_\star$ bins with edges at $[20, 50, 100, 1000, 5000]$.
We divide the samples into FIRE-2 CDM and SIDM galaxies and show the result separately as purple and green data points. 

The absolute residuals exhibit a clear decline with increasing tracer count $N_\star$.
This trend is similar to that observed with \vmax and \mvirp in Figures~\ref{fig:vmax-residual} and \ref{fig:m200-residual}, which is expected given the strong correlations between $N_\star$ and the galaxy mass (e.g. the stellar-to-halo-mass relation).
For \vmax, residuals are approximately $20\%$ at low tracer counts ($N_\star < 100$), decreasing to a plateau of roughly $10\%$ for $N_\star > 100$.
For \vmax, residuals are approximately $20\%$ at low tracer counts ($N_\star < 100$), decreasing to a plateau of roughly $10\%$ for $N_\star > 100$.
Notably, both \vmax and \mvirp exhibit considerably larger residual scatter in the low tracer count regime.
Finally, we observe that \gnn exhibits marginally superior performance on SIDM galaxies at high tracer counts and on CDM galaxies at low tracer counts, though these differences are small compared to the residual scatter.

\subsection{Tidal disruption}
\label{app:tidal}

\begin{figure*}
    \centering
    \includegraphics[width=\linewidth]{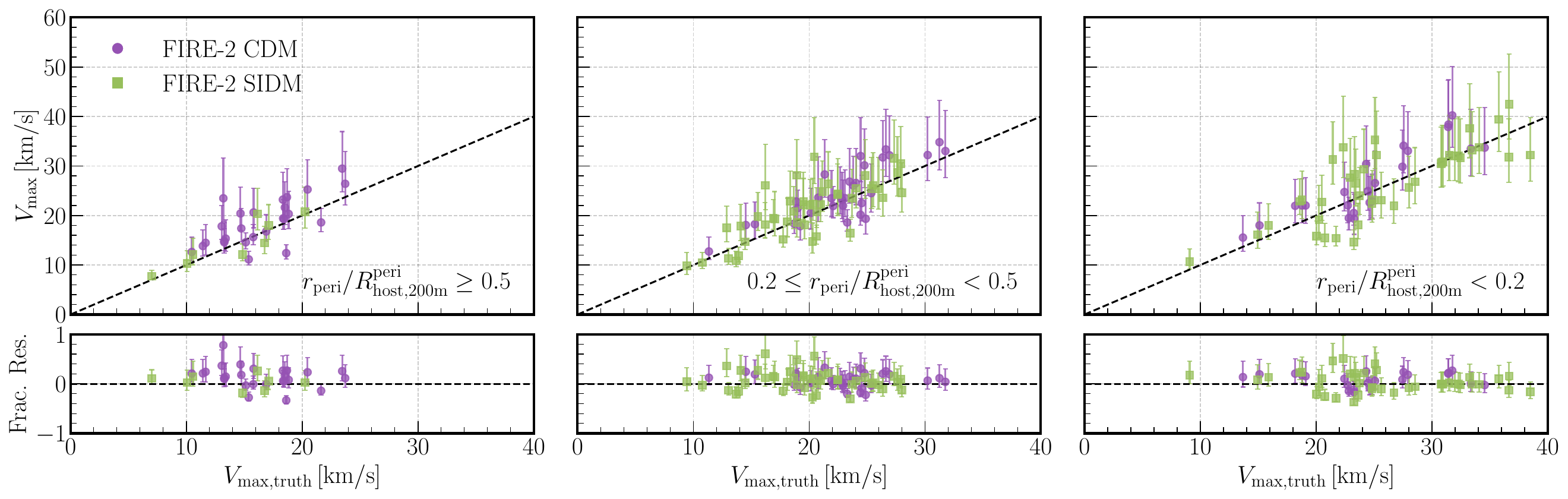}
    \includegraphics[width=\linewidth]{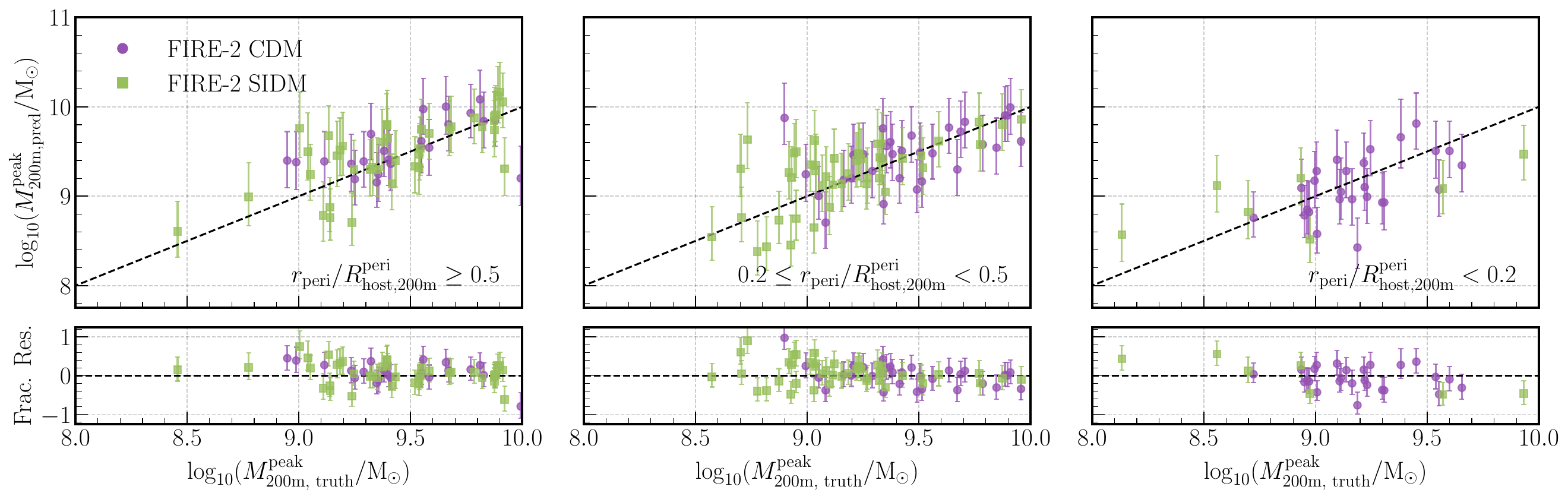}
    \caption{    
    Recovery of the peak circular velocity \vmax and the peak virial mass \mvir of dwarf galaxies in the FIRE dataset.
    Galaxies are grouped by the last pericentric distances \rperi, in units of the host $R_\mathrm{200m}$, with each column corresponding to a different grouping.  
    Panels are the same as Figure~\ref{fig:vmax-nperi}.
    }
    \label{fig:vmax-m200-rperi}
\end{figure*}

\begin{figure*}
    \centering
    \includegraphics[width=\linewidth]{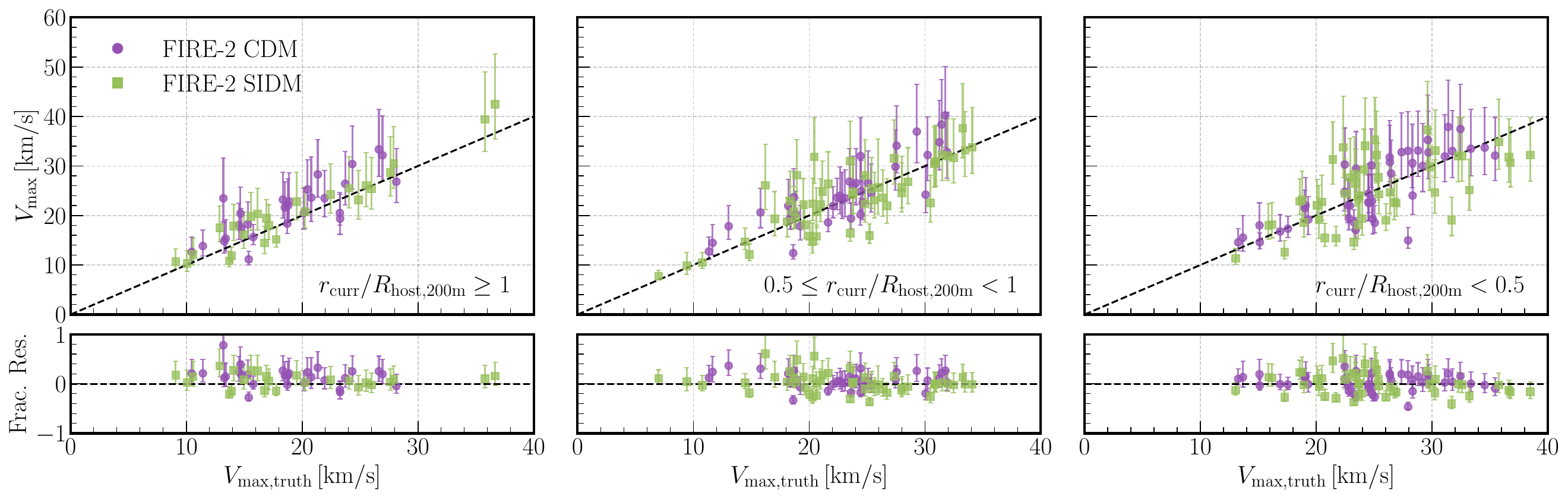}
    \includegraphics[width=\linewidth]{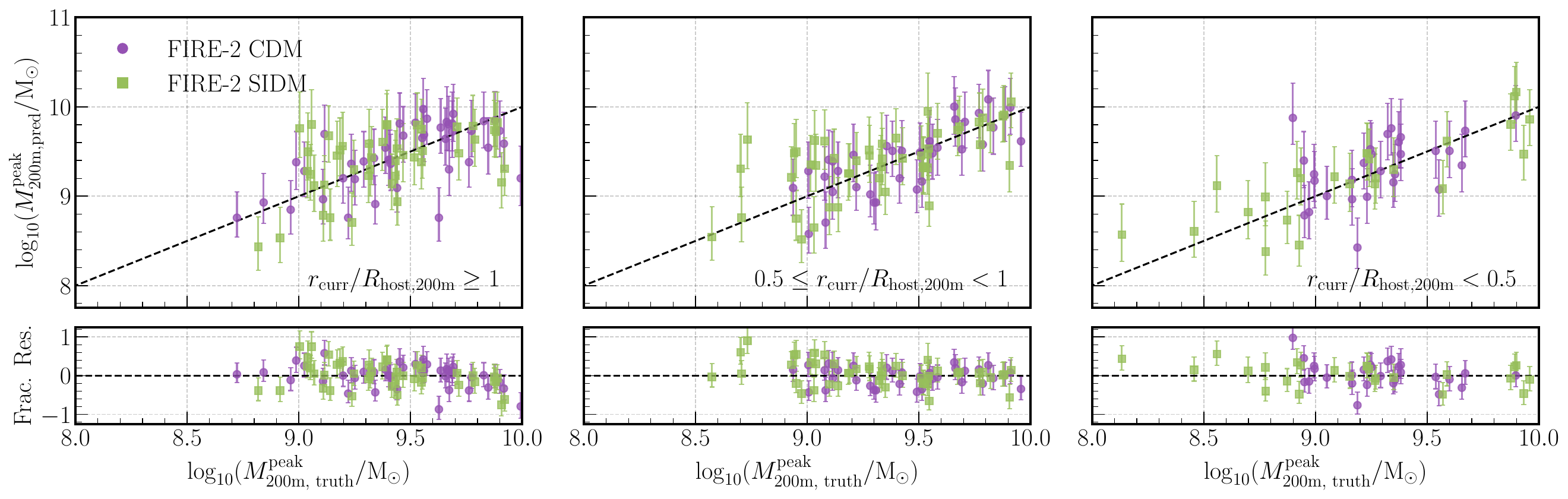}
    \caption{    
    Recovery of the peak circular velocity \vmax and the peak virial mass \mvir of dwarf galaxies in the FIRE dataset.
    Galaxies are grouped by the current distance \rcurr, in units of the host $R_\mathrm{200m}$, with each column corresponding to a different grouping.  
    Panels are the same as Figure~\ref{fig:vmax-nperi}.
    }
    \label{fig:vmax-m200-rcurr}
\end{figure*}

\begin{figure*}
    \centering
    \includegraphics[width=\linewidth]{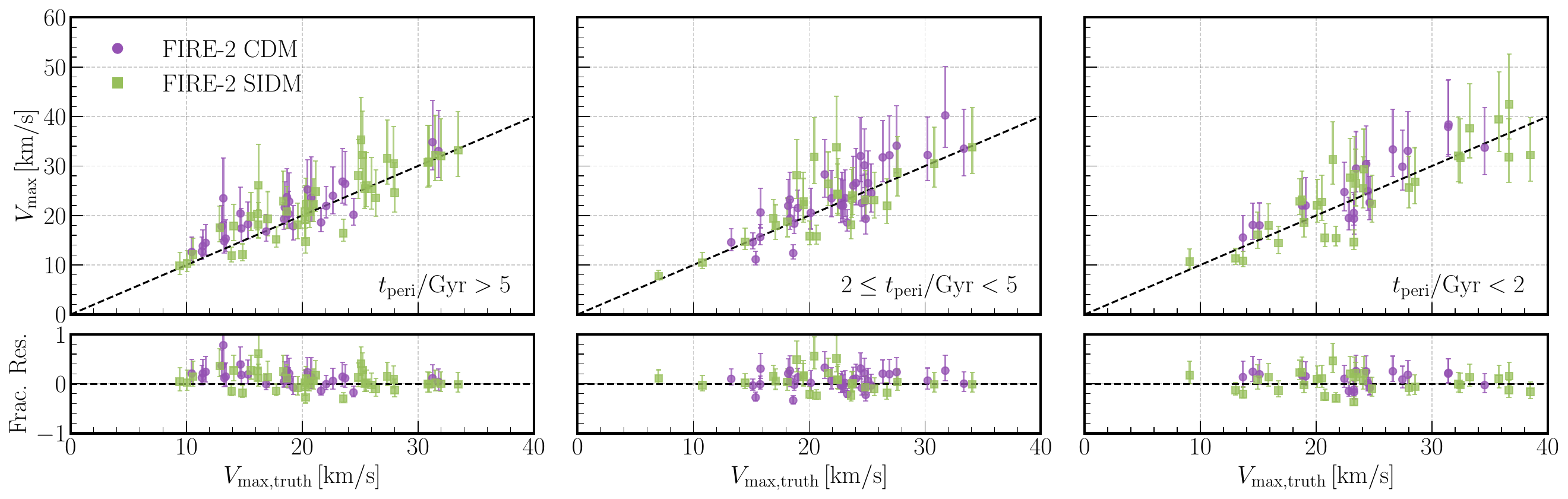}
    \includegraphics[width=\linewidth]{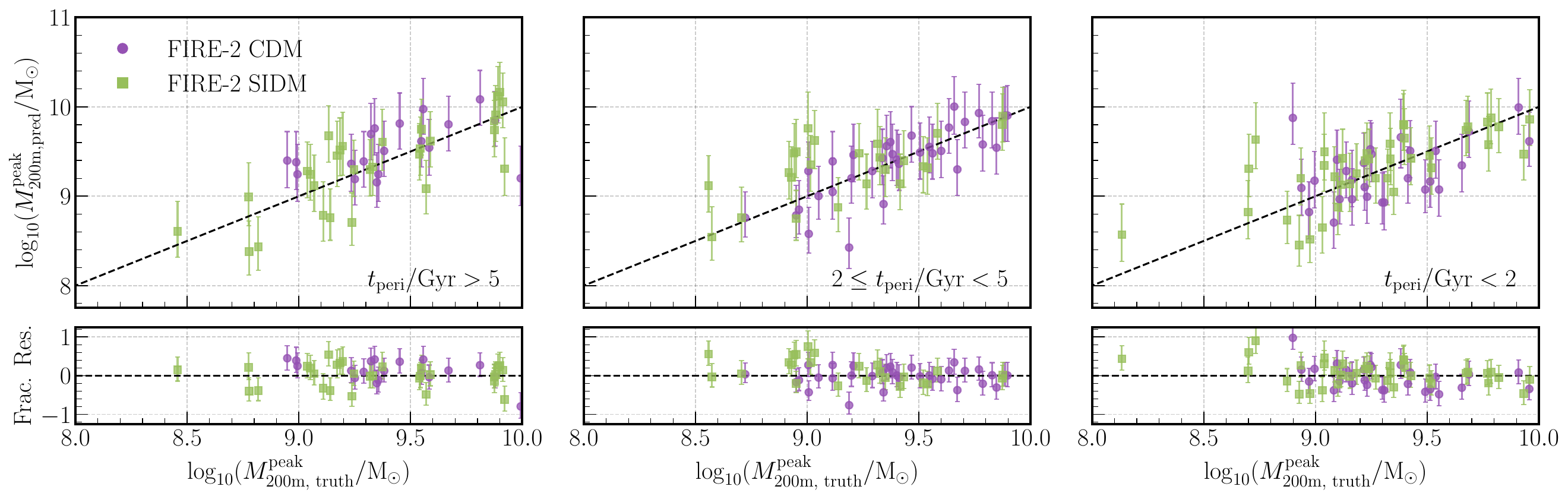}
    \caption{    
    Recovery of the peak circular velocity \vmax and the peak virial mass \mvir of dwarf galaxies in the FIRE dataset.
    Galaxies are grouped by the time since last pericenter \tperi with each column corresponding to a different grouping.  
    Panels are the same as Figure~\ref{fig:vmax-nperi}.
    }
    \label{fig:vmax-m200-tperi}
\end{figure*}

In Section~\ref{section:tidal_nperi}, we show the predicted versus true values for the peak circular velocity \vmax and the peak virial mass \mvir
across three bins of the number of pericentric passages \nperi. 
Here, we show results for three additional orbital parameters: the last pericentric distance \rperi (Figure~\ref{fig:vmax-m200-rperi}), the current distance \rcurr (Figure~\ref{fig:vmax-m200-rcurr}), and the time since last pericenter \tperi (Figure~\ref{fig:vmax-m200-tperi}).

